\let\oldvec\vec
\documentclass[english]{article}    
\let\vec\oldvec

\usepackage[T1]{fontenc}
\usepackage[utf8]{inputenc} 
\usepackage[pdfusetitle,
 bookmarks=true,bookmarksnumbered=false,bookmarksopen=false,
 breaklinks=false,pdfborder={0 0 1},backref=false,
 colorlinks=true,linkcolor=blue,citecolor=blue]
 {hyperref}
\usepackage{geometry}
\usepackage{array}
\usepackage{float}
\usepackage{amsmath}
\usepackage{amssymb}
\usepackage{graphicx}

\makeatletter

\usepackage[numbers,sort&compress]{natbib}

\newcommand\myref{\refstepcounter{equation}\theequation}
\newcommand{\refmyref}[1]{\newcounter{#1}\setcounter{#1}{\theequation}}

\usepackage{booktabs}

\makeatother

\begin{document}

\title{TMD gluon distributions for multiparton processes}

\date{}

\author{M. Bury${}^a$, P. Kotko${}^a$, K. Kutak${}^{a,b}$ \vspace{5pt}\\{\it \small ${}^a$ Institute of Nuclear Physics Polish Academy of Sciences} \\{\it \small   PL-31342 Krakow, Poland}\\
{\it \small ${}^b$ Theoretical Physics Department, CERN,}\\
{\it \small 1211 Geneva 23, Switzerland}
}

\maketitle

\vspace{-25em}
\begin{flushright}
  IFJPAN-IV-2018-17\\
  CERN-TH-2018-208
\end{flushright}
\vspace{25em}

\begin{abstract}
We derive gauge invariant operators entering definitions of the Transverse
Momentum Dependent (TMD) gluon distributions, for all five and six
parton processes. Our calculations utilize color decomposition of
amplitudes in the color flow basis. In addition, we find the general
result for multi-gluon process (with arbitrary number of gluons) at
large $N_{c}$. On phenomenological ground our results may be used
for multi-jet production in the small-$x$ regime, where the TMD gluon
distributions can be derived from the Color Glass Condensate effective theory. 
\end{abstract}

\newpage

\tableofcontents

\newpage

\section{Introduction}

Nowadays, it is an ordinary fact that most processes occurring at high energies
do not involve just one large energy scale. Consequently, the standard
collinear factorization often is not sufficient
or even does not apply. Example of a wide class of such processes
are those involving large \textsl{measurable} internal transverse
momenta of partons. A consistent
theoretical treatment of such processes was initiated by Diakonov,
Dokshitzer and Troyan \citep{Dokshitzer1980} and turned into the
Transverse Momentum Dependent (TMD) factorization \citep{Collins1985,Collins:2011zzd}
(for a recent review see \cite{Angeles-Martinez:2015sea}). This concept,
not only resumes the large logarithms, but also defines,
within the QCD theory, more general (and interesting) objects than
usual collinear parton distribution functions (PDFs) -- the TMD parton
distribution functions. The high energy factorization (or $k_{T}$-factorization)
\citep{Gribov1983,Catani:1990eg,Catani:1994sq,Collins1991} and Color
Glass Condensate (CGC) effective theory \citep{McLerran1994,McLerran2003,Gelis2010,Kovchegov:2012mbw} 
also address transverse momentum dependence of gluons in a hadron, although in somewhat different kinematic regime,
namely in the so-called small-$x$ limit, where the gluonic degrees of freedom dominate due to large logarithms of energy that have to be resummed. 
The collinear-factorization-based Monte Carlo event generators
like Pythia \citep{Sjostrand:2006za} or Herwig \citep{Bahr2008}
are also capable of simulating the semi-hard processes by constructing
explicit parton branching mechanisms, based on the Dokshitzer-Gribov-Lipatov-Altarelli-Parisi
(DGLAP) evolution kernel, \textit{i.e.} the Sudakov form factors (although
they have to implement several model-dependent mechanisms to maintain
the momentum conservation, regulate singularities, etc...). The Cascade
Monte Carlo event generator \citep{Jung:2010si} attacks similar problems
employing the evolution equations which resum also the logarithms
enhanced at \mbox{small $x$}.

Although the applicability of the strict TMD factorization theorems
is limited to few processes only (like Drell-Yan or semi-inclusive
DIS), the basic objects appearing in the formalism -- the TMD parton
distributions -- can be studied in the broader context. They are
defined as the Fourier transforms of the hadronic matrix elements
of bilocal field operators with non-light-like separation. To ensure
the gauge invariance the Wilson links connecting the two space-time
points must be inserted. For the gauge invariance itself the shape
of the links is not relevant. In the TMD factorization however, the
shape of the links is determined by the hard process accompanying
the TMD parton distribution. This happens because the collinear gluons
(to the incoming hadron), which couple to various components of the hard
process have to be considered as a part of the nonperturbative wave
function. They can be resummed into the Wilson links attached to each
external leg by means of the Ward identity. Since the external legs
are connected by certain color matrix, so are the pieces of Wilson
links and this is how the process dependence enters (see \citep{Bomhof:2006dp,Bomhof:2007xt}
for details). For simple processes like Drell-Yan pairs production,
the color flow in the hard process is rather simple because of only
two colored partons. Consequently, the resulting TMD parton distribution
has also simple structure. On the contrary, for processes with several
colored partons one gets multiple nonequivalent structures (including
Wilson loops) which cannot be eliminated by a gauge choice. 

Although, as mentioned, the strict all-order factorization theorems fail for more than two colored partons participating
in the hard collision \citep{Rogers:2010dm}, in the nonlinear small-$x$ regime  
the lowest order TMDs are of great phenomenological importance. 
In Ref.~\citep{Dominguez:2011wm}
a leading power limit of the expressions for dijet production in $pA$
collisions within the CGC was studied. They found that the correlators
of Wilson lines averaged over color sources according to the CGC theory
correspond exactly to the TMD gluon distributions for $2\rightarrow2$
processes, provided the hadronic matrix elements are traded for the
color source averages. Not only the correlators agree, but also the
hard factors. Although it is not known whether this correspondence
survives beyond the leading order, it opened new phenomenological
opportunities to study with better theoretical control semi-hard jets in the gluon saturation domain,
see \citep{Kotko:2015ura,VanHameren2016a,Marquet2016,Kotko2017b,Marquet:2017xwy}.
In particular, in Ref. \citep{Kotko:2015ura} a beyond-leading-power
extension of the TMD factorization for forward dijets in $pA$ collisions
was proposed, such that it coincides with the leading power of CGC
in the dense nucleus regime, and with the all-power high energy factorization
in the dilute nucleus limit. One should understand the notion 'factorization'
here in the following sense. First, the overall kinematic conditions
justify so called hybrid approach \citep{Dumitru:2005gt}, \textit{i.e.} where the
projectile proton is treated as a dilute state so that an average parton coming
from it is a large-$x$ parton modeled from ordinary collinear parton
distribution functions. The target nucleus is probed in the dense state,
so it is modeled basing on the small-$x$ dynamics (note that the
operator definitions of the TMD distributions formally are valid also
at small $x$). Second, it is a generalized factorization, \textit{i.e.} the
formulae involve several TMD gluon distributions for nucleus. 
In the formal leading-power TMD factorization, even the generalized factorization breaks,
because one is unable to define the separate correlators whilst more than two colored partons are present \citep{Rogers:2010dm}. In the small-$x$ approach for dilute-dense collisions described above, however, 
 there is only one correlator with transverse separation. Therefore the complications leading
to the lack of possibility to separate Wilson links into TMD operators, formally do not appear
here. Outside the small-$x$ limit for dilute-dense collisions these results might also be useful: for example to assess the factorization breaking effects.

In the TMD factorization formalism the TMD parton distributions
have operator definitions and evolve according to the renormalization
group equations \citep{Collins:2011zzd}. In the small $x$ regime with gluon saturation playing significant role, which is of main interest in the context of this work, the evolution equations are nonlinear and thus more complicated than the equations at moderate $x$ \citep{Balitsky:2015qba,Balitsky:2016dgz,Kovchegov:2018zeq}. In addition, the program of obtaining the renormalization group evolution equations for all possible TMD operators is nowhere near the end.
 Hopefully, the correspondence of the small-$x$ TMD gluon distributions and CGC correlators \citep{Dominguez:2011wm} allows for 
a treatment of evolution in the strict small $x$ limit using the Balitsky-Jalilian-Marian-Iancu-McLerran-Weigert-Leonidov-Kovner (B-JIMWLK) equations \citep{Balitsky1996,Jalilian-Marian1997,JalilianMarian:1997gr,Iancu:2000hn,Iancu:2001ad,Ferreiro:2001qy,Weigert:2000gi} following Ref.~\citep{Marquet2016}.
At small $x$, but in the linear regime, where the saturation scale is much smaller  than the typical scale of the internal transverse momenta, it seems that the various TMD gluon distributions converge to one universal distribution, which may be identified with the so-called unintegrated gluon distribution \citep{VanHameren2016a,Marquet2016,Marquet:2017xwy}. This object is much better understood  and constrained from data. There are several approaches to their evolution. First, there are extensions of the original Balitsky-Fadin-Kuraev-Lipatov (BFKL) equation (see \textit{e.g.} \citep{Lipatov:1996ts} for a review) like the Catani-Ciafaloni-Fiorani-Marchesini (CCFM) equation \citep{Ciafaloni:1987ur,Catani:1989sg,Catani:1989yc,CCFMd}, Kwieci\'nski-Martin-Sta\'sto (KMS) equation \citep{Kwiecinski:1997ee} or the Kimber-Martin-Ryskin (KMR) approach \citep{Kimber:2001sc}. As the linear evolution can be solved through the explicit branching process, it allows for a natural determination of the unintegrated PDFs from Monte Carlo simulations \citep{Hautmann:2017xtx,Hautmann:2017fcj,Martinez:2018jxt}.
The complete set of evolution equations in the linear regime can be also derived through the projector method \citep{Curci1980}, see \citep{Gituliar:2015agu,Hentschinski:2017ayz}
for a recent approach. 
There have been many
calculations attacking various processes where the usage of unintegrated parton distributions is important, see for example \citep{Deak:2009xt,Deak2010,Deak:2011ga,Kutak:2012rf,Nefedov:2013ywa,VanHameren2013,vanHameren:2014lna,vanHameren:2014ala,Dooling:2014kia,VanHameren2015,Bury:2017jxo,Deak:2018obv} where mostly forward jet observables in hadroproduction were addressed.
These calculations, however, use universal unintegrated gluon distributions,
the same for any color flow. While in the linear regime or in certain
phase space regions, this is a good approximation, it is definitely
not the case in the region where the gluon saturation may dominate \citep{Marquet:2007vb,Dominguez:2011wm}.

Motivated by the phenomenological usability of the non-universal TMD
gluon distributions discussed above (and demonstrated in \citep{VanHameren2016a}), we will
present explicit results for the operator structures for all five
and six colored parton processes. Instead of working with particular
Feynman diagrams and calculating the corresponding operator structure,
we choose to work with color decomposition of amplitudes (see \textit{e.g.}
\citep{Mangano:1990by}). This is motivated simply by the way the
amplitudes are calculated at present in practice. Such a procedure
for the operator structures in the TMD gluon distributions was for
the first time used in \citep{Kotko:2015ura} for four parton processes.

The paper is organized as follows. We will start with definitions
of the TMD parton distribution functions and summary of color decomposition
of scattering amplitudes (Section~\ref{sec:Prelim}). Next, in Section~\ref{sec:ColorFlowFR},
we will introduce the color flow diagrams for the operators appearing
in the definitions of the TMD distributions. Basing on these rules,
we list all structures appearing in arbitrary process in Section~\ref{sec:Operators}.
The explicit results for four, five and six parton processes will
be given in Section~\ref{sec:Results}. We will summarize our work
in Section~\ref{sec:Summary}.

\section{Preliminaries }

\label{sec:Prelim}

We start off by providing necessary definitions and conventions for
the TMD gluon distributions and color decomposition. 

We adopt the light cone basis defined using two null four vectors
$n=\left(1,0,0,-1\right)/\sqrt{2}$ and $\tilde{n}=\left(1,0,0,1\right)/\sqrt{2}$.
They define the 'plus' and 'minus' components of a four vector $v$:
$v^{+}=n\cdot v$, $v^{-}=\tilde{n}\cdot v$, so that the four vector
has a decomposition
\begin{equation}
v^{\mu}=v^{+}\,\tilde{n}^{\mu}+v^{-}\,n^{\mu}+v_{T}^{\mu}\,.
\end{equation}
The light-cone coordinates are $\left(v^{+},v^{-},\vec{v}_{T}\right)$,
where the Euclidean transverse vector is defined (in canonical coordinates)
as $v_{T}^{\mu}=\left(0,\vec{v}_{T},0\right)$.

We consider $n$-parton processes with a gluon in the initial state
\begin{equation}
g\left(k_{1}\right)+b_{n}\left(k_{n}\right)\rightarrow b_{2}\left(k_{2}\right)+\dots+b_{n-1}\left(k_{n-1}\right)\,,
\end{equation}
where the partons $b_{i}$ can be quarks or gluons (restricted by
the flavor number conservation of course). The initial state gluon
with momentum $k_{1}$ carries an $x$ fraction of the parent hadron
with momentum $P^{\mu}=P^{+}\tilde{n}^{\mu}$:
\begin{equation}
k_{1}^{\mu}\simeq xP^{\mu}+k_{T}^{\mu}\,.
\end{equation}
Above, the minus component is suppressed as it is neglected in the
hard part. The transverse component is also neglected within the leading
twist collinear and TMD factorization. In the more general case, the gluon
may be off-shell and a suitable redefinition of the hard process is
required to maintain the gauge invariance (see \textit{e.g.} \citep{VanHameren2012,VanHameren2013a,Kotko2014a,vanHameren:2015bba,vanHameren:2017hxx}).
Even then, at least formally, the principles to obtain the TMD distributions
still hold, therefore we shall not distinguish these situations here. 

\subsection{TMD gluon distributions}

\label{subsec:TMD_distr}

In the present work we will be concerned with the \textit{gluon} TMD
distributions as explained in the Introduction. A generic distribution
is defined as the following matrix element: 
\begin{equation}
\mathcal{F}\left(x,k_{T}\right)=2\int\frac{d\xi^{-}d^{2}\xi_{T}}{\left(2\pi\right)^{3}P^{+}}\,e^{ixP^{+}\xi^{-}-i\vec{k}_{T}\cdot\vec{\xi}_{T}}\,\left\langle P\right|\mathrm{Tr}\left\{ \hat{F}^{i+}\left(0\right)\mathcal{U}_{C_{1}}\hat{F}^{i+}\left(\xi^{+}=0,\xi^{-},\vec{\xi}_{T}\right)\mathcal{U}_{C_{2}}\right\} \left|P\right\rangle \,,\label{eq:GenericTMD}
\end{equation}
where $\left|P\right\rangle $ is a hadron state, $\hat{F}^{\mu\nu}\left(x\right)=F_{a}^{\mu\nu}\left(x\right)t^{a}$
is the $\mathrm{SU}\left(N_{c}\right)$ algebra-valued field strength
tensor (we use $\mathrm{Tr}\left(t^{a}t^{b}\right)=T_{F}\delta^{ab}$,
$T_{F}=1/2$ convention for the generators), $\mathcal{U}_{C_{1}}$, $\mathcal{U}_{C_{2}}$
are certain fundamental representation Wilson lines joining space-time
points $\left(\xi^{+}=0,\xi^{-}=0,\vec{\xi}_{T}=\vec{0}_{T}\right)$
and $\left(\xi^{+}=0,\xi^{-},\vec{\xi}_{T}\right)$, multiplied by
possible traces of Wilson loops. The exact shape of Wilson lines will
depend on the hard process coupled to the TMD. Their calculation for
multiple partons is the main goal of the present work.

The above generic definition represents a bare TMD distribution. In QCD there are divergences, in particular the rapidity divergence that have to be regulated. In the present work we will not be considering the renormalization of these operators. Recent studies of that matter in the small-$x$ limit which mainly motivates the present work are given in \citep{Balitsky:2015qba,Balitsky:2016dgz,Kovchegov:2018zeq}.

A generic Wilson line joining $x$ and $y$ through a path $C$ is
defined as
\begin{equation}
\mathcal{U}_{C}=\mathcal{P}\exp\left\{ -ig\int_{C}dz_{\mu}\hat{A}^{\mu}\left(z\right)\right\} \,.\label{eq:WilsonLineDef}
\end{equation}
The Wilson line can be defined also in the adjoint representation,
by replacing generators $t^{a}$ by $\left(T^{a}\right)_{bc}=-if^{abc}$.
In the case where the path is a straight line segment, we will use
the following notation
\begin{equation}
\mathcal{U}_{C}\equiv\left[x,y\right]\,.
\end{equation}

\subsection{Color decomposition}

\label{subsec:ColorDecomp}

The calculation of the operator structure entering the TMD distributions
is nicely systematized not by considering a particular diagrams, but
rather by considering various color flows in the amplitude (squared)
under consideration. Such systematization is achieved by using gauge
invariant decomposition of amplitudes into so-called color-ordered
amplitudes (called also partial, or dual amplitudes). Here we are
presenting only necessary definitions and properties, see \textit{e.g.}
\citep{Mangano:1990by} for a complete review. 

Let us start with pure gluonic tree-level amplitudes. For the sake
of this section we assume that all the momenta are outgoing (later,
it will become necessary to distinguish incoming and outgoing legs).
The most standard decomposition reads
\begin{equation}
\mathcal{M}^{a_{1}\dots a_{n}}\left(k_{1},\dots,k_{n}\right)=\sum_{\pi\in S_{n}/Z_{n}}\mathrm{Tr}\left(t^{a_{\pi\left(1\right)}}\dots t^{a_{\pi\left(n\right)}}\right)\,\mathcal{A}\left(\pi\left(1\right),\dots,\pi\left(n\right)\right)\,,\label{eq:glue_color_decomp_fund}
\end{equation}
where the sum runs over all noncyclic permutations $\pi$ of an $n$-element
set. Three important properties of the above decomposition are: i)
the partial amplitudes $\mathcal{A}$ are gauge invariant, ii) the
partial amplitudes contain only planar diagrams; consequently the
full amplitude squared satisfies $\left|\mathcal{M}\right|^{2}=C\sum_{S_{n-1}}\left|\mathcal{A}\left(1,\mathcal{\pi}\left(2\right),\dots,\pi\left(n\right)\right)\right|^{2}+\mathcal{O}\left(1/N_{c}^{2}\right)$,
with $C$ being a color factor, iii) the amplitudes $\mathcal{A}$
satisfy so-called Ward identities: $\mathcal{A}\left(1,\dots,n\right)+\mathcal{A}\left(1,\dots,n,n-1\right)+\dots+\mathcal{A}\left(1,n,2,\dots\right)=0$
(and similar for other partial amplitudes). Because of the last property,
sometimes more desirable is a decomposition which utilizes only $\left(n-2\right)!$
independent partial amplitudes, instead of $\left(n-1\right)!$ as
in the fundamental-representation (\ref{eq:glue_color_decomp_fund}).
Such decomposition uses the adjoint generators \citep{DelDuca1999}:
\begin{equation}
\mathcal{M}^{a_{1}\dots a_{n}}\left(k_{1},\dots,k_{n}\right)=\frac{1}{2}\sum_{\pi\in S_{n-2}}\left(T^{a_{\pi\left(2\right)}}\dots T^{a_{\pi\left(n-1\right)}}\right)_{a_{1}a_{n}}\,\mathcal{A}\left(1,\pi\left(2\right),\dots,\pi\left(n-1\right),n\right)\,,\label{eq:glue_color_decomp_adjoint}
\end{equation}
with $\left(T^{a}\right)_{bc}=-if^{abc}$. The partial amplitudes
above are the same as in the fundamental-representation decomposition.

Finally, let us recall the so-called color flow decomposition \citep{Maltoni:2002mq}.
It will be useful especially for processes with quarks as it treats
gluons and quarks on equal footing. The basic idea is to work with
the gluon fields as the elements of the $\mathrm{SU}\left(N_{c}\right)$
algebra, \textit{i.e.} matrices $\hat{A}_{j}^{i}\equiv A_{a}\left(t^{a}\right)_{j}^{i}$.
That is, a gluon is characterized by a pair of fundamental and anti-fundamental
representation indices $i,j=\left\{ 1,\dots,N_{c}\right\} $. In this
representation, the amplitude can be decomposed as
\begin{equation}
\mathcal{M}_{j_{1}\dots j_{n}}^{i_{1}\dots i_{n}}\left(k_{1},\dots,k_{n}\right)=2^{-n/2}\sum_{\pi\in S_{n-1}}\delta_{j_{\pi\left(2\right)}}^{i_{1}}\delta_{j_{\pi\left(3\right)}}^{i_{\pi\left(2\right)}}\delta_{j_{\pi\left(4\right)}}^{i_{\pi\left(3\right)}}\dots\delta_{j_{1}}^{i_{\pi\left(n\right)}}\mathcal{A}\left(1,\pi\left(2\right),\dots,\pi\left(n\right)\right),\label{eq:glue_color_flow_decomp}
\end{equation}
again with exactly the same partial amplitudes as in the other two
representations.

For processes with quarks, we use the color flow decomposition as
it treats the quarks and gluons uniformly, and is best for easy calculation
of the TMD operator structures. The decomposition for a process with
one quark--anti-quark pair, 
\[
g\left(k_{1}\right)q\left(k_{2}\right)g\left(k_{3}\right)\dots g\left(k_{n-1}\right)\bar{q}\left(k_{n}\right)\rightarrow\emptyset\,,
\]
reads:
\begin{multline}
\mathcal{M}_{j_{1}j_{3}\dots j_{n-1}j_{n}^{\bar{q}}}^{i_{1}i_{2}^{q}i_{3}\dots i_{n-1}}\left(k_{1},\dots,k_{n}\right)=2^{-(n-2)/2}\sum_{\pi\in S_{n-2}}\delta_{j_{\pi\left(1\right)}}^{i_{2}^{q}}\delta_{j_{\pi\left(3\right)}}^{i_{\pi\left(1\right)}}\delta_{j_{\pi\left(4\right)}}^{i_{\pi\left(3\right)}}\dots\delta_{j_{n}^{\bar{q}}}^{i_{\pi\left(n-1\right)}}\\
\mathcal{A}\left(2^{q},\pi\left(1\right),\pi\left(3\right),\dots,\pi\left(n-1\right),n^{\bar{q}}\right)\,.\label{eq:qq_color_flow_decomp}
\end{multline}
Above we have put superscripts $q$, $\bar{q}$ to remind which
indices belong to a quark (anti-quark). The decomposition for a process
with two quark--anti-quark pairs, 
\[
g\left(k_{1}\right)q\left(k_{2}\right)\bar{q}\left(k_{3}\right)q\left(k_{4}\right)g\left(k_{5}\right)\dots g\left(k_{n-1}\right)\bar{q}\left(k_{n}\right)\rightarrow\emptyset\,,
\]
reads:
\begin{multline}
\mathcal{M}_{j_{1}j_{3}^{\bar{q}}\dots j_{n-1}j_{n}^{\bar{q}}}^{i_{1}i_{2}^{q}i_{4}^{q}i_{5}\dots i_{n-1}}\left(k_{1},\dots,k_{n}\right)=2^{-(n-4)/2}\bigg[\sum_{\pi\in S_{n-3}}\delta_{j_{\pi\left(1\right)}}^{i_{2}^{q}}\delta_{j_{\pi\left(5\right)}}^{i_{\pi\left(1\right)}}\delta_{j_{\pi\left(6\right)}}^{i_{\pi\left(5\right)}}\dots\delta_{j_{\{n}^{\bar{q}}}^{i_{\pi\left(n-1\right)}}\delta_{j_{3}^{\bar{q}}}^{i_{4\}}^{q}}\\\mathcal{A}\left(2^{q},\pi\left(1\right),\pi\left(5\right),\dots,\pi\left(n-1\right),\pi\left(\left\{ n^{\bar{q}},4^{q}\right\} \right),3^{\bar{q}}\right)\\-\frac{1}{N_{c}}\sum_{\pi\in S_{n-3}}\delta_{j_{\pi\left(1\right)}}^{i_{2}^{q}}\delta_{j_{\pi\left(5\right)}}^{i_{\pi\left(1\right)}}\delta_{j_{\pi\left(6\right)}}^{i_{\pi\left(5\right)}}\dots\delta_{j_{\{3}^{\bar{q}}}^{i_{\pi\left(n-1\right)}}\delta_{j_{n}^{\bar{q}}}^{i_{4\}}^{q}}\mathcal{A}\left(2^{q},\pi\left(1\right),\pi\left(5\right),\dots,\pi\left(n-1\right),\pi\left(\left\{ 3^{\bar{q}},4^{q}\right\} \right),n^{\bar{q}}\right)\bigg]\,.
\label{eq:qqqq_color_flow_decomp}
\end{multline}
In the decomposition above, the first sum runs over all permutations
of the $n-4$ gluons and a quark--anti-quark pair (the curly brackets
in deltas denote that the enclosed indices should be permuted together, according to the permutation $\pi$),
while the second sum runs over various partitions of the two quark--anti-quark
pairs with gluon insertions. The second sum is genuinely suppressed
by $1/N_{c}$ in case of distinct quark--anti-quark pairs; for identical
pairs subleading terms will contribute to both sums in the partial
amplitudes. In the present work, we shall explicitly consider processes
with up to 6 partons, thus we do not give decomposition for more quark--anti-quark
pairs.

In a sense, there is a price for the simplicity of the color flow
decomposition. Namely, to each final state gluon we have to apply
the projector
\begin{equation}
\mathcal{P}_{jj'}^{ii'}=\delta^{ii'}\delta_{jj'}-\frac{1}{N_{c}}\delta_{j}^{i}\delta_{j'}^{i'}\,,\label{eq:Projector}
\end{equation}
which removes the redundant degrees of freedom from the sum over colors.
For pure gluon amplitude they are actually not needed, but must be
applied to the quark amplitudes.

\section{Color flow Feynman rules for TMD operators}

\label{sec:ColorFlowFR}

The color flow Feynman rules (see \textit{e.g.} \citep{Maltoni:2002mq})
are useful for calculating color factors. It turns out that they are
also very useful in the context of calculation of the structure of
the TMD operators in (\ref{eq:GenericTMD}), especially, when quarks
are involved. We shall supplement the standard color flow rules for
color-ordered diagrams (see Table~\ref{tab:color-flow-FR}) with
a set of additional rules which are simple color flow representations
of the rules derived in \citep{Bomhof:2006dp} for calculation of
a TMD operator structure in an arbitrary process. 

\begin{table}
\begin{centering}
\begin{tabular}{c|c|c}
\hline 
\toprule[1pt]\midrule[0.3pt] triple gluon vertex & %
\begin{minipage}[c]{3cm}%
\[
\sim\delta_{j_{1}}^{i_{2}}\delta_{j_{2}}^{i_{3}}\delta_{j_{3}}^{i_{1}}
\]
\end{minipage} & %
\begin{minipage}[c][4cm]{4cm}%
\includegraphics[width=4cm]{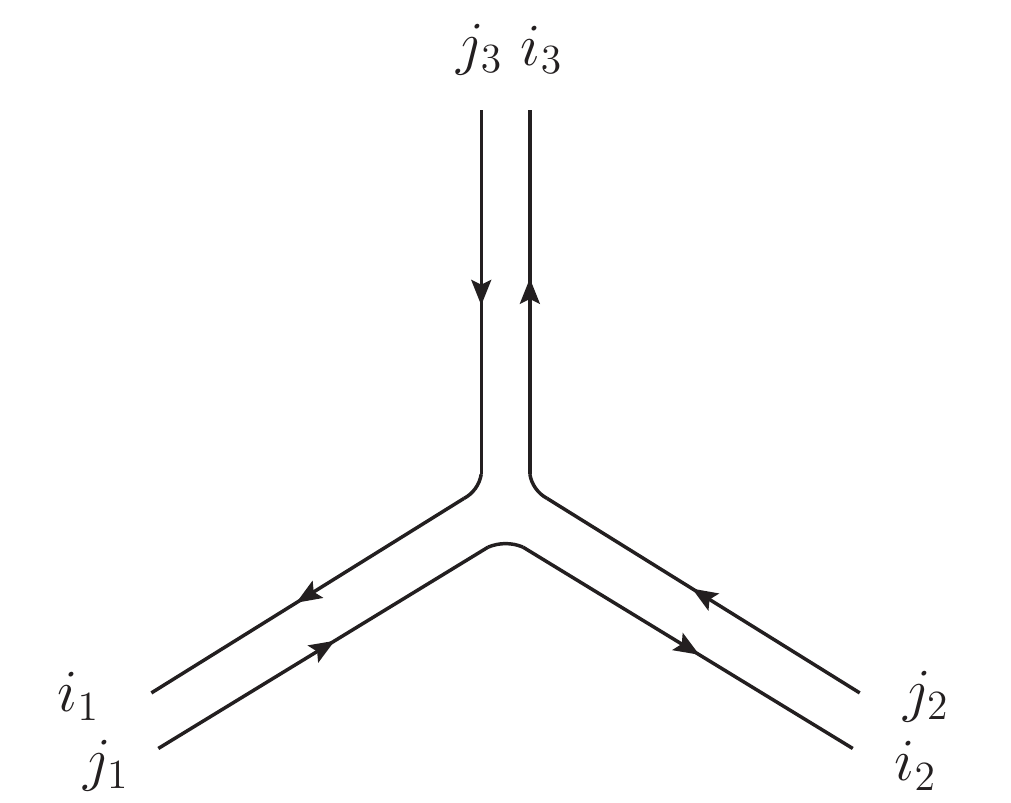}%
\end{minipage}\tabularnewline
\hline 
\hline 
four-gluon vertex & %
\begin{minipage}[c]{3cm}%
\[
\sim\delta_{j_{1}}^{i_{2}}\delta_{j_{2}}^{i_{3}}\delta_{j_{3}}^{i_{4}}\delta_{j_{4}}^{i_{1}}
\]
\end{minipage} & %
\begin{minipage}[c][4.5cm]{4.5cm}%
\begin{center}
\includegraphics[width=4cm]{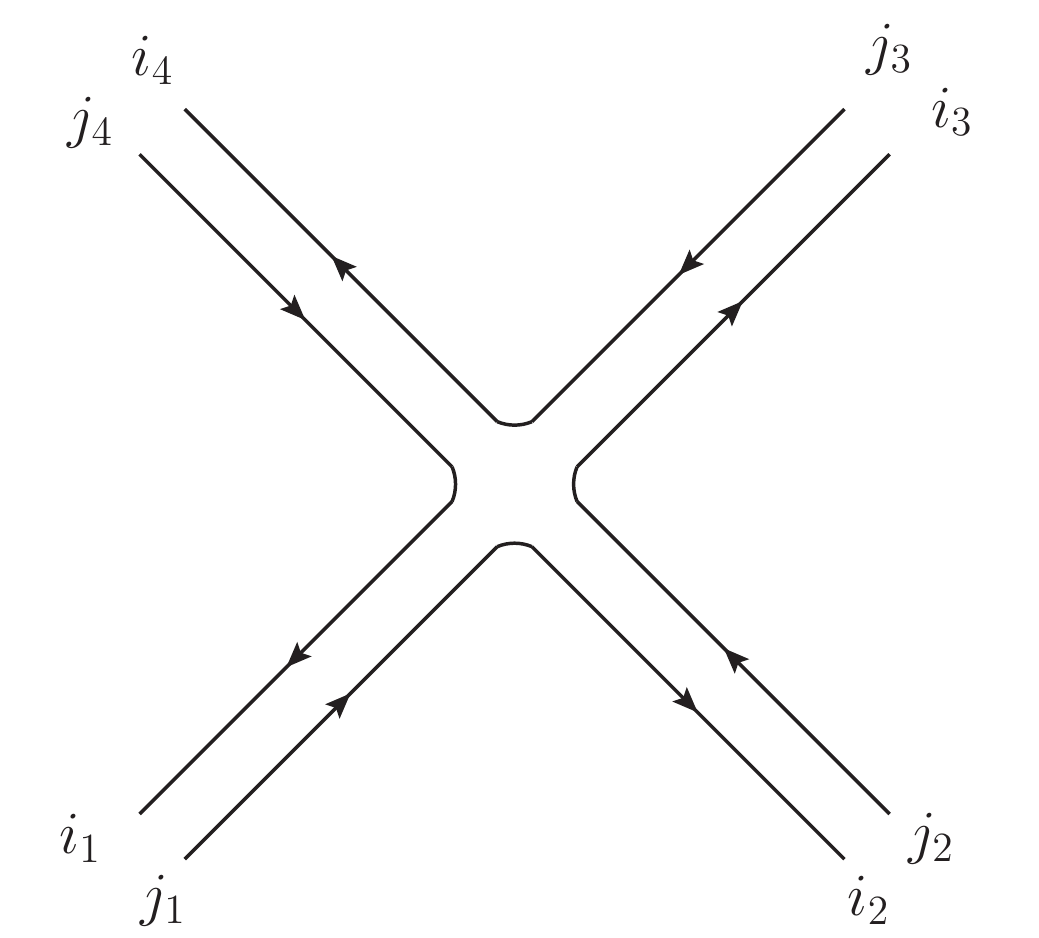}
\par\end{center}%
\end{minipage}\tabularnewline
\hline 
\hline 
quark-gluon vertex & %
\begin{minipage}[c]{3cm}%
\[
\sim\delta_{j_{1}}^{i^{q}}\delta_{j^{q}}^{i_{1}}
\]
\end{minipage} & %
\begin{minipage}[c][4cm]{4cm}%
\includegraphics[width=4cm]{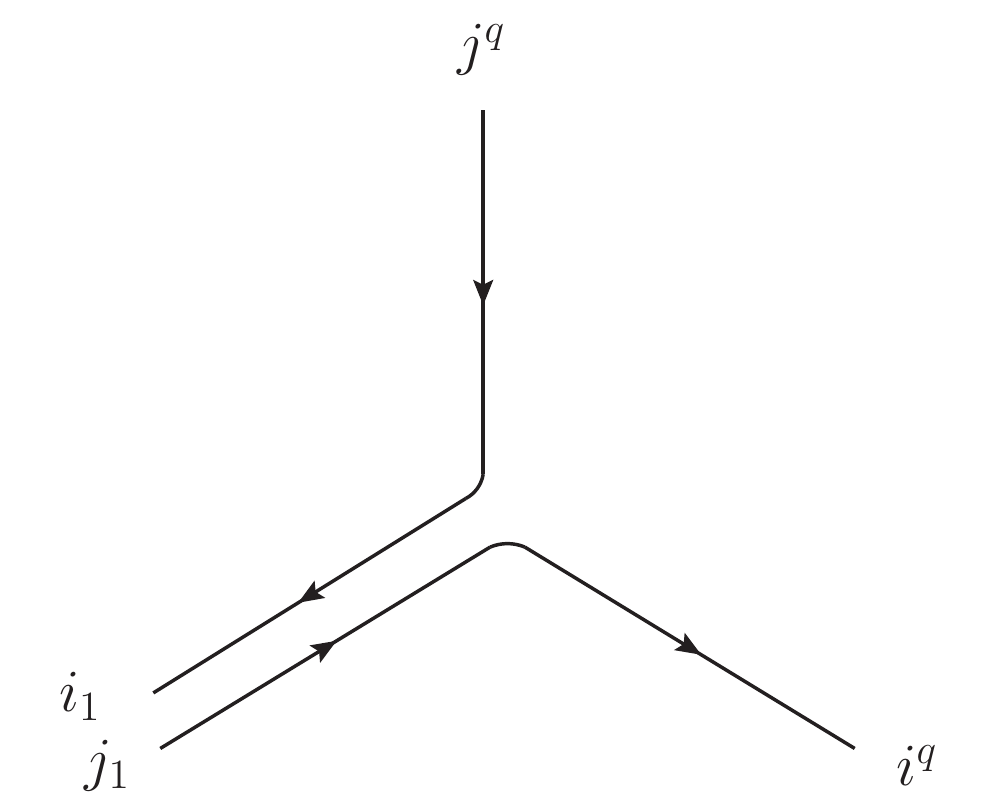}%
\end{minipage}\tabularnewline
\hline 
\hline 
gluon propagator & %
\begin{minipage}[c]{3cm}%
\begin{multline*}
\sim\Big(\delta^{ii'}\delta_{jj'}\\
-\frac{1}{N_{c}}\delta_{j}^{i}\delta_{j'}^{i'}\Big)
\end{multline*}
\end{minipage} & %
\begin{minipage}[c][4cm]{4cm}%
\includegraphics[width=4cm]{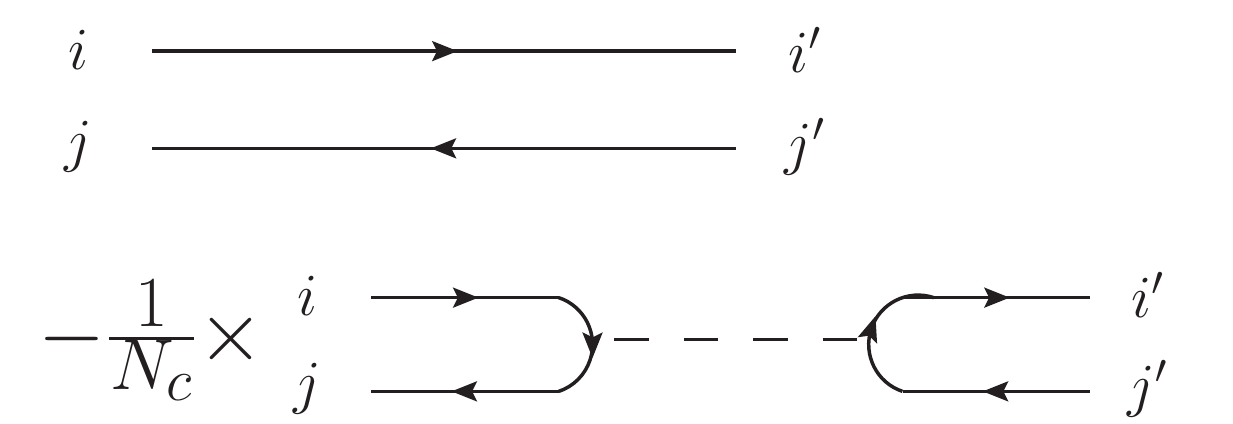}%
\end{minipage}\tabularnewline
\hline 
\end{tabular}
\par\end{centering}
\caption{Standard color flow Feynman rules for partial amplitudes. All momenta
are outgoing. In the middle column we show the color part only. \label{tab:color-flow-FR}}
\end{table}

The original procedure effectively leads to the following recipe.
For each \textit{final state} we assign the gauge link $\mathcal{U}^{\left[+\right]}$,
which joins the points $0$ and $\xi$ (see Subsection \ref{subsec:TMD_distr})
through the point in $+\infty$:
\begin{multline}
\mathcal{U}^{\left[+\right]}=\left[\left(0^{+},0^{-},\vec{0}_{T}\right),\left(0^{+},\infty^{-},\vec{0}_{T}\right)\right]\\
\left[\left(0^{+},\infty^{-},\vec{0}_{T}\right),\left(0^{+},\infty^{-},\vec{\xi}_{T}\right)\right]\left[\left(0^{+},\infty^{-},\vec{\xi}_{T}\right),\left(0^{+},\xi^{-},\vec{\xi}_{T}\right)\right]\,.
\end{multline}
In case of gluons the gauge link is to be defined in adjoint representation.
The Wilson link replaces the deltas for color summation when the amplitude
is squared: $\delta_{i'i}\rightarrow\left(\mathcal{U}^{\left[+\right]}\right)_{i'i}$
for quarks, $\delta^{jj'}\rightarrow\left(\mathcal{U}^{\left[+\right]\dagger}\right)^{jj'}$
for anti-quarks and $\delta_{a'a}\rightarrow\left(\mathcal{U}^{\left[+\right]}\right)_{a'a}$
for gluons (here and in what follows $i,j,k,\dots$ are fundamental
color indices, while $a,b,c,\dots$ are adjoint). For the \textit{initial
state} (not connected to the TMD gluon distribution), the resummation
of the initial state interactions leads to the Wilson line extending
to $-\infty$:
\begin{multline}
\mathcal{U}^{\left[-\right]}=\left[\left(0^{+},0^{-},\vec{0}_{T}\right),\left(0^{+},-\infty^{-},\vec{0}_{T}\right)\right]\\
\left[\left(0^{+},-\infty^{-},\vec{0}_{T}\right),\left(0^{+},-\infty^{-},\vec{\xi}_{T}\right)\right]\left[\left(0^{+},-\infty^{-},\vec{\xi}_{T}\right),\left(0^{+},\xi^{-},\vec{\xi}_{T}\right)\right]\,.
\end{multline}
Similar to final states, one needs to replace the color deltas for
initial states by the matrix elements of  $\,\mathcal{U}^{\left[-\right]}$.
The remaining initial state (connected to the TMD) is attached to
$F_{a}^{i+}\left(\xi\right)$ in the amplitude and to $F_{a'}^{i+}\left(0\right)$
in the conjugate amplitude. The rest of the procedure is similar to
calculating color factors: one extracts the color structure of the
pertinent amplitude and makes all the contractions (here with Wilson
lines and field strength tensors instead of deltas). In the end one
needs to divide-out the color factor for a process without gauge links.

Passing to the color flow representation is straightforward. Nothing
really is to be done for quarks and anti-quarks. For gluons, we first
need to make a connection of the adjoint Wilson line with the trace
of fundamental-representation instances of the same Wilson line, and
next project it onto the fundamental color quantum numbers with the
help of the Fierz identity. All rules with graphical representation
are collected in Table~\ref{tab:color-flow-GL}. The procedure of
calculating the TMD operator structures is now reduced to considering
all possible color flows and applying the rules. Although, in principle,
we could consider all standard Feynman diagrams, draw them in the
color flow representation and calculate TMD operator structures,
fortunately, we do not need to do this. Instead we can just use the color flow
decomposition described in Subsection~\ref{subsec:ColorDecomp}.
This will also ensure, that we work with gauge invariant sets from
the start.

When constructing the TMD operators, the initial and final states
are treated differently, \textit{i.e.} they are assigned different gauge links.
Therefore, we have to adjust the color flow decomposition (\ref{eq:glue_color_flow_decomp})-(\ref{eq:qqqq_color_flow_decomp})
to take into account the fact, that two legs are incoming (recall,
that these decomposition are within the standard convention of all
outgoing partons). This is fixed by making the replacement $i_{1}\longleftrightarrow j_{1}$,
$i_{n}\longleftrightarrow j_{n}$, as in our convention always the
first and the last partons are incoming.

Below, we present some examples to better illustrate the procedure.

\begin{table}
\begin{centering}
\begin{tabular}{c|c|c}
\hline 
\toprule[1pt]\midrule[0.3pt] outgoing gluon & %
\begin{minipage}[c]{3.7cm}%
\begin{multline*}
\left(\mathcal{U}^{\left[+\right]}\right)_{i'i}\left(\mathcal{U}^{\left[+\right]\dagger}\right)^{jj'}\\
-\frac{1}{N_{c}}\delta_{i}^{j}\delta_{i'}^{j'}
\end{multline*}
\end{minipage} & %
\begin{minipage}[c][3.2cm]{4.5cm}%
\includegraphics[width=4.5cm]{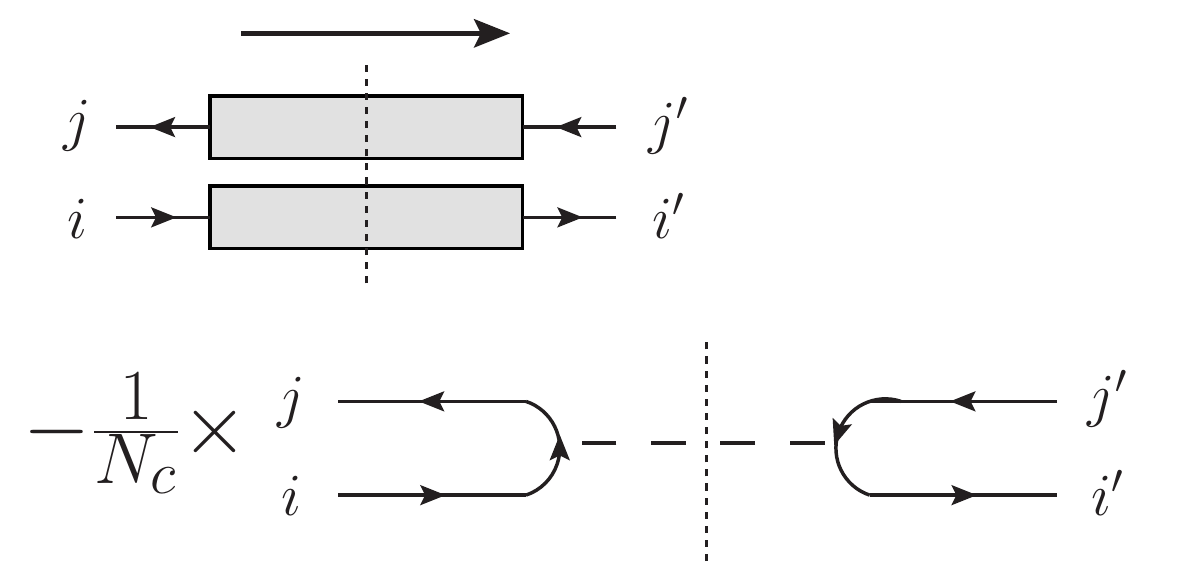}%
\end{minipage}\tabularnewline
\hline 
\hline 
incoming gluon & %
\begin{minipage}[c]{3.7cm}%
\begin{multline*}
\left(\mathcal{U}^{\left[-\right]\dagger}\right)^{ii'}\left(\mathcal{U}^{\left[-\right]}\right)_{j'j}\\
-\frac{1}{N_{c}}\delta_{j}^{i}\delta_{j'}^{i'}
\end{multline*}
\end{minipage} & %
\begin{minipage}[c][3.2cm]{4.5cm}%
\includegraphics[width=4.5cm]{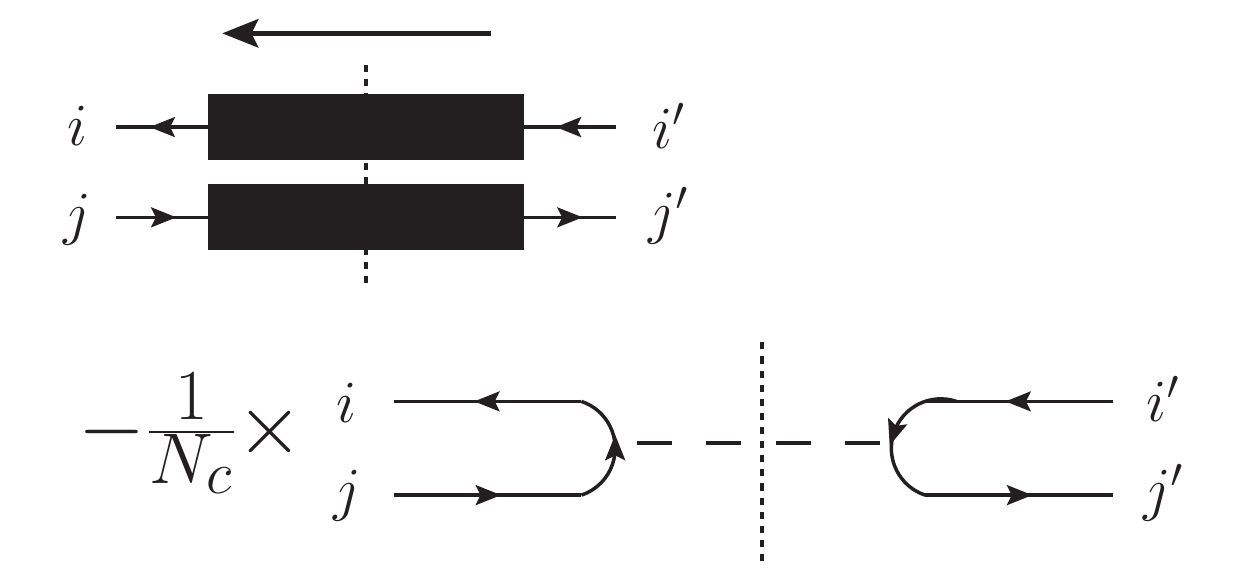}%
\end{minipage}\tabularnewline
\hline 
\hline 
outgoing quark & %
\begin{minipage}[c]{2.7cm}%
\[
\left(\mathcal{U}^{\left[+\right]}\right)_{i'i}
\]
\end{minipage} & %
\begin{minipage}[c][2.2cm]{4.5cm}%
\begin{center}
\includegraphics[width=3cm]{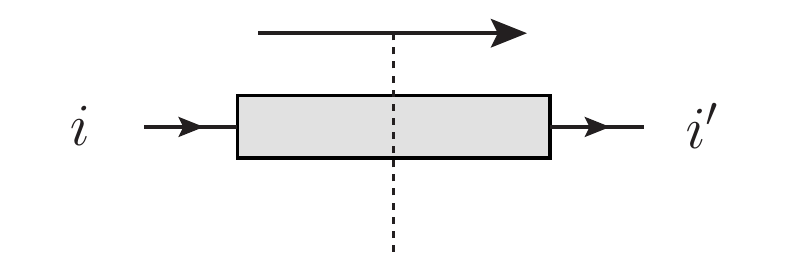}
\par\end{center}%
\end{minipage}\tabularnewline
\hline 
\hline 
incoming quark & %
\begin{minipage}[c]{2.7cm}%
\[
\left(\mathcal{U}^{\left[-\right]\dagger}\right)^{ii'}
\]
\end{minipage} & %
\begin{minipage}[c][2.2cm]{4.5cm}%
\begin{center}
\includegraphics[width=3cm]{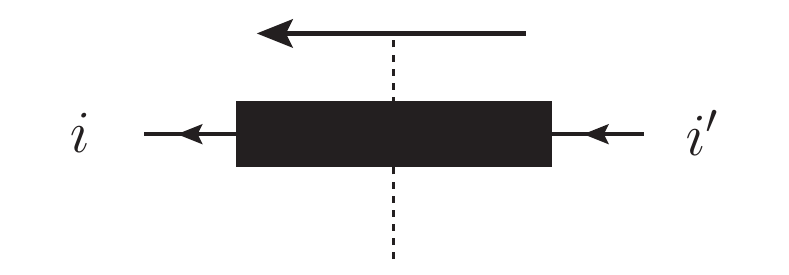}
\par\end{center}%
\end{minipage}\tabularnewline
\hline 
\hline 
outgoing anti-quark & %
\begin{minipage}[c]{2.7cm}%
\[
\left(\mathcal{U}^{\left[+\right]\dagger}\right)^{jj'}
\]
\end{minipage} & %
\begin{minipage}[c][2.2cm]{4.5cm}%
\begin{center}
\includegraphics[width=3cm]{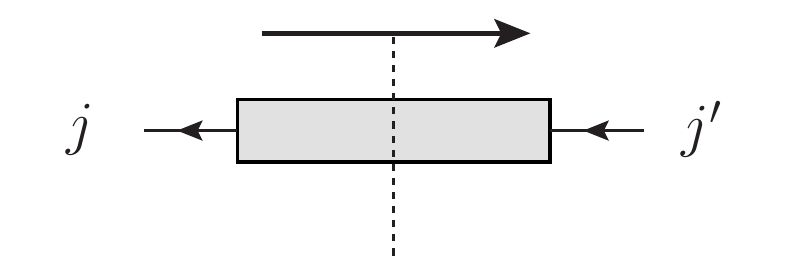}
\par\end{center}%
\end{minipage}\tabularnewline
\hline 
\hline 
incoming anti-quark & %
\begin{minipage}[c]{2.7cm}%
\[
\left(\mathcal{U}^{\left[-\right]}\right)_{j'j}
\]
\end{minipage} & %
\begin{minipage}[c][2.2cm]{4.5cm}%
\begin{center}
\includegraphics[width=3cm]{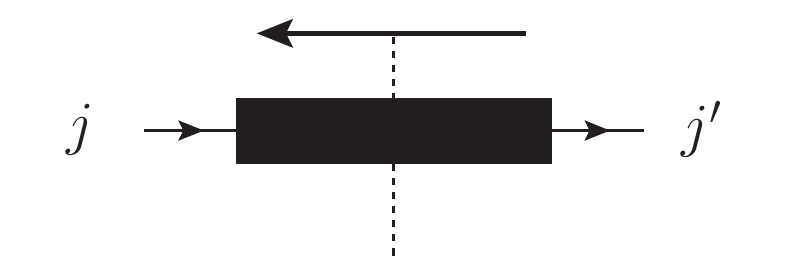}
\par\end{center}%
\end{minipage}\tabularnewline
\hline 
\hline 
field strength operators & %
\begin{minipage}[c]{2.7cm}%
\[\!\!\!\!\!\!\!\!\! 
2\left(\hat{F}^{+i}\left(\xi\right)\right)_{i}^{j}\left(\hat{F}^{+i}\left(0\right)\right)_{i'}^{j'}
\]
\end{minipage} & %
\begin{minipage}[c][2.2cm]{4.5cm}%
\begin{center}
\includegraphics[width=3cm]{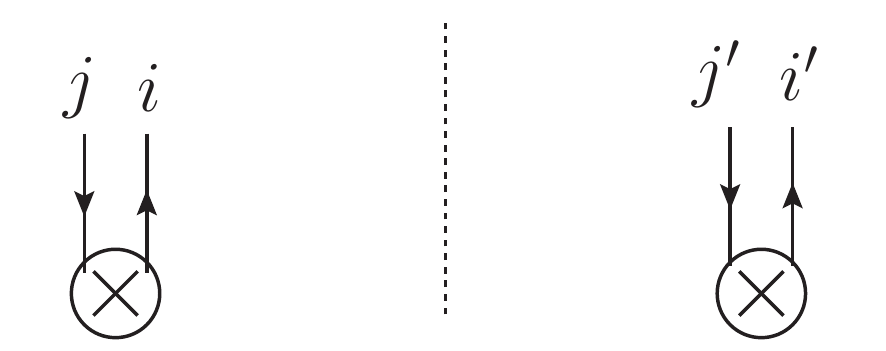}
\par\end{center}%
\end{minipage}\tabularnewline
\hline 
\end{tabular}
\par\end{centering}
\caption{Color flow Feynman rules for the gauge links. The diagrams correspond
to the cut lines, as denoted by the vertical dotted line. The routing
in the color loops is clock-wise. \label{tab:color-flow-GL}}
\end{table}

\subsection{Examples}

\label{subsec:FRexamples}

Let us first illustrate the usage of color flow Feynman rules to calculate
the structure of the TMD operator for the following diagram:
\begin{flushleft}
\begin{tabular}{>{\centering}m{0.87\columnwidth}>{\centering}m{0.05\columnwidth}}
\centering{}\smallskip{}
\includegraphics[width=3cm]{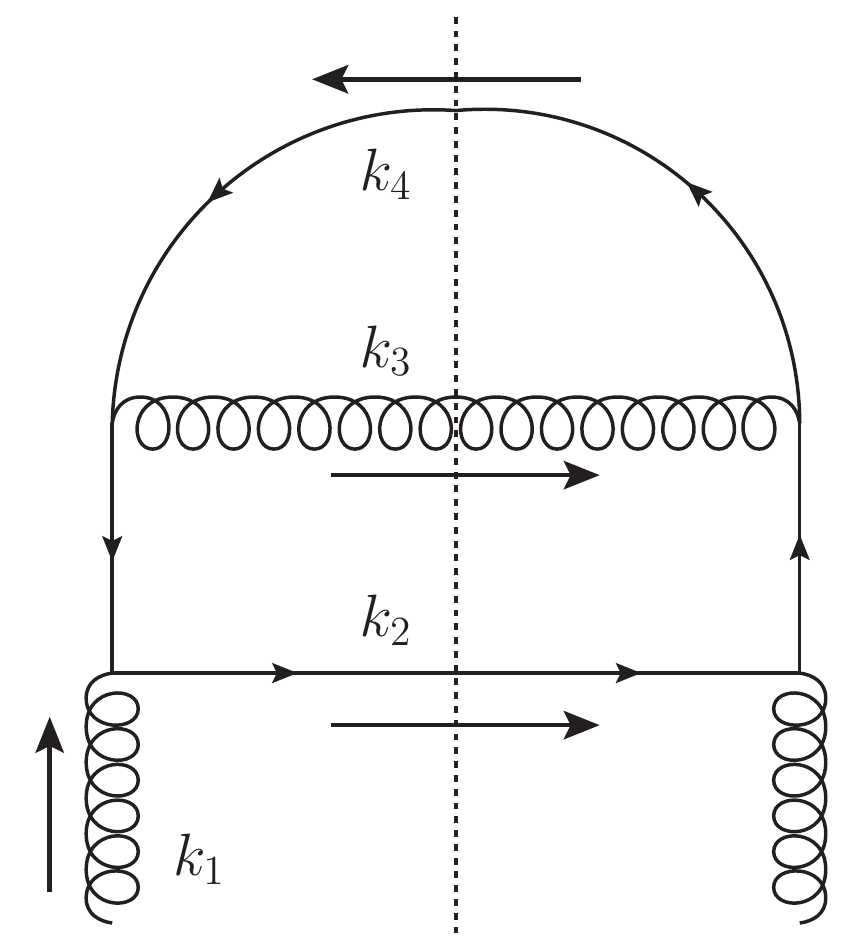} & \centering{}\centering{(\myref )}
\refmyref{example1_diag1}\tabularnewline
\end{tabular}
\par\end{flushleft}

\noindent  This diagram contributes to the process $g\left(k_{1}\right)q\left(k_{4}\right)\rightarrow q\left(k_{2}\right)g\left(k_{3}\right)$
and represents the diagram squared and summed over final/initial colors
(except insertions of the field operators). The arrows indicate whether the line is incoming/outgoing. Let us stress, that considering
particular diagrams is not the way we will ultimately proceed; instead
we will consider various color flows as defined in Eqs.~(\ref{eq:glue_color_flow_decomp})-(\ref{eq:qqqq_color_flow_decomp}).
The structure of the TMD operator for this diagram was calculated
in \citep{Bomhof:2006dp}. In the color flow representation we have
to consider two diagrams:
\begin{flushleft}
\begin{tabular}{>{\centering}m{0.87\columnwidth}>{\centering}m{0.05\columnwidth}}
\centering{}\smallskip{}
\includegraphics[width=7cm]{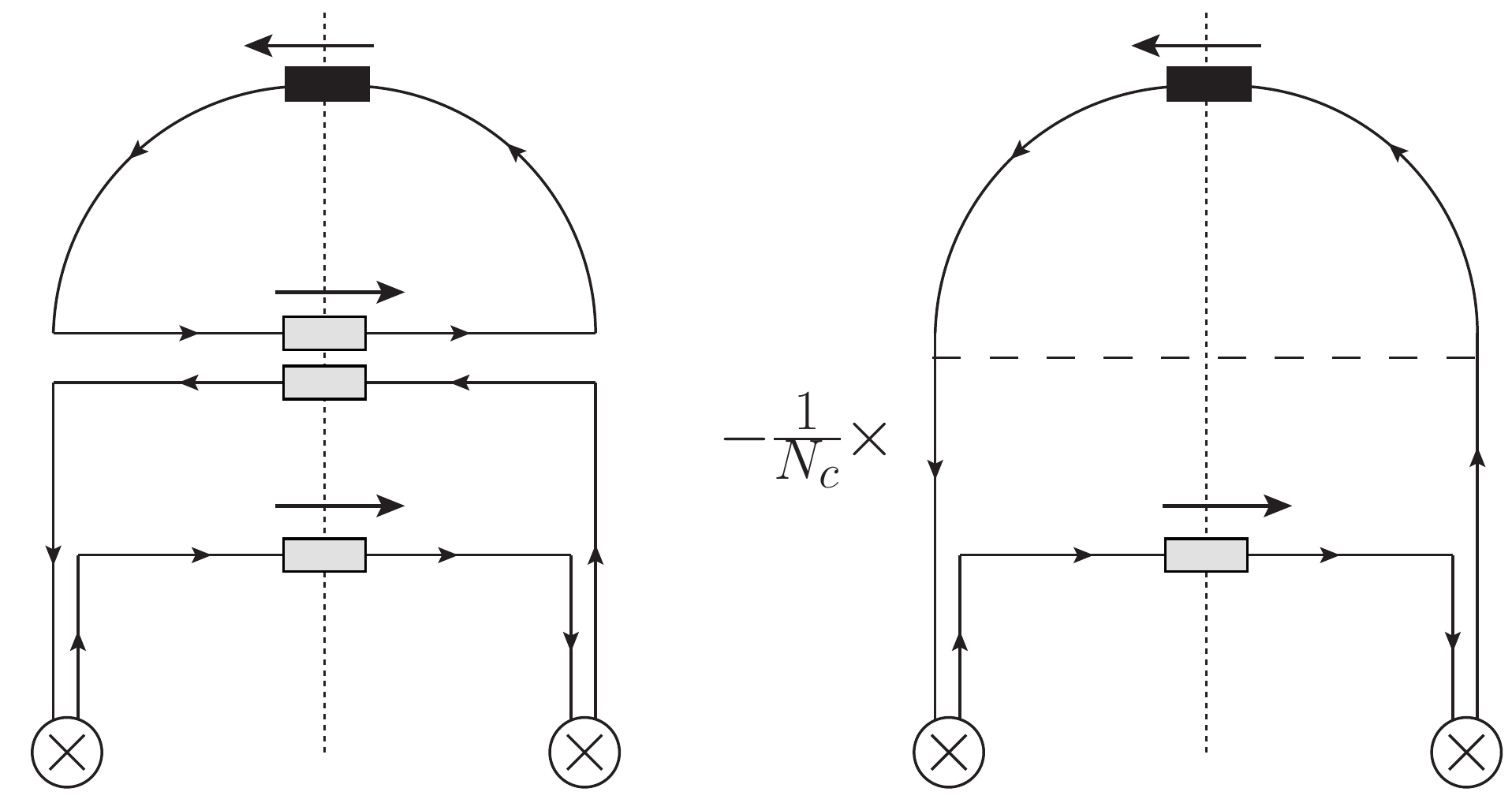} & \centering{}\centering{(\myref )}
\refmyref{example1_diag2}\tabularnewline
\end{tabular}
\par\end{flushleft}

\noindent The diagram with dashed line represents an exchange of
the $U\left(1\right)$ gluon (a colorless gluon). To calculate the
diagrams we simply look for the closed quark loops and make the trace
of the objects appearing in the loop. The direction of the trace is
clockwise. The dashed lines carry no color, thus they do not make
any traces (they also always accompany $1/N_{c}$ factors). Note,
we calculate only color part (with possible $\mathrm{SU}\left(N_{c}\right)$
matrix insertions) -- we are not concerned with any kinematic factors.
For the first diagram, we have
\begin{equation}
\mathrm{Tr}\left\{ F\left(\xi\right)\mathcal{U}^{\left[+\right]\dagger}F\left(0\right)\mathcal{U}^{\left[+\right]}\right\} \mathrm{Tr}\left\{ \mathcal{U}^{\left[\square\right]}\right\} \,,
\end{equation}
where the first trace corresponds to the bottom loop, the second to
the top loop. Above, we defined the Wilson loop \citep{Bomhof:2006dp}
\begin{equation}
\mathcal{U}^{\left[\square\right]}=\mathcal{U}^{\left[-\right]\dagger}\mathcal{U}^{\left[+\right]}\,.\label{eq:WilsonLoopDef}
\end{equation}
We also use shorthand notation $F\left(\xi\right)\equiv\hat{F}^{i+}\left(\xi^{+}=0,\xi^{-},\vec{\xi}_{T}\right)$.
The second diagram reads
\begin{equation}
-\frac{1}{N_{c}}\,\mathrm{Tr}\left(F\left(\xi\right)\mathcal{U}^{\left[-\right]\dagger}F\left(0\right)\mathcal{U}^{\left[+\right]}\right)\,.
\end{equation}
To get the final result, the sum of the two contributions must be
divided by the sum of the color factors (without the Wilson lines),
with open indices where the field operators are attached:
\begin{flushleft}
\begin{tabular}{>{\centering}m{0.87\columnwidth}>{\centering}m{0.05\columnwidth}}
\centering{}\smallskip{}
\includegraphics[width=7cm]{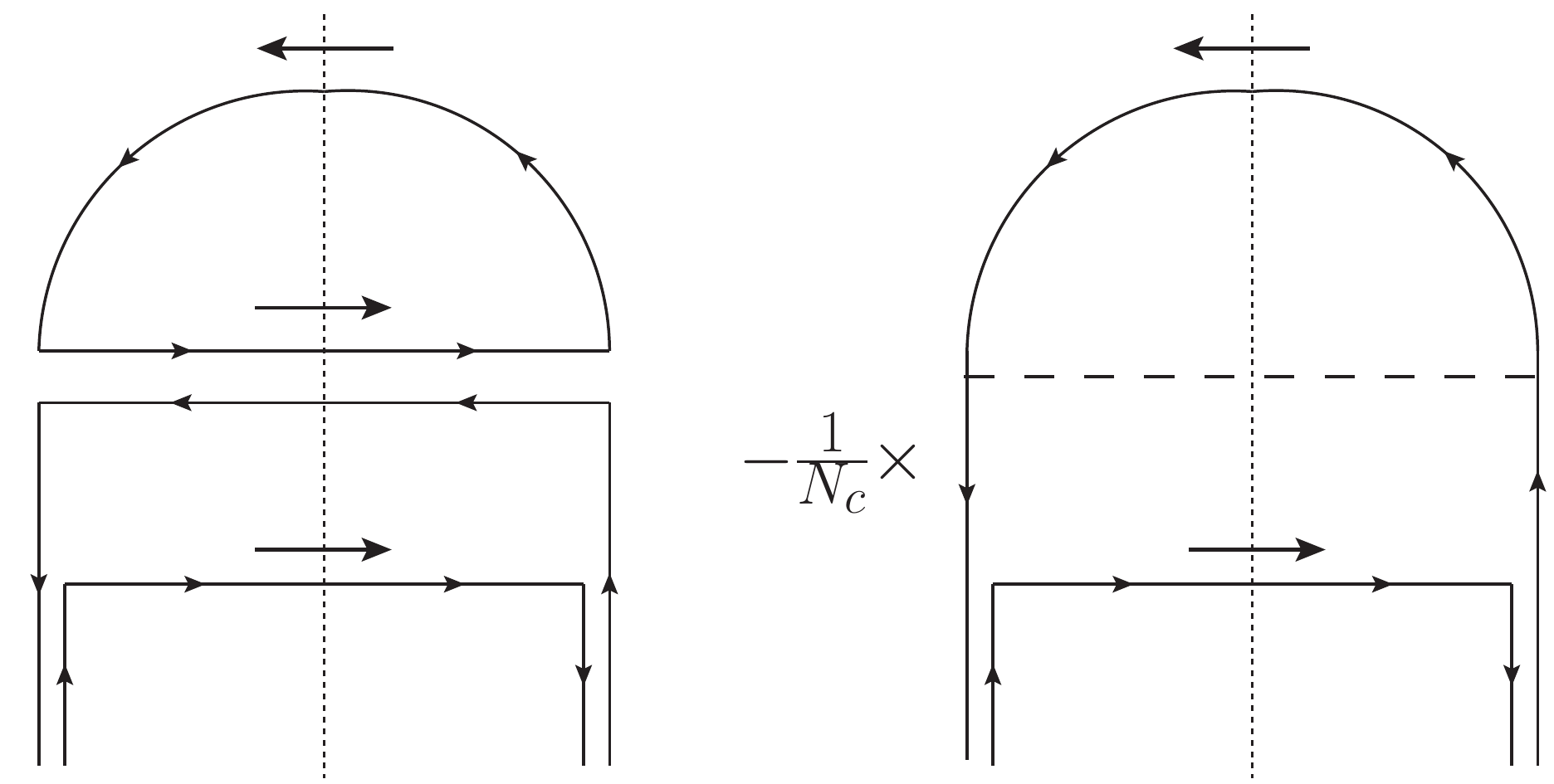} & \centering{}\centering{(\myref )}
\refmyref{example1_diag3}\tabularnewline
\end{tabular}
\par\end{flushleft}

\noindent The part multiplying the open indices reads
\begin{equation}
N_{c}-\frac{1}{N_{c}}=\frac{N_{c}^{2}-1}{N_{c}}\,.
\end{equation}
Thus the TMD operator reads
\begin{equation}
\mathrm{Tr}\left\{ F\left(\xi\right)\left[\frac{N_{c}^{2}}{N_{c}^{2}-1}\,\frac{\mathrm{Tr}\mathcal{U}^{\left[\square\right]}}{N_{c}}\,\mathcal{U}^{\left[+\right]\dagger}-\frac{1}{N_{c}^{2}-1}\mathcal{U}^{\left[-\right]\dagger}\right]F\left(0\right)\mathcal{U}^{\left[+\right]}\right\} \,,
\end{equation}
which exactly agrees with the result quoted in \citep{Bomhof:2006dp}.

As an illustration of a more complicated structure, let us consider
an example contribution to the process $gg\rightarrow q\bar{q}gg$:

\begin{tabular}{>{\centering}m{0.87\columnwidth}>{\centering}m{0.05\columnwidth}}
\centering{}\bigskip{}
\includegraphics[width=10cm]{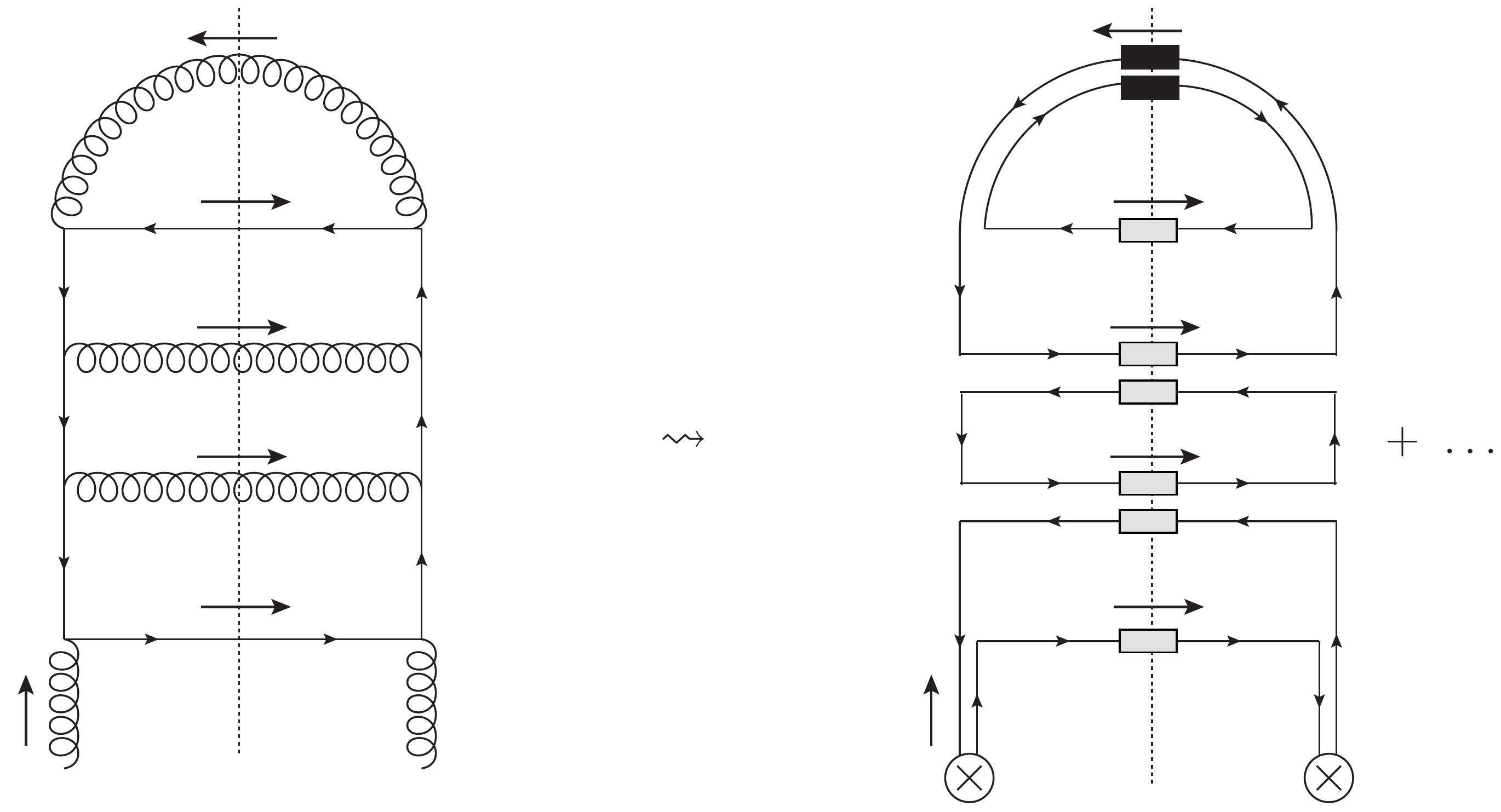} & \centering{}\centering{(\myref )}
\refmyref{example2_diag1}\tabularnewline
\end{tabular}

\noindent Applying the color flow rules gives immediately the operator
structure for the leading color flow displayed on the r.h.s.:
\begin{equation}
N_{c}\mathrm{Tr}\left\{ F\left(\xi\right)\mathcal{U}^{\left[+\right]\dagger}F\left(0\right)\mathcal{U}^{\left[+\right]}\right\} \,\mathrm{Tr}\mathcal{U}^{\left[\square\right]}\,\mathrm{Tr}\mathcal{U}^{\left[\square\right]\dagger}\,.
\end{equation}
Above, the $N_{c}$ factor comes from the second loop from the bottom,
$\mathrm{Tr}\left\{ \mathcal{U}^{\left[+\right]\dagger}\mathcal{U}^{\left[+\right]}\right\} =\mathrm{Tr}\mathbf{1}=N_{c}$.

To close this section, let us stress, that the problem of proliferation
of color flow diagrams compared to ordinary diagrams, will not concern
us at all. As mentioned, we shall use the color flow decomposition,
which sets the color flow without need to consider particular diagrams.

\section{The operator basis for arbitrary TMD gluon distribution}

\label{sec:Operators}

Using the color flow Feynman rules from the previous section we can
easily determine all possible 'basis' operators, from which a TMD
gluon distribution for arbitrary process can be constructed. Alternatively,
one can think about 'basis' TMD gluon distributions.

Plenty of different operators already appear for processes with four
colored partons considered in \citep{Bomhof:2006dp}. In order to
find all of them, we use the following facts. First, there are at
most two $\mathcal{U}^{\left[-\right]}$ Wilson lines. This is the case for initial state gluons where $\mathcal{U}^{\left[-\right]}$ and $\mathcal{U}^{\left[-\right]\dagger}$ appear. Thus, we can built at most two Wilson loops (\ref{eq:WilsonLoopDef}),
when they are looped with $\mathcal{U}^{\left[+\right]}$ or $\mathcal{U}^{\left[+\right]\dagger}$
(see the last example in Section~\ref{sec:ColorFlowFR}). Second,
any color flow loop will contribute trace of at most first power of
$\mathcal{U}^{\left[\pm\right]}$, $\mathcal{U}^{\left[\pm\right]\dagger}$
(and $F\left(\xi\right)$, $F\left(0\right)$, or both), in addition
to mentioned Wilson loops (at most $\mathcal{U}^{\left[\square\right]}$
and $\mathcal{U}^{\left[\square\right]\dagger}$). This is because
for a color flow loop with many Wilson lines (contributed by many
final states), most of the Wilson lines will collapse to unity, $\mathcal{U}^{\left[+\right]\dagger}\mathcal{U}^{\left[+\right]}=\mathbf{1}$,
leaving only at most single instances of $\mathcal{U}^{\left[\pm\right]}$, $\mathcal{U}^{\left[\pm\right]\dagger}$, $\mathcal{U}^{\left[\square\right]}$, $\mathcal{U}^{\left[\square\right]\dagger}$.

Basing on the above, below we list all 'basis' TMD gluon distributions,
from which an arbitrary TMD is given as a linear combination. We assume
here, that the \textit{correlators are real} valued functions.

\begin{equation}
\begin{aligned}\mathcal{F}_{qg}^{(1)}\left(x,k_{T}\right) & =2\int\frac{d\xi^{-}d^{2}\xi_{T}}{\left(2\pi\right)^{3}P^{+}}\,e^{ixP^{+}\xi^{-}-i\vec{k}_{T}\cdot\vec{\xi}_{T}}\,\left<\mathrm{Tr}\left[\hat{F}^{i+}\left(\xi\right)\mathcal{U}^{[-]\dagger}\hat{F}^{i+}\left(0\right)\mathcal{U}^{[+]}\right]\right>\,\\
 & =2\int\frac{d\xi^{-}d^{2}\xi_{T}}{\left(2\pi\right)^{3}P^{+}}\,e^{ixP^{+}\xi^{-}-i\vec{k}_{T}\cdot\vec{\xi}_{T}}\,\left<\mathrm{Tr}\left[\hat{F}^{i+}\left(\xi\right)\mathcal{U}^{[+]\dagger}\hat{F}^{i+}\left(0\right)\mathcal{U}^{[-]}\right]\right>\,,
\end{aligned}
\label{eq:Fqg1}
\end{equation}

\begin{equation}
\begin{aligned}\mathcal{F}_{qg}^{(2)}\left(x,k_{T}\right) & =2\int\frac{d\xi^{-}d^{2}\xi_{T}}{\left(2\pi\right)^{3}P^{+}}\,e^{ixP^{+}\xi^{-}-i\vec{k}_{T}\cdot\vec{\xi}_{T}}\,\left<\frac{\mathrm{Tr}\left[\mathcal{U}^{[\square]}\right]}{N_{c}}\mathrm{Tr}\left[\hat{F}^{i+}\left(\xi\right)\mathcal{U}^{[+]\dagger}\hat{F}^{i+}\left(0\right)\mathcal{U}^{[+]}\right]\right>\,\\
 & =2\int\frac{d\xi^{-}d^{2}\xi_{T}}{\left(2\pi\right)^{3}P^{+}}\,e^{ixP^{+}\xi^{-}-i\vec{k}_{T}\cdot\vec{\xi}_{T}}\,\left<\frac{\mathrm{Tr}\left[\mathcal{U}^{[\square]\dagger}\right]}{N_{c}}\mathrm{Tr}\left[\hat{F}^{i+}\left(\xi\right)\mathcal{U}^{[+]\dagger}\hat{F}^{i+}\left(0\right)\mathcal{U}^{[+]}\right]\right>\,,
\end{aligned}
\end{equation}

\begin{equation}
\begin{aligned}\mathcal{F}_{qg}^{(3)}\left(x,k_{T}\right) & =2\int\frac{d\xi^{-}d^{2}\xi_{T}}{\left(2\pi\right)^{3}P^{+}}\,e^{ixP^{+}\xi^{-}-i\vec{k}_{T}\cdot\vec{\xi}_{T}}\,\left<\mathrm{Tr}\left[\hat{F}^{i+}\left(\xi\right)\mathcal{U}^{[+]\dagger}\hat{F}^{i+}\left(0\right)\mathcal{U}^{[\square]}\mathcal{U}^{[+]}\right]\right>\,\\
 & =2\int\frac{d\xi^{-}d^{2}\xi_{T}}{\left(2\pi\right)^{3}P^{+}}\,e^{ixP^{+}\xi^{-}-i\vec{k}_{T}\cdot\vec{\xi}_{T}}\,\left<\mathrm{Tr}\left[\hat{F}^{i+}\left(\xi\right)\mathcal{U}^{[\square]\dagger}\mathcal{U}^{[+]\dagger}\hat{F}^{i+}\left(0\right)\mathcal{U}^{[+]}\right]\right>\,,
\end{aligned}
\label{eq:Fqg3}
\end{equation}

\begin{equation}
\begin{aligned}\mathcal{F}_{gg}^{(1)}\left(x,k_{T}\right) & =2\int\frac{d\xi^{-}d^{2}\xi_{T}}{\left(2\pi\right)^{3}P^{+}}\,e^{ixP^{+}\xi^{-}-i\vec{k}_{T}\cdot\vec{\xi}_{T}}\,\left<\frac{\mathrm{Tr}\left[\mathcal{U}^{[\square]\dagger}\right]}{N_{c}}\mathrm{Tr}\left[\hat{F}^{i+}\left(\xi\right)\mathcal{U}^{[-]\dagger}\hat{F}^{i+}\left(0\right)\mathcal{U}^{[+]}\right]\right>\,\\
 & =2\int\frac{d\xi^{-}d^{2}\xi_{T}}{\left(2\pi\right)^{3}P^{+}}\,e^{ixP^{+}\xi^{-}-i\vec{k}_{T}\cdot\vec{\xi}_{T}}\,\left<\frac{\mathrm{Tr}\left[\mathcal{U}^{[\square]}\right]}{N_{c}}\mathrm{Tr}\left[\hat{F}^{i+}\left(\xi\right)\mathcal{U}^{[+]\dagger}\hat{F}^{i+}\left(0\right)\mathcal{U}^{[-]}\right]\right>\,,
\end{aligned}
\label{eq:Fgg1}
\end{equation}

\begin{equation}
\begin{aligned}\mathcal{F}_{gg}^{(2)}\left(x,k_{T}\right) & =2\int\frac{d\xi^{-}d^{2}\xi_{T}}{\left(2\pi\right)^{3}P^{+}}\,e^{ixP^{+}\xi^{-}-i\vec{k}_{T}\cdot\vec{\xi}_{T}}\,\frac{1}{N_{c}}\left<\mathrm{Tr}\left[\hat{F}^{i+}\left(\xi\right)\mathcal{U}^{[\square]\dagger}\right]\mathrm{Tr}\left[\hat{F}^{i+}\left(0\right)\mathcal{U}^{[\square]}\right]\right>\,\\
 & =2\int\frac{d\xi^{-}d^{2}\xi_{T}}{\left(2\pi\right)^{3}P^{+}}\,e^{ixP^{+}\xi^{-}-i\vec{k}_{T}\cdot\vec{\xi}_{T}}\,\frac{1}{N_{c}}\left<\mathrm{Tr}\left[\hat{F}^{i+}\left(\xi\right)\mathcal{U}^{[\square]}\right]\mathrm{Tr}\left[\hat{F}^{i+}\left(0\right)\mathcal{U}^{[\square]\dagger}\right]\right>\,,
\end{aligned}
\label{eq:Fgg2}
\end{equation}

\begin{equation}
\mathcal{F}_{gg}^{(3)}\left(x,k_{T}\right)=2\int\frac{d\xi^{-}d^{2}\xi_{T}}{\left(2\pi\right)^{3}P^{+}}\,e^{ixP^{+}\xi^{-}-i\vec{k}_{T}\cdot\vec{\xi}_{T}}\,\left<\mathrm{Tr}\left[\hat{F}^{i+}\left(\xi\right)\mathcal{U}^{[+]\dagger}\hat{F}^{i+}\left(0\right)\mathcal{U}^{[+]}\right]\right>\,,
\end{equation}

\begin{equation}
\mathcal{F}_{gg}^{(4)}\left(x,k_{T}\right)=2\int\frac{d\xi^{-}d^{2}\xi_{T}}{\left(2\pi\right)^{3}P^{+}}\,e^{ixP^{+}\xi^{-}-i\vec{k}_{T}\cdot\vec{\xi}_{T}}\,\left<\mathrm{Tr}\left[\hat{F}^{i+}\left(\xi\right)\mathcal{U}^{[-]\dagger}\hat{F}^{i+}\left(0\right)\mathcal{U}^{[-]}\right]\right>\,,
\end{equation}

\begin{equation}
\mathcal{F}_{gg}^{(5)}\left(x,k_{T}\right)=2\int\frac{d\xi^{-}d^{2}\xi_{T}}{\left(2\pi\right)^{3}P^{+}}\,e^{ixP^{+}\xi^{-}-i\vec{k}_{T}\cdot\vec{\xi}_{T}}\,\left<\mathrm{Tr}\left[\hat{F}^{i+}\left(\xi\right)\mathcal{U}^{[\square]\dagger}\mathcal{U}^{[+]\dagger}\hat{F}^{i+}\left(0\right)\mathcal{U}^{[\square]}\mathcal{U}^{[+]}\right]\right>\,,
\end{equation}

\begin{equation}
\mathcal{F}_{gg}^{(6)}\left(x,k_{T}\right)=2\int\frac{d\xi^{-}d^{2}\xi_{T}}{\left(2\pi\right)^{3}P^{+}}\,e^{ixP^{+}\xi^{-}-i\vec{k}_{T}\cdot\vec{\xi}_{T}}\,\left<\frac{\mathrm{Tr}\left[\mathcal{U}^{[\square]}\right]}{N_{c}}\frac{\mathrm{Tr}\left[\mathcal{U}^{[\square]\dagger}\right]}{N_{c}}\mathrm{Tr}\left[\hat{F}^{i+}\left(\xi\right)\mathcal{U}^{[+]\dagger}\hat{F}^{i+}\left(0\right)\mathcal{U}^{[+]}\right]\right>\,,\label{eq:Fgg6}
\end{equation}

\begin{equation}
\begin{aligned}\mathcal{F}_{gg}^{(7)}\left(x,k_{T}\right) & =2\int\frac{d\xi^{-}d^{2}\xi_{T}}{\left(2\pi\right)^{3}P^{+}}\,e^{ixP^{+}\xi^{-}-i\vec{k}_{T}\cdot\vec{\xi}_{T}}\,\left<\frac{\mathrm{Tr}\left[\mathcal{U}^{[\square]}\right]}{N_{c}}\mathrm{Tr}\left[\hat{F}^{i+}\left(\xi\right)\mathcal{U}^{[\square]\dagger}\mathcal{U}^{[+]\dagger}\hat{F}^{i+}\left(0\right)\mathcal{U}^{[+]}\right]\right>\,\\
 & =2\int\frac{d\xi^{-}d^{2}\xi_{T}}{\left(2\pi\right)^{3}P^{+}}\,e^{ixP^{+}\xi^{-}-i\vec{k}_{T}\cdot\vec{\xi}_{T}}\,\left<\frac{\mathrm{Tr}\left[\mathcal{U}^{[\square]\dagger}\right]}{N_{c}}\mathrm{Tr}\left[\hat{F}^{i+}\left(\xi\right)\mathcal{U}^{[+]\dagger}\hat{F}^{i+}\left(0\right)\mathcal{U}^{[\square]}\mathcal{U}^{[+]}\right]\right>\,.
\end{aligned}
\label{eq:Fgg7}
\end{equation}
In the definitions above, the average should be understand as the
hadronic matrix elements, \textit{cf.} Eq.~(\ref{eq:GenericTMD}). Two new
structures appear in addition to those known in the literature: $\mathcal{F}_{qg}^{\left(3\right)}$
and $\mathcal{F}_{gg}^{\left(7\right)}$.

Here the subscripts refer to a partonic process to which a given TMD distribution belongs -- whether this is a pure gluonic process or a process with quarks\footnote{The notation for the above TMD gluon distributions should not be confused with the double-TMD parton distributions (see \textit{e.g.} \citep{Buffing:2017mqm}).}. This notation was first introduced in \citep{Dominguez:2011wm} in the context of  the small-$x$ limit and we stick to that notation in the present work.

The above set of basic TMD gluon distribution constitutes the basis for any TMD gluon distribution to be convoluted with a hard process, assuming there are no other TMD operators involved. As discussed in the Introduction, this assumption is meant to be used at small-$x$ where it can be justified from the CGC effective theory. It is important to note, that it is the complete basis within the rules of \citep{Bomhof:2007xt} -- it does not represent a basis for a gluon correlator with arbitrary gauge link structure. At least formally, the basis structures are independent; however, in the large $k_T$ limit they start to be degenerate (or vanish), as discussed in the Introduction.

\section{Operator structures for 5 and 6 colored partons}

\label{sec:Results}

In this section we present the main result of the present work, \textit{i.e.}
the form of the TMD gluon distributions for the processes with 5 and
6 colored partons, together with their large $N_{c}$ limit, which
might be useful phenomenologically in short run. We will start with
a derivation of 4 parton TMD operators, to demonstrate the procedure
utilizing the color decomposition and, more importantly, to introduce
the general notation we shall use for more complicated processes (the
operator structures for 4 parton processes were first obtained in
\citep{Bomhof:2006dp}, and in \citep{Kotko:2015ura} using the color
decomposition).

\subsection{Outline of the method using 4 parton example}

As the color decomposition is most straightforward for pure gluonic
amplitude, let us start with the process
\begin{equation}
g\left(k_{1}\right)g\left(k_{4}\right)\rightarrow g\left(k_{2}\right)g\left(k_{3}\right)\,.
\end{equation}
For gluons, three color decompositions can be used: the fundamental
(\ref{eq:glue_color_decomp_fund}), the color flow (\ref{eq:glue_color_flow_decomp}),
and the adjoint (\ref{eq:glue_color_decomp_adjoint}). First two involve
6 partial amplitudes, while the last one only two. As mentioned in
Section~\ref{subsec:ColorDecomp} the 6 partial amplitudes are not
independent, but their squares give the leading contribution in the
large $N_{c}$ limit -- a property which we will use in Section~\ref{sec:LargeNc}.
Here, we are interested in the full answer, thus we use the adjoint
color decomposition (for processes with quarks we will use exclusively
color flow decomposition). It reads

\begin{equation}
\mathcal{M}^{a_{1}a_{2}a_{3}a_{4}}\left(k_{1},k_{2},k_{3},k_{4}\right)=\frac{1}{2}\left(T^{a_{2}}T^{a_{3}}\right)_{a_{1}a_{4}}\,\mathcal{A}(1,2,3,4)+\frac{1}{2}\left(T^{a_{3}}T^{a_{2}}\right)_{a_{1}a_{4}}\,\mathcal{A}(1,3,2,4)\,.\label{eq:4partons_color_decomp}
\end{equation}
The square of the amplitude, summed over colors, can in general be
written in a matrix form

\begin{equation}
\left|\mathcal{M}\right|^{2}=\vec{\mathcal{A}}^{\,\dagger}\,\mathbf{C}\,\vec{\mathcal{A}}\,,
\end{equation}
where $\mathbf{C}$ is the color matrix and $\vec{\mathcal{A}}$ is a column
vector constructed from the partial amplitudes. For the present simple
case 

\begin{equation}
\mathbf{C}=\frac{1}{4}N_{c}^{2}N_{A}\left(\begin{array}{cc}
1 & \frac{1}{2}\\
\frac{1}{2} & 1
\end{array}\right)\,,\label{eq:colormatrix}
\end{equation}
and $\vec{\mathcal{A}}$ is given in Table~\ref{tab:partial_amps_4part}. 

In order to calculate the TMD operator structure, we need to insert
the appropriate gauge links instead of deltas summing over colors,
as reviewed in Section~\ref{sec:ColorFlowFR}

\begin{equation}
\mathcal{M}^{a_{1}a_{2}a_{3}a_{4}}\mathcal{M}_{a'_{1}a'_{2}a'_{3}a'_{4}}^{\dagger}\left(\mathcal{U}^{[+]}\right)^{a'_{2}a_{2}}\left(\mathcal{U}^{[+]}\right)^{a'_{3}a_{3}}\left(\mathcal{U}^{[-]\dagger}\right)^{a_{4}a'_{4}}F_{a_{1}}^{i+}\left(\xi\right)F_{a'_{1}}^{i+}\left(0\right)\,.
\end{equation}
The structure of TMD operators is most conveniently (and inevitably)
expressed in the fundamental representation. Thus the Wilson lines
are transformed to the fundamental representation using
\begin{equation}
\left(\mathcal{U}^{[\pm]}\right)_{ab}=\frac{1}{T_{F}}\mathrm{Tr}\left[t^{a}\mathcal{U}^{[\pm]}t^{b}\mathcal{U}^{[\pm]\dagger}\right]\,.
\end{equation}
Next, the decomposition (\ref{eq:4partons_color_decomp}) is used
to represent the above expression in the following general form
\begin{equation}
\vec{\mathcal{A}}^{\,\dagger}\,\boldsymbol{F}\,\vec{\mathcal{A}}\,,
\end{equation}
where $\boldsymbol{F}$ is the matrix of the TMD operators containing
implicitly the color factors of the hard process. In most cases, it
is reasonable to keep these color factors together with the hard matrix
elements. Thus, to avoid double counting, we divide the elements of
$\boldsymbol{F}$ by the corresponding color factors of the square
of the amplitude, but without the summation of indices where the field
operators are attached (this corresponds to the elements of the matrix
$\mathbf{C}$ (\ref{eq:colormatrix}) divided by $N_{A}$). This leads
to the following definition of the TMD distribution matrix
\begin{equation}
\mathbf{\Phi}=2\int\frac{d\xi^{-}d^{2}\xi_{T}}{\left(2\pi\right)^{3}P^{+}}\,e^{ixP^{+}\xi^{-}-i\vec{k}_{T}\cdot\vec{\xi}_{T}}\,\left\langle P\right|\boldsymbol{F}\oslash\left(\frac{1}{N_{A}}\mathbf{C}\right)\left|P\right\rangle \,,\label{eq:Phi_def}
\end{equation}
where the symbol $\oslash$ represents the Hadamard division, \textit{i.e.}
the element-wise division: $\left(\boldsymbol{A}\oslash\boldsymbol{B}\right)_{ij}=\boldsymbol{A}_{ij}/\boldsymbol{B}_{ij}$.
It may happen, for certain multiparticle processes, that some elements
of the color matrix $\mathbf{C}$ vanish, but the corresponding elements
of $\boldsymbol{F}$ are non-zero. In that case, we need to modify
the above prescription. We shall come back to this point when discussing
processes where this happens. An additional motivation to divide out
the color factors from the TMD operators is that one could in principle
use the results with matrix elements not represented in the color-ordered
form. 

With the above definitions, the cross section for a collinear parton
$a$ to scatter off a gluon with some internal transverse momentum
and producing certain number of colored partons, can be generically
written as
\begin{equation}
d\sigma_{ag\rightarrow X}=\int\,\vec{\mathcal{A}}^{\,\dagger}\left(\boldsymbol{C}\circ\mathbf{\Phi}_{ag\rightarrow X}\right)\vec{\mathcal{A}}\,\,d\Gamma\,,\label{eq:xsect}
\end{equation}
where $d\Gamma$ represents all pre-factors, phase space, and convolution
in $x$ and $k_{T}$. The symbol $\circ$ is the Hadamard (element-wise)
multiplication, $\left(\boldsymbol{A}\circ\boldsymbol{B}\right)_{ij}=\boldsymbol{A}_{ij}\boldsymbol{B}_{ij}$.

In the present example of four gluons, the TMD gluon distribution
matrix reads
\begin{equation}
\mathbf{\Phi}_{gg\rightarrow gg}=\left(\begin{array}{cc}
\Phi_{1} & \Phi_{2}\\
\Phi_{2} & \Phi_{1}
\end{array}\right)\,,
\end{equation}
with two independent TMD gluon distributions expressed in terms of
the basis distributions:
\begin{equation}
\Phi_{1}=\frac{1}{2N_{c}^{2}}\left(N_{c}^{2}\mathcal{F}_{\text{gg}}^{(1)}-2\mathcal{F}_{\text{gg}}^{(3)}+\mathcal{F}_{\text{gg}}^{(4)}+\mathcal{F}_{\text{gg}}^{(5)}+N_{c}^{2}\mathcal{F}_{\text{gg}}^{(6)}\right)\,,
\end{equation}

\begin{equation}
\Phi_{2}=\frac{1}{N_{c}^{2}}\left(N_{c}^{2}\mathcal{F}_{\text{gg}}^{(2)}-2\mathcal{F}_{\text{gg}}^{(3)}+\mathcal{F}_{\text{gg}}^{(4)}+\mathcal{F}_{\text{gg}}^{(5)}+N_{c}^{2}\mathcal{F}_{\text{gg}}^{(6)}\right)\,.
\end{equation}
For more complicated processes with gluons it is useful to write the
above equations in matrix form:
\begin{equation}
\left(\begin{array}{c}
\Phi_{1}\\
\vdots\\
\Phi_{k}
\end{array}\right)=\mathbf{M}\left(\begin{array}{c}
\mathcal{F}_{\text{gg}}^{(1)}\\
\mathcal{F}_{\text{gg}}^{(2)}\\
\vdots\\
\mathcal{F}_{\text{gg}}^{(7)}
\end{array}\right)\,,\label{eq:MmatrixDef}
\end{equation}
where $\mathbf{M}$ is a matrix with $k$ rows and 7 columns. For
the present case, this matrix reads
\begin{equation}
\mathbf{M}_{gg\rightarrow gg}=\left(\begin{array}{ccccccc}
\frac{1}{2} & 0 & -\frac{1}{N_{c}^{2}} & \frac{1}{2N_{c}^{2}} & \frac{1}{2N_{c}^{2}} & \frac{1}{2} & 0\\
0 & 1 & -\frac{2}{N_{c}^{2}} & \frac{1}{N_{c}^{2}} & \frac{1}{N_{c}^{2}} & 1 & 0
\end{array}\right)\,.
\end{equation}

\begin{table}
\begin{centering}
\begin{tabular}{c|c|c|c}
\toprule
$g_{1}g_{4}\rightarrow g_{2}g_{3}$ &  $g_{1}q_{4}\rightarrow g_{2}q_{3}$ & $g_{1}\bar{q}_{4}\rightarrow g_{2}\bar{q}_{3}$ & $g_{1}g_{4}\rightarrow q_{2}\bar{q}_{3}$\tabularnewline
\midrule
$\left(\begin{array}{c}
\mathcal{A}(1,2,3,4)\\
\mathcal{A}(1,3,2,4)
\end{array}\right)$ & $\left(\begin{array}{c}
\mathcal{A}(3,1,2,4)\\
\mathcal{A}(3,2,1,4)
\end{array}\right)$ & $\left(\begin{array}{c}
\mathcal{A}(4,1,2,3)\\
\mathcal{A}(4,2,1,3)
\end{array}\right)$ & $\left(\begin{array}{c}
\mathcal{A}(2,1,4,3)\\
\mathcal{A}(2,4,1,3)
\end{array}\right)$\tabularnewline
\bottomrule
\end{tabular}
\par\end{centering}
\caption{Definitions of the vector of partial amplitudes $\vec{\mathcal{A}}$
for all four-parton processes. The subscripts in the sub-process indication
correspond to the momenta enumeration. \label{tab:partial_amps_4part}}
\end{table}

In a similar fashion, one can derive the matrices $\mathbf{\Phi}$
and $\mathbf{M}$ for other 4 parton channels. The only difference
is that for processes with quarks, we always use the color flow color
decomposition of an amplitude. For the channel
\begin{equation}
g\left(k_{1}\right)q\left(k_{4}\right)\rightarrow g\left(k_{2}\right)q\left(k_{3}\right)\,,
\end{equation}
we obtain
\begin{equation}
\mathbf{\Phi}_{gq\rightarrow gq}=\left(\begin{array}{cc}
\Phi_{2} & \Phi_{1}\\
\Phi_{1} & \Phi_{1}
\end{array}\right)\,,
\end{equation}
with the $\Phi_{i}$ given in Table~\ref{tab:Mmatrices_4part}. For
a similar process with an anti-quark we get
\begin{equation}
\mathbf{\Phi}_{g\bar{q}\rightarrow g\bar{q}}=\left(\begin{array}{cc}
\Phi_{1} & \Phi_{1}\\
\Phi_{1} & \Phi_{2}
\end{array}\right)\,.
\end{equation}
Finally, for
\begin{equation}
g\left(k_{1}\right)g\left(k_{4}\right)\rightarrow q\left(k_{2}\right)\bar{q}\left(k_{3}\right)
\end{equation}
we have
\begin{equation}
\mathbf{\Phi}_{gg\rightarrow q\bar{q}}=\left(\begin{array}{cc}
\Phi_{1} & \Phi_{2}\\
\Phi_{2} & \Phi_{1}
\end{array}\right)\,.
\end{equation}
The partial amplitude vectors $\vec{\mathcal{A}}$ for the above cases
are listed in Table~\ref{tab:partial_amps_4part}.

\begin{table}
\begin{centering}
\begin{tabular}{c||c|c||c}
\toprule
\multicolumn{2}{c|}{$g_{1}g_{4}\rightarrow g_{2}g_{3}$} & \multicolumn{2}{c}{$g_{1}g_{4}\rightarrow q_{2}\bar{q}_{3}$}\tabularnewline
\midrule
\multicolumn{2}{c|}{$\left(\begin{array}{cccccc}
\frac{1}{2} & 0 & -\frac{1}{N_{c}^{2}} & \frac{1}{2N_{c}^{2}} & \frac{1}{2N_{c}^{2}} & \frac{1}{2}\\
0 & 1 & -\frac{2}{N_{c}^{2}} & \frac{1}{N_{c}^{2}} & \frac{1}{N_{c}^{2}} & 1
\end{array}\right)$} & \multicolumn{2}{c}{$\left(\begin{array}{cccccc}
\frac{N_{c}^{2}}{N_{A}} & 0 & -\frac{1}{N_{A}} & 0 & 0 & 0\\
0 & -N_{c}^{2} & 1 & 0 & 0 & 0
\end{array}\right)$}\tabularnewline
\midrule
\multicolumn{2}{c|}{$g_{1}q_{4}\rightarrow g_{2}q_{3}$} & \multicolumn{2}{c}{$g_{1}\bar{q}_{4}\rightarrow g_{2}\bar{q}_{3}$}\tabularnewline
\midrule 
\multicolumn{2}{c|}{$\left(\begin{array}{cc}
1 & 0\\
-\frac{1}{N_{A}} & \frac{N_{c}^{2}}{N_{A}}
\end{array}\right)$} & \multicolumn{2}{c}{$\left(\begin{array}{cc}
1 & 0\\
-\frac{1}{N_{A}} & \frac{N_{c}^{2}}{N_{A}}
\end{array}\right)$}\tabularnewline
\bottomrule 
\end{tabular}
\par\end{centering}
\caption{Matrices $\mathbf{M}$ of structures appearing in four-parton processes.
The subscripts in the sub-process indication correspond to the momenta
enumeration. \label{tab:Mmatrices_4part}}
\end{table}

\subsection{Five partons}

The calculation of the TMD gluon distributions with 5 colored partons
proceeds in the same fashion, but is technically more complicated.
Also, a new feature appears. Certain color factors, building up the
matrix $\mathbf{C}$, vanish for some processes. However, some of
the corresponding TMD operators do not vanish (more precisely, we
mean here corresponding elements of the $\mathbf{F}$ matrix). It
is a special property of the TMD factorization: certain color flows
would not contribute in the collinear factorization (where only the
matrix $\mathbf{C}$ appears), but they do contribute if the TMD gluon
distributions are considered. Thus we need to modify the definition
of the TMD gluon distribution matrix $\mathbf{\Phi}$ (\ref{eq:Phi_def})
and the Eq.~(\ref{eq:xsect}) for such processes. In both formulas,
instead of the matrix $\mathbf{C}$ (which has zeros), we use the
matrix $\mathbf{C}'$ with elements
\begin{equation}
\mathbf{C}'_{ij}=\begin{cases}
\mathbf{C}_{ij} & \textrm{if }\mathbf{C}_{ij}\neq0\\
1 & \textrm{if }\mathbf{C}_{ij}=0
\end{cases}\,. \label{eq:Cprime}
\end{equation}
This is a simple way to extract the hard matrix element color factors
only from those TMD operators, for which the color factor is nonzero.
For reader's convenience, the color factors for 5 parton processes
in the color-ordered-amplitude representation are collected in~\ref{sec:App_Color}
(they were cross-checked with \citep{Kuijf:1991kn},\citep{DelDuca1999}).

Below, we present the TMD gluon distribution matrices $\mathbf{\Phi}$
for various channels. The vectors $\vec{\mathcal{A}}$ of partial
amplitudes, corresponding to the entries of the matrices $\mathbf{\Phi}$,
are given in Table~\ref{tab:partial_amps_5part} in~\ref{sec:App_Avec_M}. The TMD gluon distributions
$\Phi_{i}$ building up these matrices, are expressed through the
'basis' distributions (\ref{eq:Fqg1})-(\ref{eq:Fgg7}), as given
by the $\mathbf{M}$ matrices listed in Table~\ref{tab:Mmatrices_5part} (\ref{sec:App_Avec_M}).
The $\mathbf{M}$ matrices for processes in which an incoming and
outgoing quarks are replaced by incoming and outgoing anti-quarks
are the same.

For the pure gluonic process,
\begin{equation}
g\left(k_{1}\right)g\left(k_{5}\right)\rightarrow g\left(k_{2}\right)g\left(k_{3}\right)g\left(k_{4}\right)\,,
\end{equation}
we obtain
\begin{equation}
\mathbf{\Phi}_{gg\rightarrow ggg}=\left(\begin{array}{cccccc}
\Phi_{1} & \Phi_{2} & \Phi_{2} & \Phi_{3} & \Phi_{3} & \Phi_{4}^{*}\\
\Phi_{2} & \Phi_{1} & \Phi_{3} & \Phi_{4}^{*} & \Phi_{2} & \Phi_{3}\\
\Phi_{2} & \Phi_{3} & \Phi_{1} & \Phi_{2} & \Phi_{4}^{*} & \Phi_{3}\\
\Phi_{3} & \Phi_{4}^{*} & \Phi_{2} & \Phi_{1} & \Phi_{3} & \Phi_{2}\\
\Phi_{3} & \Phi_{2} & \Phi_{4}^{*} & \Phi_{3} & \Phi_{1} & \Phi_{2}\\
\Phi_{4}^{*} & \Phi_{3} & \Phi_{3} & \Phi_{2} & \Phi_{2} & \Phi_{1}
\end{array}\right)\,.
\end{equation}
The $\Phi_{i}$ gluon distributions are listen in the first row of
Table~\ref{tab:Mmatrices_5part}. This process has the property mentioned
in the beginning of this subsection. The entries for which the color
factors are zero are marked with the asterix $^{*}$.

For
\begin{equation}
g\left(k_{1}\right)g\left(k_{5}\right)\rightarrow q\left(k_{2}\right)\bar{q}\left(k_{3}\right)g\left(k_{4}\right)\,,
\end{equation}
we get
\begin{equation}
\mathbf{\Phi}_{gg\rightarrow q\bar{q}g}=\left(\begin{array}{cccccc}
\Phi_{1} & \Phi_{2} & \Phi_{2} & \Phi_{3} & \Phi_{3} & \Phi_{4}\\
\Phi_{2} & \Phi_{2} & \Phi_{5} & \Phi_{6} & \Phi_{3} & \Phi_{3}\\
\Phi_{2} & \Phi_{5} & \Phi_{2} & \Phi_{3} & \Phi_{6} & \Phi_{3}\\
\Phi_{3} & \Phi_{6} & \Phi_{3} & \Phi_{2} & \Phi_{5} & \Phi_{2}\\
\Phi_{3} & \Phi_{3} & \Phi_{6} & \Phi_{5} & \Phi_{2} & \Phi_{2}\\
\Phi_{4} & \Phi_{3} & \Phi_{3} & \Phi_{2} & \Phi_{2} & \Phi_{1}
\end{array}\right)\,,
\end{equation}
with the TMD gluon distributions given in the second row of Table~\ref{tab:Mmatrices_5part}.

For the process with initial state quark 
\begin{equation}
g\left(k_{1}\right)q\left(k_{5}\right)\rightarrow g\left(k_{2}\right)g\left(k_{3}\right)q\left(k_{4}\right)\,,
\end{equation}
or anti-quark, we obtain, respectively
\begin{equation}
\mathbf{\Phi}_{gq\rightarrow ggq}=\left(\begin{array}{cccccc}
\Phi_{1} & \Phi_{2} & \Phi_{3} & \Phi_{4} & \Phi_{5} & \Phi_{4}\\
\Phi_{2} & \Phi_{1} & \Phi_{5} & \Phi_{4} & \Phi_{3} & \Phi_{4}\\
\Phi_{3} & \Phi_{5} & \Phi_{3} & \Phi_{4} & \Phi_{6} & \Phi_{4}\\
\Phi_{4} & \Phi_{4} & \Phi_{4} & \Phi_{4} & \Phi_{4} & \Phi_{4}\\
\Phi_{5} & \Phi_{3} & \Phi_{6} & \Phi_{4} & \Phi_{3} & \Phi_{4}\\
\Phi_{4} & \Phi_{4} & \Phi_{4} & \Phi_{4} & \Phi_{4} & \Phi_{4}
\end{array}\right)\,,
\end{equation}
and
\begin{equation}
\mathbf{\Phi}_{g\bar{q}\rightarrow gg\bar{q}}=\left(\begin{array}{cccccc}
\Phi_{4} & \Phi_{4} & \Phi_{4} & \Phi_{4} & \Phi_{4} & \Phi_{4}\\
\Phi_{4} & \Phi_{4} & \Phi_{4} & \Phi_{4} & \Phi_{4} & \Phi_{4}\\
\Phi_{4} & \Phi_{4} & \Phi_{3} & \Phi_{3} & \Phi_{6} & \Phi_{5}\\
\Phi_{4} & \Phi_{4} & \Phi_{3} & \Phi_{1} & \Phi_{5} & \Phi_{2}\\
\Phi_{4} & \Phi_{4} & \Phi_{6} & \Phi_{5} & \Phi_{3} & \Phi_{3}\\
\Phi_{4} & \Phi_{4} & \Phi_{5} & \Phi_{2} & \Phi_{3} & \Phi_{1}
\end{array}\right)\,,
\end{equation}
with the TMD gluon distributions given in the third row of Table~\ref{tab:Mmatrices_5part}.
These matrices differ by the permutations of the entries, which has
its origin in a slightly different color decomposition for quarks
and anti-quarks. Namely the order of quark--anti-quark lines (with
the outgoing-momenta convention) is reversed in one case with respect
to the other.

Finally, the processes with two quark--anti-quark pairs, with incoming
quark 
\begin{equation}
g\left(k_{1}\right)q\left(k_{5}\right)\rightarrow q\left(k_{2}\right)\bar{q}\left(k_{3}\right)q\left(k_{4}\right)\,,
\end{equation}
or anti-quark, involve respectively
\begin{equation}
\mathbf{\Phi}_{gq\rightarrow q\bar{q}q}=\left(\begin{array}{cccc}
\Phi_{1} & 0 & \Phi_{1} & \Phi_{1}\\
0 & \Phi_{2} & \Phi_{3} & \Phi_{1}\\
\Phi_{1} & \Phi_{3} & \Phi_{2} & 0\\
\Phi_{1} & \Phi_{1} & 0 & \Phi_{1}
\end{array}\right)\,,
\end{equation}
and
\begin{equation}
\mathbf{\Phi}_{g\bar{q}\rightarrow q\bar{q}\bar{q}}=\left(\begin{array}{cccc}
\Phi_{1} & 0 & \Phi_{1} & \Phi_{1}\\
0 & \Phi_{2} & \Phi_{1} & \Phi_{3}\\
\Phi_{1} & \Phi_{1} & \Phi_{1} & 0\\
\Phi_{1} & \Phi_{3} & 0 & \Phi_{2}
\end{array}\right)\,.
\end{equation}
The TMD gluon distributions appearing in these matrices are listed
in the fourth row of Table~\ref{tab:Mmatrices_5part}. Interestingly,
for this case, not only some of the color factors vanish, but also
the corresponding TMDs.

As the potential phenomenological application of the results (in short
run) concerns rather the large $N_{c}$ limit, we present the relevant
matrices in this limit in~\ref{sec:App_LargeNc}.

\subsection{Six partons}

Six parton processes do not involve new features, except more channels
and more involved calculations. The vectors $\vec{\mathcal{A}}$ of
the partial amplitudes, and the $\mathbf{M}$ matrices are given in
Tables~\ref{tab:partial_amps_6part} and \ref{tab:Mmatrices_6part1}-\ref{tab:Mmatrices_6part4} 
in~\ref{sec:App_Avec_M}. The $\mathbf{M}$ matrices for processes in which an incoming and
outgoing quarks are replaced by incoming and outgoing anti-quarks
are the same. Below, we present results for the TMD gluon distribution
matrices $\mathbf{\Phi}$ for all channels. The number of partial
amplitudes necessitates the use of block matrices to compactify the
notation.

For the six-gluon process,
\begin{equation}
g\left(k_{1}\right)g\left(k_{6}\right)\rightarrow g\left(k_{2}\right)g\left(k_{3}\right)g\left(k_{4}\right)g\left(k_{5}\right)\,,
\end{equation}
the $\mathbf{\Phi}$ matrix is

\begin{equation}
\mathbf{\Phi}_{gg\rightarrow gggg}=\left(\begin{array}{cccc}
T_{1} & T_{2} & T_{3} & T_{4}\\
T_{2} & T_{1} & T_{5} & T_{6}\\
T_{3}^{\intercal} & T_{5} & T_{1} & T_{7}\\
T_{4}^{\intercal} & T_{6}^{\intercal} & T_{7} & T_{1}
\end{array}\right)\,,
\end{equation}
where $T_{i}$ are 6$\times$6 block matrices given by equations~(\ref{eq:6g_T1})-(\ref{eq:6g_T7}) in~\ref{ssec:6g_Ts}.

In the present case we have two TMD operators, for which the color
factor vanishes -- $\Phi_{4}^{*}$ and $\Phi_{8}^{*}$ (we remind,
that we mark these matrix elements with an asterix). The full list
of the TMD gluon distributions is given in the Table~\ref{tab:Mmatrices_6part1}.

Next consider the process

\begin{equation}
g\left(k_{1}\right)g\left(k_{6}\right)\rightarrow q\left(k_{2}\right)\bar{q}\left(k_{3}\right)g\left(k_{4}\right)g\left(k_{5}\right)\,.
\end{equation}
The TMD matrix reads

\begin{equation}
\mathbf{\Phi}_{gg\rightarrow q\bar{q}gg}=\left(\begin{array}{cccc}
T_{1} & T_{2} & T_{3} & T_{4}\\
T_{2}^{\intercal} & T_{5} & T_{6} & T_{7}\\
T_{3}^{\intercal} & T_{6} & T_{5} & T_{8}\\
T_{4}^{\intercal} & T_{7}^{\intercal} & T_{8}^{\intercal} & T_{9}
\end{array}\right)\,,
\end{equation}
where the blocks gathered in equations~(\ref{eq:gg_qqbgg_T1}) - (\ref{eq:gg_qqbgg_T9}) in~\ref{ssec:gg_qqbgg_Ts}.
The TMD gluon distributions are given in the Table~\ref{tab:Mmatrices_6part2}.

For the process
\begin{equation}
g\left(k_{1}\right)q\left(k_{6}\right)\rightarrow g\left(k_{2}\right)g\left(k_{3}\right)g\left(k_{4}\right)q\left(k_{5}\right)\,,
\end{equation}
the TMD matrix reads 

\begin{equation}
\mathbf{\Phi}_{gq\rightarrow gggq}=\left(\begin{array}{cccc}
T_{1} & T_{2} & T_{3} & T_{4}\\
T_{2}^{\intercal} & T_{5} & T_{6} & T_{7}\\
T_{3}^{\intercal} & T_{6} & T_{5} & T_{8}\\
T_{4}^{\intercal} & T_{7}^{\intercal} & T_{8} & T_{5}
\end{array}\right)\,,
\end{equation}
with the blocks expressed by equations~(\ref{eq:gq_gggq_T1}) - (\ref{eq:gq_gggq_T8}) in~\ref{ssec:gq_gggq_Ts}.
The TMD distributions are in Table~\ref{tab:Mmatrices_6part3}.

Similarly, for the process with the anti-quark

\begin{equation}
g\left(k_{1}\right)\bar{q}\left(k_{6}\right)\rightarrow g\left(k_{2}\right)g\left(k_{3}\right)g\left(k_{4}\right)\bar{q}\left(k_{5}\right)\,,
\end{equation}
we get

\begin{equation}
\mathbf{\Phi}_{g\bar{q}\rightarrow ggg\bar{q}}=\left(\begin{array}{cccc}
T_{1} & T_{1} & T_{1} & T_{1}\\
T_{1} & T_{2} & T_{3} & T_{4}\\
T_{1} & T_{3} & T_{2} & T_{5}\\
T_{1} & T_{4}^{\intercal} & T_{5} & T_{2}
\end{array}\right)\,,
\end{equation}
with the blocks given by equations~(\ref{eq:gqb_gggqb_T1}) - (\ref{eq:gqb_gggqb_T5}) in~\ref{ssec:gqb_gggqb_Ts}.
The TMD distributions are in Table~\ref{tab:Mmatrices_6part3}.

Processes with two quark--anti-quark pairs have smaller number of
partial amplitudes. For the process

\begin{equation}
g\left(k_{1}\right)g\left(k_{6}\right)\rightarrow q\left(k_{2}\right)\bar{q}\left(k_{3}\right)q\left(k_{4}\right)\bar{q}\left(k_{5}\right)\,,
\end{equation}
we obtain

\begin{equation}
\mathbf{\Phi}_{gg\rightarrow q\bar{q}q\bar{q}}=\left(\begin{array}{cc}
T_{1} & T_{2}\\
T_{2} & T_{1}
\end{array}\right)\,,
\end{equation}
with only two blocks:

\begin{equation}
T_{1}=\left(\begin{array}{cccccc}
\Phi_{1} & 0 & \Phi_{2} & \Phi_{3} & 0 & \Phi_{2}\\
0 & \Phi_{4} & 0 & 0 & \Phi_{5} & 0\\
\Phi_{2} & 0 & \Phi_{1} & \Phi_{2} & 0 & \Phi_{3}\\
\Phi_{3} & 0 & \Phi_{2} & \Phi_{1} & 0 & \Phi_{2}\\
0 & \Phi_{5} & 0 & 0 & \Phi_{4} & 0\\
\Phi_{2} & 0 & \Phi_{3} & \Phi_{2} & 0 & \Phi_{1}
\end{array}\right),\,\,\,\,T_{2}=\left(\begin{array}{cccccc}
\Phi_{1} & \Phi_{3} & \Phi_{1} & \Phi_{3} & \Phi_{1} & \Phi_{3}\\
\Phi_{3} & \Phi_{6} & \Phi_{1} & \Phi_{1} & \Phi_{6} & \Phi_{3}\\
\Phi_{1} & \Phi_{1} & \Phi_{1} & \Phi_{3} & \Phi_{3} & \Phi_{3}\\
\Phi_{3} & \Phi_{1} & \Phi_{3} & \Phi_{1} & \Phi_{3} & \Phi_{1}\\
\Phi_{1} & \Phi_{6} & \Phi_{3} & \Phi_{3} & \Phi_{6} & \Phi_{1}\\
\Phi_{3} & \Phi_{3} & \Phi_{3} & \Phi_{1} & \Phi_{1} & \Phi_{1}
\end{array}\right).
\end{equation}
The TMD distributions are given in the Table~\ref{tab:Mmatrices_6part4}.

For the process 

\begin{equation}
g\left(k_{1}\right)q\left(k_{6}\right)\rightarrow g\left(k_{2}\right)q\left(k_{3}\right)\bar{q}\left(k_{4}\right)q\left(k_{5}\right)\,,
\end{equation}
we have

\begin{equation}
\mathbf{\Phi}_{gq\rightarrow gq\bar{q}q}=\left(\begin{array}{cc}
T_{1} & T_{2}\\
T_{2}^{\intercal} & T_{3}
\end{array}\right)\,,
\end{equation}
with three different blocks

\begin{equation}
T_{1}=\left(\begin{array}{cccccc}
\Phi_{1} & 0 & \Phi_{2} & \Phi_{3} & 0 & \Phi_{2}\\
0 & \Phi_{4} & \Phi_{5}^{*} & 0 & \Phi_{3} & \Phi_{6}^{*}\\
\Phi_{2} & \Phi_{5}^{*} & \Phi_{4} & \Phi_{3} & 0 & \Phi_{4}\\
\Phi_{3} & 0 & \Phi_{3} & \Phi_{3} & 0 & \Phi_{3}\\
0 & \Phi_{3} & 0 & 0 & \Phi_{3} & 0\\
\Phi_{2} & \Phi_{6}^{*} & \Phi_{4} & \Phi_{3} & 0 & \Phi_{4}
\end{array}\right),\,\,\,\,T_{2}=\left(\begin{array}{cccccc}
\Phi_{1} & \Phi_{3} & \Phi_{1} & \Phi_{3} & \Phi_{1} & \Phi_{3}\\
\Phi_{2} & \Phi_{3} & \Phi_{1} & \Phi_{2} & \Phi_{7} & \Phi_{3}\\
\Phi_{2} & \Phi_{3} & \Phi_{1} & \Phi_{2} & \Phi_{2} & \Phi_{3}\\
\Phi_{3} & \Phi_{3} & \Phi_{3} & \Phi_{3} & \Phi_{3} & \Phi_{3}\\
\Phi_{3} & \Phi_{3} & \Phi_{3} & \Phi_{3} & \Phi_{3} & \Phi_{3}\\
\Phi_{2} & \Phi_{3} & \Phi_{3} & \Phi_{7} & \Phi_{2} & \Phi_{3}
\end{array}\right),
\end{equation}

\begin{equation}
T_{3}=\left(\begin{array}{cccccc}
\Phi_{4} & 0 & \Phi_{2} & \Phi_{4} & \Phi_{8}^{*} & \Phi_{3}\\
0 & \Phi_{3} & 0 & 0 & \Phi_{3} & 0\\
\Phi_{2} & 0 & \Phi_{1} & \Phi_{2} & 0 & \Phi_{3}\\
\Phi_{4} & 0 & \Phi_{2} & \Phi_{4} & \Phi_{9}^{*} & \Phi_{3}\\
\Phi_{8}^{*} & \Phi_{3} & 0 & \Phi_{9}^{*} & \Phi_{4} & 0\\
\Phi_{3} & 0 & \Phi_{3} & \Phi_{3} & 0 & \Phi_{3}
\end{array}\right).
\end{equation}
Note, that in this process there appear both the vanishing structures
for vanishing color factors and non-vanishing structures for vanishing
color factors. The list of the TMD distributions is given in the Table~\ref{tab:Mmatrices_6part4}. Similarly for the process
with an anti-quark, we get:

\begin{equation}
\mathbf{\Phi}_{g\bar{q}\rightarrow gq\bar{q}\bar{q}}=\left(\begin{array}{cc}
T_{1} & T_{2}\\
T_{2} & T_{3}
\end{array}\right)\,,
\end{equation}
with

\begin{equation}
T_{1}=\left(\begin{array}{cccccc}
\Phi_{3} & 0 & \Phi_{3} & \Phi_{3} & 0 & \Phi_{3}\\
0 & \Phi_{4} & \Phi_{6}^{*} & 0 & \Phi_{3} & \Phi_{5}^{*}\\
\Phi_{3} & \Phi_{6}^{*} & \Phi_{4} & \Phi_{2} & 0 & \Phi_{4}\\
\Phi_{3} & 0 & \Phi_{2} & \Phi_{1} & 0 & \Phi_{2}\\
0 & \Phi_{3} & 0 & 0 & \Phi_{3} & 0\\
\Phi_{3} & \Phi_{5}^{*} & \Phi_{4} & \Phi_{2} & 0 & \Phi_{4}
\end{array}\right),\,\,\,T_{2}=\left(\begin{array}{cccccc}
\Phi_{3} & \Phi_{3} & \Phi_{3} & \Phi_{3} & \Phi_{3} & \Phi_{3}\\
\Phi_{3} & \Phi_{7} & \Phi_{2} & \Phi_{1} & \Phi_{3} & \Phi_{2}\\
\Phi_{3} & \Phi_{2} & \Phi_{7} & \Phi_{3} & \Phi_{3} & \Phi_{2}\\
\Phi_{3} & \Phi_{1} & \Phi_{3} & \Phi_{1} & \Phi_{3} & \Phi_{1}\\
\Phi_{3} & \Phi_{3} & \Phi_{3} & \Phi_{3} & \Phi_{3} & \Phi_{3}\\
\Phi_{3} & \Phi_{2} & \Phi_{2} & \Phi_{1} & \Phi_{3} & \Phi_{2}
\end{array}\right),
\end{equation}

\begin{equation}
T_{3}=\left(\begin{array}{cccccc}
\Phi_{3} & 0 & \Phi_{3} & \Phi_{3} & 0 & \Phi_{3}\\
0 & \Phi_{4} & \Phi_{9}^{*} & 0 & \Phi_{3} & \Phi_{8}^{*}\\
\Phi_{3} & \Phi_{9}^{*} & \Phi_{4} & \Phi_{2} & 0 & \Phi_{4}\\
\Phi_{3} & 0 & \Phi_{2} & \Phi_{1} & 0 & \Phi_{2}\\
0 & \Phi_{3} & 0 & 0 & \Phi_{3} & 0\\
\Phi_{3} & \Phi_{8}^{*} & \Phi_{4} & \Phi_{2} & 0 & \Phi_{4}
\end{array}\right)\,.
\end{equation}

The large $N_{c}$ limits of gluon distributions for 6 parton processes
were gathered in Tables~\ref{tab:Mlimits_6part1}-\ref{tab:Mlimits_6part3}
in~\ref{sec:App_LargeNc}. Additionally, we collect the
color factors for all processes in~\ref{sec:App_Color}.

\section{Large $N_{c}$ analysis for arbitrary number of gluons}

\label{sec:LargeNc}

In this section, we shall utilize the color flow method to give the
large $N_{c}$ results for a process with $n$ gluons
\begin{equation}
g\left(k_{1}\right)g\left(k_{n}\right)\rightarrow g\left(k_{2}\right)\dots g\left(k_{n-1}\right)\,.
\end{equation}

We shall use the fact that the color flow decomposition (\ref{eq:glue_color_flow_decomp})
involves all $\left(n-1\right)!$ partial amplitudes which are the
same as in the fundamental decomposition (\ref{eq:glue_color_decomp_fund}).
Therefore, the leading $N_{c}$ contribution is given by the partial
amplitudes squared (the interference terms are subleading) \citep{Mangano:1987xk}
\begin{equation}
\left|\mathcal{M}\right|^{2}=\mathcal{C}\sum_{\pi\in S_{n}/Z_{n}}\left\{ \left|\mathcal{A}\left(\pi\left(1\right),\dots,\pi\left(n\right)\right)\right|^{2}+\mathcal{O}\left(\frac{1}{N_{c}^{2}}\right)\right\} \,,\label{eq:ColorSupp}
\end{equation}
with $\mathcal{C}$ being some color coefficient. Note, that if we
used the adjoint color decomposition to reduce the number of partial
amplitudes only to the linearly independent ones, as we did in the
previous section, we would not be able to claim (\ref{eq:ColorSupp}).
Consequently, the general analysis of large $N_{c}$ would be very
difficult. Therefore, there is a trade off: switching to a general
argumentation requires giving up the advantage of using minimal number
of amplitudes. In practice, however, any partial amplitude can be
easily calculated numerically, so the real loss is not so big.

Based on the above, the idea is to calculate first the diagonal elements
of matrix $\mathbf{\Phi}$, as they will definitely contribute in
the large $N_{c}$ limit. This would be the final answer, if there
is no enhancement of powers of $N_{c}^{2}$ for some of the non-diagonal
elements. In fact, as we shall see, the enhancement indeed occurs,
but still the TMD gluon distribution appearing off the diagonal is
numerically small.

Let us start with calculating the diagonal elements of the TMD gluon
distribution matrix $\mathbf{\Phi}$. It is sufficient to consider
only the following diagrams:

\begin{tabular}{>{\centering}m{0.87\columnwidth}>{\centering}m{0.05\columnwidth}}
\centering{}\bigskip{}
\includegraphics[width=10cm]{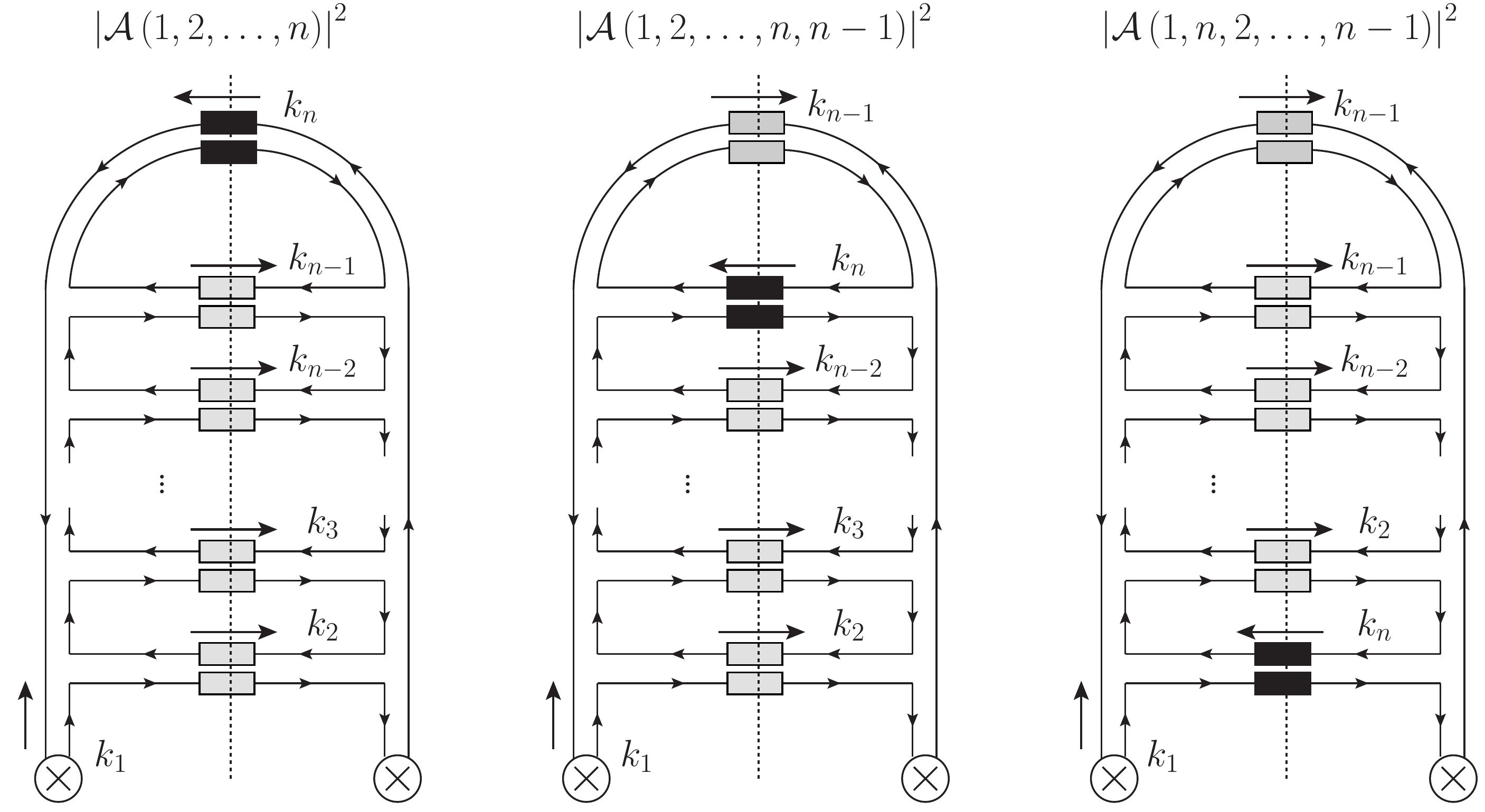} & \centering{}\centering{(\myref )}
\refmyref{largeNc_diag1}\tabularnewline
\end{tabular}

\noindent The first diagram from the left corresponds to the partial
amplitude squared $\left|\mathcal{A}\left(1,2,\dots,n\right)\right|^{2}$
and the TMD operator reads (after dividing by the corresponding color
factor)
\begin{equation}
\frac{N_{c}^{n-3}}{N_{c}^{n-2}}\mathrm{Tr}\left\{ F\left(\xi\right)\mathcal{U}^{\left[-\right]\dagger}F\left(0\right)\mathcal{U}^{\left[+\right]}\right\} \mathrm{Tr}\mathcal{U}^{\left[\square\right]\dagger}\,\rightsquigarrow\,\mathcal{F}_{gg}^{\left(1\right)}\,,\label{eq:largeNc_r1}
\end{equation}
\textit{i.e.} it corresponds to the TMD $\mathcal{F}_{gg}^{\left(1\right)}$,
Eq.~(\ref{eq:Fgg1}). However, any permutation of the following $\left(n-2\right)$
final state legs will give the same contribution, thus, the set
\begin{equation}
\left\{ \left.\left|\mathcal{A}\left(1,\pi\left(2\right),\pi\left(3\right),\dots,\pi\left(n-1\right),n\right)\right|^{2}\right|\pi\in S_{n-2}\right\} \rightsquigarrow\,\mathcal{F}_{gg}^{\left(1\right)}\,.
\end{equation}
The second diagram, corresponding to $\left|\mathcal{A}\left(1,2,\dots,n,n-1\right)\right|^{2}$,
gives 
\begin{equation}
\frac{N_{c}^{n-4}}{N_{c}^{n-2}}\mathrm{Tr}\left\{ F\left(\xi\right)\mathcal{U}^{\left[+\right]\dagger}F\left(0\right)\mathcal{U}^{\left[+\right]}\right\} \mathrm{Tr}\mathcal{U}^{\left[\square\right]\dagger}\mathrm{Tr}\mathcal{U}^{\left[\square\right]}\rightsquigarrow\,\mathcal{F}_{gg}^{\left(6\right)}\,.
\end{equation}
Not only any permutation of final states will give the same result,
but also any diagram with leg $k_{n}$ permuted with $\left\{ 3,\dots,n-2\right\} $.
Thus
\begin{multline}
\left\{ \left.\left|\mathcal{A}\left(1,\pi\left(2\right),\dots,\pi\left(n-2\right),n,\pi\left(n-1\right)\right)\right|^{2}\right|\pi\in S_{n-2}\right\} \\
\cup\left\{ \left.\left|\mathcal{A}\left(1,\pi\left(2\right),\dots,\pi\left(n-3\right),n,\pi\left(n-2\right),\pi\left(n-1\right)\right)\right|^{2}\right|\pi\in S_{n-2}\right\} \\
\dots\\
\cup\left\{ \left.\left|\mathcal{A}\left(1,\pi\left(2\right),n,\pi\left(3\right),\dots,\pi\left(n-2\right),\pi\left(n-1\right)\right)\right|^{2}\right|\pi\in S_{n-2}\right\} \\
\rightsquigarrow\,\mathcal{F}_{gg}^{\left(6\right)}\,.
\end{multline}
Finally, the third diagram, gives complex conjugate of the operator
in (\ref{eq:largeNc_r1}), thus also $\mathcal{F}_{gg}^{\left(1\right)}$,
because of our assumption of the reality of the correlators. We get
therefore
\[
\left\{ \left.\left|\mathcal{A}\left(1,n,\pi\left(2\right),\pi\left(3\right),\dots,\pi\left(n-1\right)\right)\right|^{2}\right|\pi\in S_{n-2}\right\} \rightsquigarrow\,\mathcal{F}_{gg}^{\left(1\right)}\,.
\]

Now let us put together the above results, using the matrix notation
as in Section~\ref{sec:Results}. Let us define the partial amplitude
vector so that it preserves the block structure emerging above:
\begin{equation}
\vec{\mathcal{A}}=\left(\begin{array}{c}
\mathcal{A}(\widehat{1},2,\dots,n-2,n-1,\widehat{n})\\
\vdots\\
\mathcal{A}(\widehat{1},2,3,\dots,n-2,\widehat{n},n-1)\\
\vdots\\
\mathcal{A}(\widehat{1},2,3,\dots,\widehat{n},n-2,n-1)\\
\vdots\\
\mathcal{A}(\widehat{1},2,\widehat{n},3,\dots,n-2,n-1)\\
\vdots\\
\mathcal{A}(\widehat{1},\widehat{n},3,\dots,n-2,n-1)\\
\vdots
\end{array}\right)\,,
\end{equation}
where we have used hats to denote momenta with fixed position in a
given group (the actual ordering in each group doesn't matter). For
this choice of the vector $\vec{\mathcal{A}}$, the diagonal contribution
to the matrix $\mathbf{\Phi}$ at large $N_{c}$ reads
\begin{equation}
\mathbf{\Phi}_{\mathrm{diag}}=\left(\begin{array}{ccc}
T_{1} & 0 & 0\\
0 & T_{2} & 0\\
0 & 0 & T_{1}
\end{array}\right)\,,
\end{equation}
where 
\begin{equation}
T_{1}=\mathcal{F}_{gg}^{\left(1\right)}\boldsymbol{1}_{\left(n-2\right)!},\,\,\,\,\,\,T_{2}=\mathcal{F}_{gg}^{\left(6\right)}\boldsymbol{1}_{(n-3)\left(n-2\right)!}\,.
\end{equation}
For example, for $n=4$, we have explicitly
\begin{equation}
\mathbf{\Phi}_{\mathrm{diag}}=\left(\begin{array}{cccccc}
\mathcal{F}_{gg}^{\left(1\right)} & 0 & 0 & 0 & 0 & 0\\
0 & \mathcal{F}_{gg}^{\left(1\right)} & 0 & 0 & 0 & 0\\
0 & 0 & \mathcal{F}_{gg}^{\left(6\right)} & 0 & 0 & 0\\
0 & 0 & 0 & \mathcal{F}_{gg}^{\left(6\right)} & 0 & 0\\
0 & 0 & 0 & 0 & \mathcal{F}_{gg}^{\left(1\right)} & 0\\
0 & 0 & 0 & 0 & 0 & \mathcal{F}_{gg}^{\left(1\right)}
\end{array}\right)\,.
\end{equation}

Now, let us consider the nondiagonal elements. As said above, these
elements will be convoluted with partial amplitudes (interference
terms) whose color factors are suppressed by at least $1/N_{c}^{2}$
(to say it differently, the non-diagonal elements of the color matrix
$\mathbf{C}$ in (\ref{eq:xsect}), if it is calculated in fundamental
color decomposition, are subleading of at least $1/N_{c}^{2}$). Therefore,
they do not contribute in large $N_{c}$, unless some off-diagonal
TMD gluon distribution is enhanced by at least $N_{c}^{2}$. This
still might not be enough, but is a sign that a careful analysis has
to be carried.

The most suspicious non-diagonal elements are those, which correspond
to color flow diagrams with least number of loops. This is slightly
counter-intuitive, but we have to keep in mind that, by definition,
we divide the color factors out of the TMD (there are no vanishing
color factors for gluons in the color flow representation, unlike
for the adjoint representation). Thus, the enhancement may happen
if the diagrams with Wilson lines have much more loops than the pure
color factor diagrams. It is best to illustrate this by an explicit
example. Consider a 4 gluon process and the following interference
term:
\begin{equation}
\mathcal{A}\left(1,2,3,4\right)\mathcal{A}^{*}\left(1,4,2,3\right)\,.
\end{equation}
We have the following two leading diagrams for the color factor:

\begin{tabular}{>{\centering}m{0.87\columnwidth}>{\centering}m{0.05\columnwidth}}
\centering{}\bigskip{}
\includegraphics[width=6cm]{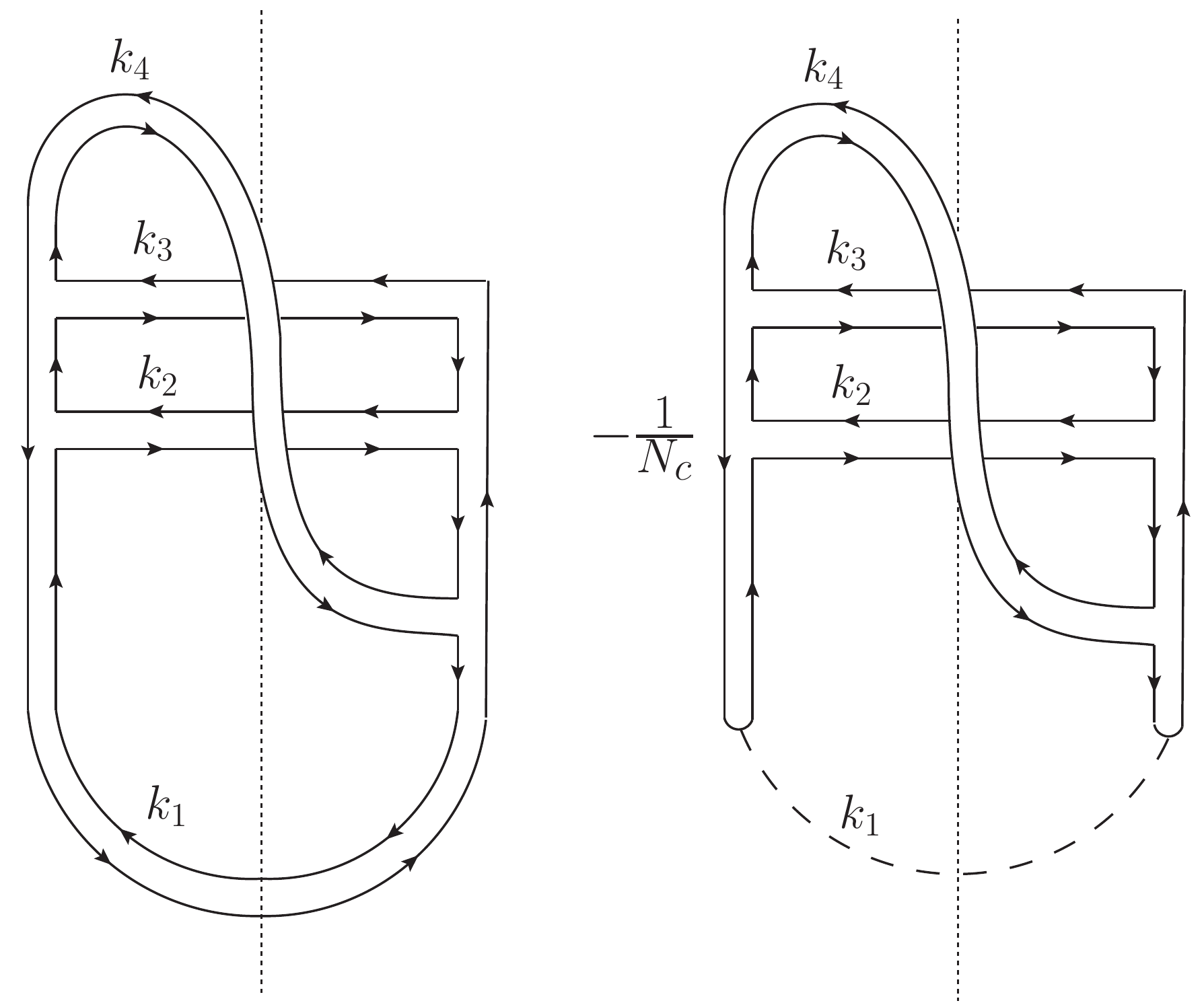} & \centering{}\centering{(\myref )}
\refmyref{largeNc_4g_1}\tabularnewline
\end{tabular}

\noindent The $U\left(1\right)$ colorless propagator for the $k_{1}$
leg stems from the projectors that have to be inserted for the final
state gluons, \textit{cf.} (\ref{eq:Projector}). Recall, that in general
these color factor diagrams have to be divided by $N_{A}$ to get
the color factor with 'open indices'. The first, M\"obius-loop-like
diagram, cancels with the second:
\begin{equation}
\frac{1}{N_{A}}\left(N_{c}^{2}-\frac{1}{N_{c}}N_{c}^{3}\right)=0\,.
\end{equation}
(We called the first diagram 'M\"obius-loop-like diagram' because
one of the internal loops shares its border with the external loop.)
Similar cancellation happens for the diagrams, where the $U\left(1\right)$
gluon appears for legs $k_{2}$ and $k_{3}$. The sub-leading diagrams
are those where $U\left(1\right)$ colorless gluon is $k_{4}$, \textit{i.e.}
it crosses the other legs:

\begin{tabular}{>{\centering}m{0.87\columnwidth}>{\centering}m{0.05\columnwidth}}
\centering{}\bigskip{}
\includegraphics[width=6cm]{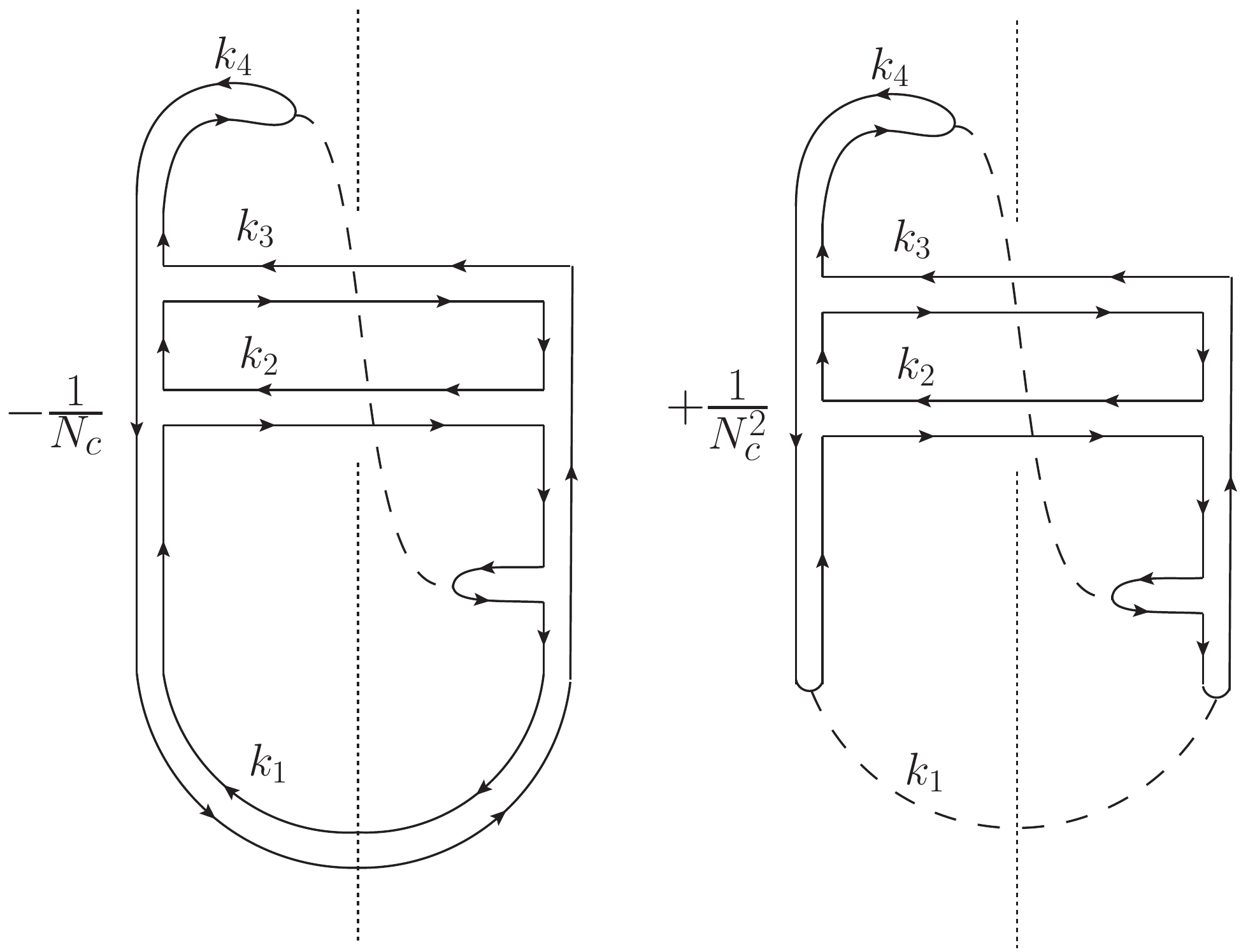} & \centering{}\centering{(\myref )}
\refmyref{largeNc_4g_2}\tabularnewline
\end{tabular}

\noindent In this case, we get
\begin{equation}
\frac{1}{N_{A}}\left(-\frac{1}{N_{c}}N_{c}^{3}+\frac{1}{N_{c}^{2}}N_{c}^{2}\right)=-1.
\end{equation}
Now, let us look at the leading diagram for the TMD operator:

\begin{tabular}{>{\centering}m{0.87\columnwidth}>{\centering}m{0.05\columnwidth}}
\centering{}\bigskip{}
\includegraphics[width=3.5cm]{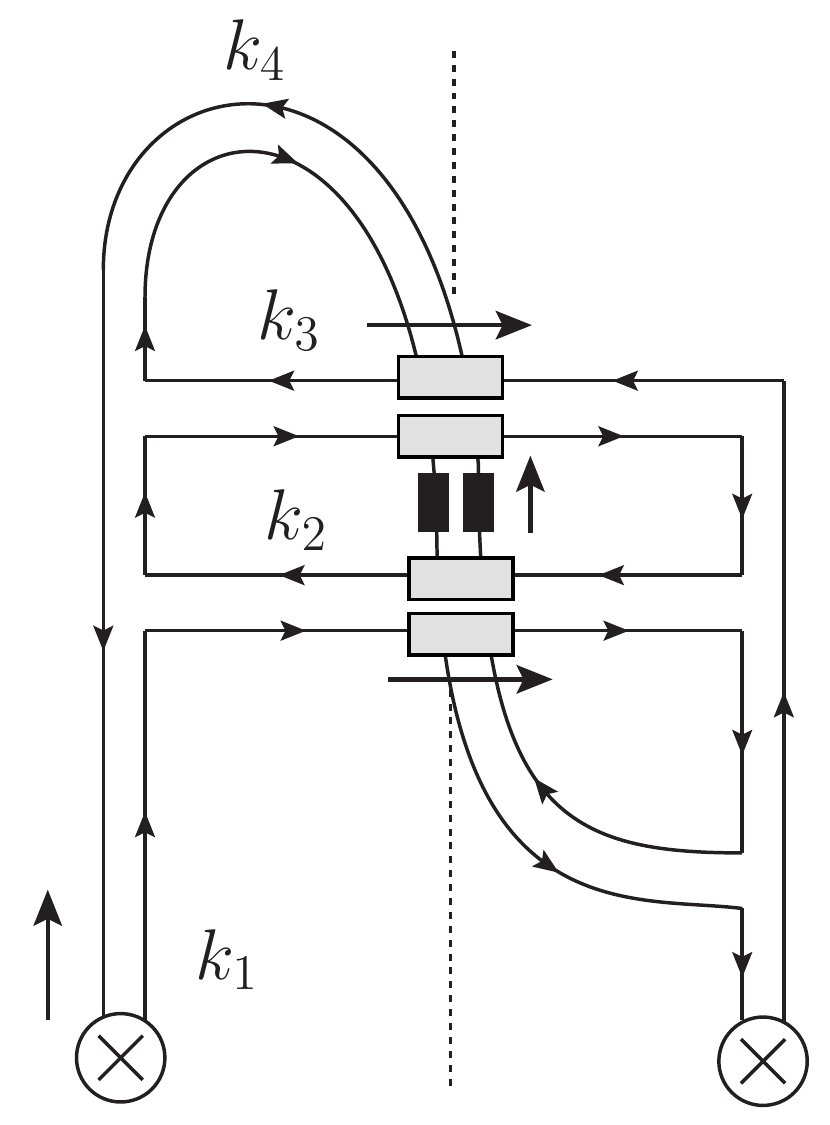} & \centering{}\centering{(\myref )}
\refmyref{largeNc_4g_TMD}\tabularnewline
\end{tabular}

\noindent It reads
\begin{equation}
N_{c}\mathrm{Tr}\left\{ F\left(\xi\right)\mathcal{U}^{\left[\square\right]}\right\} \mathrm{Tr}\left\{ F\left(0\right)\mathcal{U}^{\left[\square\right]\dagger}\right\} =N_{c}^{2}\mathcal{F}_{gg}^{\left(2\right)}\,.
\end{equation}
Dividing by the leading color factor (divided by $N_{A}$) we get
finally the following non-diagonal element of the $\mathbf{\Phi}$
matrix in the large $N_{c}$ limit: 
\begin{equation}
\left(\mathbf{\Phi}\right)_{15}=-N_{c}^{2}\mathcal{F}_{gg}^{\left(2\right)}\,.
\end{equation}
As the color factor for $\mathcal{A}\left(1,2,3,4\right)\mathcal{A}^{*}\left(1,4,3,2\right)$
is suppressed with respect to diagonal elements by $1/N_{c}^{2}$,
the TMD gluon distribution $-\mathcal{F}_{gg}^{\left(2\right)}$ indeed
contributes in the large $N_{c}$ limit. In a similar manner, but
considering much more complicated diagrams with maximal number of
crossed lines, one can deduce, that this will be always the case for
some non-diagonal elements for any multi gluon process. For example,
after a similar but tedious calculation for 5 gluon process, we find
that the dominant non-diagonal element is $-N_{c}^{4}\mathcal{F}_{gg}^{\left(2\right)}/4$.
We will always get the $\mathcal{F}_{gg}^{\left(2\right)}$ TMD gluon
distribution, because of the M\"obius-loop-like structure, which
gives the two traces appearing in the definition (\ref{eq:Fgg2}).

While perhaps it is possible to derive the answer for the non-diagonal
leading $N_{c}$ elements for any $n$, let us note that $\mathcal{F}_{gg}^{\left(2\right)}$
gives numerically rather small contribution to the cross section, compared to the other gluon distributions
\citep{VanHameren2016a,Marquet2016}. Indeed, it vanishes very
quickly with $k_{T}$, so that it is small for transverse momenta around the saturation scale. Moreover, it does not survive the collinear
limit. Therefore, in possible phenomenological studies of multigluon
production, it is safe to set
\begin{equation}
\mathbf{\Phi}_{gg\rightarrow g\dots g}=\mathbf{\Phi}_{\mathrm{diag}}\,.
\end{equation}

The study of large $N_{c}$ limit for multiparton processes with quarks,
and for gluons without the approximation described above, is left
for a separate work.

\section{Summary}

\label{sec:Summary}

In the present paper we have faced the task of calculating the TMD
gluon distributions for processes with five and six colored partons, 
following the procedure of \citep{Bomhof:2007xt}.
So far, in the literature, processes with four partons were considered.
Although it is known that within the formal TMD factorization 
the generalized factorization fails for processes
with more than two colored partons, it was argued from the CGC theory
that at small-$x$ for dilute-dense collisions exactly such structures appear.
At leading order, our results are sufficient to calculate three and
four jet production in the gluon saturation regime in dilute-dense collisions, provided the two
new basic TMD gluon distributions, $\mathcal{F}_{qg}^{\left(3\right)}$
and $\mathcal{F}_{gg}^{\left(7\right)}$, defined respectively in
(\ref{eq:Fqg3}) and (\ref{eq:Fgg7}), are determined. This can be
done using the B-JIMWLK equation, as in \citep{Marquet2016}. At tree
level, the hard matrix elements can be easily obtained from available
software for automatic calculations, for example KaTie \citep{vanHameren:2016kkz},
which can deal with on-shell and off-shell initial states.

Instead of calculating the structure of the operators for particular
Feynman diagrams, we have used the color decomposition, which is the
most efficient way of dealing with the multi-particle QCD amplitudes.
In particular, for processes with quarks, we have used the color flow
decomposition, which treats quarks and gluons on equal footing. In
addition, we formulated straightforward color flow Feynman rules for
the gauge links, which allow immediate derivation of the TMD operator
for a given color flow.

The color flow Feynman rules are particularly convenient for large
$N_{c}$ analysis. In the present work, as a first step towards this goal, 
we attacked multigluon processes with arbitrary number of legs.
We find a general answer, but in a certain approximation, motivated
by known numerical studies of a particular TMD gluon distribution.
In the large $N_{c}$ limit, we find that only two structures contribute,
for any number of legs. This is similar to the conclusion made in
\citep{Dominguez:2012ad}, where the universality at large $N_{c}$
was found in the multiparton production in the Color Glass Condensate:
only dipoles and quadrupoles contribute.

Finally, it would be very interesting to compare the TMD factorization
formulae for three jet production (by factorization we mean the approach
described in detail in the Introduction) with the leading power of
the corresponding CGC result, which was recently derived in \citep{2018arXiv180905526I}.

\section*{Acknowledgments}

The authors would like to thank Andreas van Hameren for numerous discussions.
This work was partially supported by NCN grant DEC-2017/27/B/ST2/01985.
The diagrams were drawn using the Jaxodraw \citep{Binosi:2003yf}.

\appendix

\section{Amplitude vectors $\vec{\mathcal{A}}$ and matrices $\mathbf{M}$}
\label{sec:App_Avec_M}

\begin{table}[H]
\begin{centering}
\begin{tabular}{c|c|c|c}
\toprule
$g_{1}g_{5}\rightarrow g_{2}g_{3}g_{4}$ & $g_{1}g_{5}\rightarrow q_{2}\bar{q}_{3}g_{4}$ & $g_{1}q_{5}\rightarrow g_{2}g_{3}q_{4}$  & $g_{1}\bar{q}_{5}\rightarrow g_{2}g_{3}\bar{q}_{4}$\tabularnewline
\midrule
$\left(\begin{array}{c}
\mathcal{A}(1,2,3,4,5)\\
\mathcal{A}(1,2,4,3,5)\\
\mathcal{A}(1,3,2,4,5)\\
\mathcal{A}(1,3,4,2,5)\\
\mathcal{A}(1,4,2,3,5)\\
\mathcal{A}(1,4,3,2,5)
\end{array}\right)$ & $\left(\begin{array}{c}
\mathcal{A}(2,1,4,5,3)\\
\mathcal{A}(2,1,5,4,3)\\
\mathcal{A}(2,4,1,5,3)\\
\mathcal{A}(2,4,5,1,3)\\
\mathcal{A}(2,5,1,4,3)\\
\mathcal{A}(2,5,4,1,3)
\end{array}\right)$ & $\left(\begin{array}{c}
\mathcal{A}(4,1,2,3,5)\\
\mathcal{A}(4,1,3,2,5)\\
\mathcal{A}(4,2,1,3,5)\\
\mathcal{A}(4,2,3,1,5)\\
\mathcal{A}(4,3,1,2,5)\\
\mathcal{A}(4,3,2,1,5)
\end{array}\right)$ & $\left(\begin{array}{c}
\mathcal{A}(5,1,2,3,4)\\
\mathcal{A}(5,1,3,2,4)\\
\mathcal{A}(5,2,1,3,4)\\
\mathcal{A}(5,2,3,1,4)\\
\mathcal{A}(5,3,1,2,4)\\
\mathcal{A}(5,3,2,1,4)
\end{array}\right)$\tabularnewline
\midrule
\multicolumn{2}{c|}{$g_{1}q_{5}\rightarrow q_{2}\bar{q}_{3}q_{4}$ } & \multicolumn{2}{c}{$g_{1}\bar{q}_{5}\rightarrow q_{2}\bar{q}_{3}\bar{q}_{4}$}\tabularnewline
\midrule
\multicolumn{2}{c|}{$\left(\begin{array}{c}
\mathcal{A}(2,3,4,1,5)\\
\mathcal{A}(2,1,3,4,5)\\
\mathcal{A}(2,5,4,1,3)\\
\mathcal{A}(2,1,5,4,3)
\end{array}\right)$} & \multicolumn{2}{c}{$\left(\begin{array}{c}
\mathcal{A}(2,3,5,1,4)\\
\mathcal{A}(2,1,3,5,4)\\
\mathcal{A}(2,4,5,1,3)\\
\mathcal{A}(2,1,4,5,3)
\end{array}\right)$}\tabularnewline
\bottomrule
\end{tabular}
\par\end{centering}
\caption{Definitions of the vector of partial amplitudes $\vec{\mathcal{A}}$
for all five-parton processes. The subscripts in the sub-process indication
correspond to the momenta enumeration. \label{tab:partial_amps_5part}}
\end{table}

\begin{table}[H]
\begin{centering}
\begin{tabular}{c|c}
\toprule
$g_{1}g_{5}\rightarrow g_{2}g_{3}g_{4}$  & $\left(\begin{array}{ccccccc}
\frac{1}{2N_{c}^{2}}+\frac{1}{4} & -\frac{1}{N_{c}^{2}} & -\frac{1}{N_{c}^{2}} & 0 & 0 & \frac{3}{4} & \frac{1}{2N_{c}^{2}}\\
\frac{1}{N_{c}^{2}} & -\frac{2}{N_{c}^{2}} & -\frac{1}{N_{c}^{2}} & -\frac{1}{2N_{c}^{2}} & -\frac{1}{2N_{c}^{2}} & 1 & \frac{1}{N_{c}^{2}}\\
\frac{4}{N_{c}^{2}} & -\frac{2}{N_{c}^{2}} & -\frac{1}{N_{c}^{2}} & \frac{1}{N_{c}^{2}} & -\frac{2}{N_{c}^{2}} & 1 & -\frac{2}{N_{c}^{2}}\\
N_{c} & -\frac{1}{4}N_{c}\left(N_{c}^{2}+2\right) & \frac{N_{c}}{4} & 0 & -\frac{3N_{c}}{4} & 0 & -\frac{N_{c}}{2}
\end{array}\right)$\tabularnewline
\midrule
$g_{1}g_{5}\rightarrow q_{2}\bar{q}_{3}g_{4}$  & $\left(\begin{array}{ccccccc}
-\frac{N_{c}^{2}}{N_{A}^{2}} & 0 & -\frac{1}{N_{A}} & 0 & 0 & \frac{N_{c}^{4}}{N_{A}^{2}} & 0\\
\frac{N_{c}^{2}}{N_{A}} & 0 & -\frac{1}{N_{A}} & 0 & 0 & 0 & 0\\
0 & -N_{c}^{2} & 1 & 0 & 0 & 0 & 0\\
0 & -\frac{N_{c}^{2}}{F} & \frac{1}{F} & 0 & \frac{N_{c}^{2}}{F} & 0 & 0\\
-N_{c}^{2} & 0 & 1 & N_{c}^{2} & 0 & 0 & 0\\
0 & -\frac{N_{c}^{2}}{F} & \frac{1}{F} & \frac{N_{c}^{2}}{F} & 0 & 0 & 0
\end{array}\right)$\tabularnewline
\midrule
$g_{1}q_{5}\rightarrow g_{2}g_{3}q_{4}$  & $\left(\begin{array}{ccc}
\frac{1}{N_{A}^{2}} & \frac{DN_{c}^{2}}{N_{A}^{2}} & 0\\
-\frac{F}{N_{A}} & \frac{2N_{c}^{2}}{N_{A}} & 0\\
-\frac{1}{N_{A}} & \frac{N_{c}^{2}}{N_{A}} & 0\\
1 & 0 & 0\\
1 & -N_{c}^{2} & N_{c}^{2}\\
\frac{1}{F} & 0 & \frac{N_{c}^{2}}{F}
\end{array}\right)$\tabularnewline
\midrule
$g_{1}q_{5}\rightarrow q_{2}\bar{q}_{3}q_{4}$  & $\left(\begin{array}{ccc}
1 & 0 & 0\\
0 & 1 & 0\\
0 & 0 & 1
\end{array}\right)$\tabularnewline
\bottomrule
\end{tabular}
\par\end{centering}
\caption{Matrices $\mathbf{M}$ of structures appearing in the five-parton
processes ($D=N_{c}^{2}-2$, $F=N_{c}^{2}+1$). The subscripts in
the sub-process indication correspond to the momenta enumeration.
\label{tab:Mmatrices_5part}}
\end{table}

\begin{table}[H]
\begin{centering}
\begin{tabular}{c|c|c|c}
\toprule
$g_{1}g_{6}\rightarrow g_{2}g_{3}g_{4}g_{5}$  & $g_{1}g_{6}\rightarrow q_{2}\bar{q}_{3}g_{4}g_{5}$ & $g_{1}q_{6}\rightarrow g_{2}g_{3}g_{4}q_{5}$ & $g_{1}\bar{q}_{6}\rightarrow g_{2}g_{3}g_{4}\bar{q}_{5}$\tabularnewline
\midrule
$\left(\begin{array}{c}
\mathcal{A}(1,2,3,4,5,6)\\
\mathcal{A}(1,2,3,5,4,6)\\
\mathcal{A}(1,2,4,3,5,6)\\
\mathcal{A}(1,2,4,5,3,6)\\
\mathcal{A}(1,2,5,3,4,6)\\
\mathcal{A}(1,2,5,4,3,6)\\
\mathcal{A}(1,3,2,4,5,6)\\
\mathcal{A}(1,3,2,5,4,6)\\
\mathcal{A}(1,3,4,2,5,6)\\
\mathcal{A}(1,3,4,5,2,6)\\
\mathcal{A}(1,3,5,2,4,6)\\
\mathcal{A}(1,3,5,4,2,6)\\
\mathcal{A}(1,4,2,3,5,6)\\
\mathcal{A}(1,4,2,5,3,6)\\
\mathcal{A}(1,4,3,2,5,6)\\
\mathcal{A}(1,4,3,5,2,6)\\
\mathcal{A}(1,4,5,2,3,6)\\
\mathcal{A}(1,4,5,3,2,6)\\
\mathcal{A}(1,5,2,3,4,6)\\
\mathcal{A}(1,5,2,4,3,6)\\
\mathcal{A}(1,5,3,2,4,6)\\
\mathcal{A}(1,5,3,4,2,6)\\
\mathcal{A}(1,5,4,2,3,6)\\
\mathcal{A}(1,5,4,3,2,6)
\end{array}\right)$ & $\left(\begin{array}{c}
\mathcal{A}(2,1,4,5,6,3)\\
\mathcal{A}(2,1,4,6,5,3)\\
\mathcal{A}(2,1,5,4,6,3)\\
\mathcal{A}(2,1,5,6,4,3)\\
\mathcal{A}(2,1,6,4,5,3)\\
\mathcal{A}(2,1,6,5,4,3)\\
\mathcal{A}(2,4,1,5,6,3)\\
\mathcal{A}(2,4,1,6,5,3)\\
\mathcal{A}(2,4,5,1,6,3)\\
\mathcal{A}(2,4,5,6,1,3)\\
\mathcal{A}(2,4,6,1,5,3)\\
\mathcal{A}(2,4,6,5,1,3)\\
\mathcal{A}(2,5,1,4,6,3)\\
\mathcal{A}(2,5,1,6,4,3)\\
\mathcal{A}(2,5,4,1,6,3)\\
\mathcal{A}(2,5,4,6,1,3)\\
\mathcal{A}(2,5,6,1,4,3)\\
\mathcal{A}(2,5,6,4,1,3)\\
\mathcal{A}(2,6,1,4,5,3)\\
\mathcal{A}(2,6,1,5,4,3)\\
\mathcal{A}(2,6,4,1,5,3)\\
\mathcal{A}(2,6,4,5,1,3)\\
\mathcal{A}(2,6,5,1,4,3)\\
\mathcal{A}(2,6,5,4,1,3)
\end{array}\right)$ & $\left(\begin{array}{c}
\mathcal{A}(5,1,2,3,4,6)\\
\mathcal{A}(5,1,2,4,3,6)\\
\mathcal{A}(5,1,3,2,4,6)\\
\mathcal{A}(5,1,3,4,2,6)\\
\mathcal{A}(5,1,4,2,3,6)\\
\mathcal{A}(5,1,4,3,2,6)\\
\mathcal{A}(5,2,1,3,4,6)\\
\mathcal{A}(5,2,1,4,3,6)\\
\mathcal{A}(5,2,3,1,4,6)\\
\mathcal{A}(5,2,3,4,1,6)\\
\mathcal{A}(5,2,4,1,3,6)\\
\mathcal{A}(5,2,4,3,1,6)\\
\mathcal{A}(5,3,1,2,4,6)\\
\mathcal{A}(5,3,1,4,2,6)\\
\mathcal{A}(5,3,2,1,4,6)\\
\mathcal{A}(5,3,2,4,1,6)\\
\mathcal{A}(5,3,4,1,2,6)\\
\mathcal{A}(5,3,4,2,1,6)\\
\mathcal{A}(5,4,1,2,3,6)\\
\mathcal{A}(5,4,1,3,2,6)\\
\mathcal{A}(5,4,2,1,3,6)\\
\mathcal{A}(5,4,2,3,1,6)\\
\mathcal{A}(5,4,3,1,2,6)\\
\mathcal{A}(5,4,3,2,1,6)
\end{array}\right)$ & $\left(\begin{array}{c}
\mathcal{A}(6,1,2,3,4,5)\\
\mathcal{A}(6,1,2,4,3,5)\\
\mathcal{A}(6,1,3,2,4,5)\\
\mathcal{A}(6,1,3,4,2,5)\\
\mathcal{A}(6,1,4,2,3,5)\\
\mathcal{A}(6,1,4,3,2,5)\\
\mathcal{A}(6,2,1,3,4,5)\\
\mathcal{A}(6,2,1,4,3,5)\\
\mathcal{A}(6,2,3,1,4,5)\\
\mathcal{A}(6,2,3,4,1,5)\\
\mathcal{A}(6,2,4,1,3,5)\\
\mathcal{A}(6,2,4,3,1,5)\\
\mathcal{A}(6,3,1,2,4,5)\\
\mathcal{A}(6,3,1,4,2,5)\\
\mathcal{A}(6,3,2,1,4,5)\\
\mathcal{A}(6,3,2,4,1,5)\\
\mathcal{A}(6,3,4,1,2,5)\\
\mathcal{A}(6,3,4,2,1,5)\\
\mathcal{A}(6,4,1,2,3,5)\\
\mathcal{A}(6,4,1,3,2,5)\\
\mathcal{A}(6,4,2,1,3,5)\\
\mathcal{A}(6,4,2,3,1,5)\\
\mathcal{A}(6,4,3,1,2,5)\\
\mathcal{A}(6,4,3,2,1,5)
\end{array}\right)$\tabularnewline
\midrule
$g_{1}g_{6}\rightarrow q_{2}\bar{q}_{3}q_{4}\bar{q}_{5}$  & \multicolumn{2}{c|}{$g_{1}q_{6}\rightarrow g_{2}q_{3}\bar{q}_{4}q_{5}$} & $g_{1}\bar{q}_{6}\rightarrow g_{2}q_{3}\bar{q}_{4}\bar{q}_{5}$\tabularnewline
\midrule
$\left(\begin{array}{c}
\mathcal{A}(2,3,4,1,6,5)\\
\mathcal{A}(2,1,3,4,6,5)\\
\mathcal{A}(2,1,6,3,4,5)\\
\mathcal{A}(2,3,4,6,1,5)\\
\mathcal{A}(2,6,3,4,1,5)\\
\mathcal{A}(2,6,1,3,4,5)\\
\mathcal{A}(2,5,4,1,6,3)\\
\mathcal{A}(2,1,5,4,6,3)\\
\mathcal{A}(2,1,6,5,4,3)\\
\mathcal{A}(2,5,4,6,1,3)\\
\mathcal{A}(2,6,5,4,1,3)\\
\mathcal{A}(2,6,1,5,4,3)
\end{array}\right)$ & \multicolumn{2}{c|}{$\left(\begin{array}{c}
\mathcal{A}(3,4,5,1,2,6)\\
\mathcal{A}(3,1,4,5,2,6)\\
\mathcal{A}(3,1,2,4,5,6)\\
\mathcal{A}(3,4,5,2,1,6)\\
\mathcal{A}(3,2,4,5,1,6)\\
\mathcal{A}(3,2,1,4,5,6)\\
\mathcal{A}(3,6,5,1,2,4)\\
\mathcal{A}(3,1,6,5,2,4)\\
\mathcal{A}(3,1,2,6,5,4)\\
\mathcal{A}(3,6,5,2,1,4)\\
\mathcal{A}(3,2,6,5,1,4)\\
\mathcal{A}(3,2,1,6,5,4)
\end{array}\right)$} & $\left(\begin{array}{c}
\mathcal{A}(3,4,6,1,2,5)\\
\mathcal{A}(3,1,4,6,2,5)\\
\mathcal{A}(3,1,2,4,6,5)\\
\mathcal{A}(3,4,6,2,1,5)\\
\mathcal{A}(3,2,4,6,1,5)\\
\mathcal{A}(3,2,1,4,6,5)\\
\mathcal{A}(3,5,6,1,2,4)\\
\mathcal{A}(3,1,5,6,2,4)\\
\mathcal{A}(3,1,2,5,6,4)\\
\mathcal{A}(3,5,6,2,1,4)\\
\mathcal{A}(3,2,5,6,1,4)\\
\mathcal{A}(3,2,1,5,6,4)
\end{array}\right)$\tabularnewline
\bottomrule
\end{tabular}
\par\end{centering}
\caption{Definition of the vector of partial amplitudes $\vec{\mathcal{A}}$
for all six-parton processes. The subscripts in the sub-process indication
correspond to the momenta enumeration. \label{tab:partial_amps_6part}}
\end{table}

\begin{table}[H]
\begin{centering}
\begin{tabular}{c}
\toprule
$g_{1}g_{6}\rightarrow g_{2}g_{3}g_{4}g_{5}$\tabularnewline
\midrule
 $\left(\begin{array}{ccccccc}
\frac{1}{4N_{c}^{2}}+\frac{1}{8} & 0 & -\frac{F}{N_{c}^{4}} & \frac{1}{2N_{c}^{4}} & \frac{1}{2N_{c}^{4}} & \frac{1}{2N_{c}^{2}}+\frac{7}{8} & \frac{1}{4N_{c}^{2}}\\
0 & \frac{1}{N_{c}^{2}} & -\frac{N_{c}^{2}+2}{N_{c}^{4}} & \frac{1}{N_{c}^{4}} & \frac{1}{N_{c}^{4}} & \frac{1}{N_{c}^{2}}+1 & 0\\
\frac{3}{N_{c}^{2}} & \frac{1}{N_{c}^{2}} & -\frac{N_{c}^{2}+2}{N_{c}^{4}} & \frac{1}{N_{c}^{4}} & \frac{1}{N_{c}^{4}} & \frac{1}{N_{c}^{2}}+1 & -\frac{3}{N_{c}^{2}}\\
\frac{N_{c}^{2}}{2} & \frac{3N_{c}^{2}}{4} & -\frac{1}{2} & \frac{1}{8}\left(N_{c}^{2}+2\right) & \frac{1}{8}\left(N_{c}^{2}+2\right) & \frac{N_{c}^{2}}{4} & -N_{c}^{2}\\
-\frac{1}{N_{c}^{2}} & \frac{4}{N_{c}^{2}} & -\frac{N_{c}^{2}+4}{N_{c}^{4}} & \frac{N_{c}^{2}+4}{2N_{c}^{4}} & \frac{N_{c}^{2}+4}{2N_{c}^{4}} & \frac{2}{N_{c}^{2}}+1 & -\frac{1}{N_{c}^{2}}\\
\frac{6}{N_{c}^{2}} & \frac{4}{N_{c}^{2}} & -\frac{N_{c}^{2}+4}{N_{c}^{4}} & \frac{N_{c}^{2}+2}{N_{c}^{4}} & \frac{2}{N_{c}^{4}} & \frac{2}{N_{c}^{2}}+1 & -\frac{8}{N_{c}^{2}}\\
\frac{4}{N_{c}^{2}} & \frac{4}{N_{c}^{2}} & -\frac{N_{c}^{2}+4}{N_{c}^{4}} & \frac{2}{N_{c}^{4}} & \frac{N_{c}^{2}+2}{N_{c}^{4}} & \frac{2}{N_{c}^{2}}+1 & -\frac{6}{N_{c}^{2}}\\
\frac{N_{c}^{2}}{4} & \frac{3N_{c}^{2}}{4} & -\frac{1}{2} & \frac{1}{4} & \frac{F}{4} & \frac{N_{c}^{2}}{4} & -\frac{3N_{c}^{2}}{4}\\
\frac{8}{N_{c}^{2}+12} & -\frac{2}{N_{c}^{2}+12} & -\frac{N_{c}^{2}+8}{N_{c}^{2}\left(N_{c}^{2}+12\right)} & \frac{N_{c}^{2}+4}{N_{c}^{2}\left(N_{c}^{2}+12\right)} & \frac{2\left(N_{c}^{2}+2\right)}{N_{c}^{2}\left(N_{c}^{2}+12\right)} & \frac{N_{c}^{2}+4}{N_{c}^{2}+12} & -\frac{2}{N_{c}^{2}+12}\\
\frac{1}{3} & 0 & -\frac{2}{3N_{c}^{2}} & \frac{1}{3N_{c}^{2}} & \frac{1}{3}\left(\frac{1}{N_{c}^{2}}+1\right) & \frac{1}{3} & 0\\
0 & \frac{1}{12}\left(N_{c}^{2}+2\right) & -\frac{N_{c}^{2}+8}{12N_{c}^{2}} & \frac{1}{3N_{c}^{2}} & \frac{1}{3N_{c}^{2}}+\frac{7}{12} & \frac{1}{3} & \frac{1}{6}
\end{array}\right)$\tabularnewline
\bottomrule
\end{tabular}
\par\end{centering}
\caption{Matrices $\mathbf{M}$ of structures appearing in the six-parton processes
(part I) ($F=N_{c}^{2}+1$). The subscripts in the sub-process indication
correspond to the momenta enumeration. \label{tab:Mmatrices_6part1}}
\end{table}

\begin{table}[H]
\begin{centering}
\begin{tabular}{c}
\toprule
$g_{1}g_{6}\rightarrow q_{2}\bar{q}_{3}g_{4}g_{5}$\tabularnewline
\midrule
$\left(\begin{array}{ccccccc}
\frac{N_{c}^{2}}{N_{A}^{3}} & 0 & -\frac{1}{N_{A}} & 0 & 0 & \frac{DN_{c}^{4}}{N_{A}^{3}} & 0\\
-\frac{N_{c}^{2}}{N_{A}^{2}} & 0 & -\frac{1}{N_{A}} & 0 & 0 & \frac{N_{c}^{4}}{N_{A}^{2}} & 0\\
-\frac{FN_{c}^{2}}{N_{A}^{2}} & 0 & -\frac{1}{N_{A}} & 0 & 0 & \frac{2N_{c}^{4}}{N_{A}^{2}} & 0\\
\frac{N_{c}^{2}}{N_{A}} & 0 & -\frac{1}{N_{A}} & 0 & 0 & -\frac{N_{c}^{4}}{N_{A}} & \frac{N_{c}^{4}}{N_{A}}\\
\frac{N_{c}^{2}}{N_{A}} & 0 & -\frac{1}{N_{A}} & 0 & 0 & 0 & 0\\
\frac{N_{c}^{2}}{N_{c}^{4}-1} & 0 & -\frac{1}{N_{A}} & 0 & 0 & 0 & \frac{N_{c}^{4}}{N_{c}^{4}-1}\\
0 & -N_{c}^{2} & 1 & 0 & 0 & 0 & 0\\
0 & -\frac{N_{c}^{2}}{F} & \frac{1}{F} & 0 & \frac{N_{c}^{2}}{F} & 0 & 0\\
-N_{c}^{2} & 0 & 1 & N_{c}^{2} & 0 & 0 & 0\\
0 & -\frac{N_{c}^{2}}{F} & \frac{1}{F} & \frac{N_{c}^{2}}{F} & 0 & 0 & 0\\
-N_{c}^{2} & -N_{c}^{4} & F & 0 & 0 & 0 & 0\\
\frac{N_{c}^{4}}{L} & \frac{N_{c}^{2}}{L} & -\frac{F}{L} & 0 & 0 & 0 & 0\\
0 & \frac{N_{c}^{2}}{K} & -\frac{F}{K} & 0 & -\frac{N_{c}^{2}}{K} & 0 & \frac{N_{c}^{4}}{K}\\
-FN_{c}^{2} & 0 & 1 & N_{c}^{2} & 0 & N_{c}^{4} & 0\\
-\frac{N_{c}^{2}}{F} & -\frac{N_{c}^{4}}{F} & 1 & \frac{N_{c}^{2}}{F} & 0 & 0 & 0\\
\frac{N_{c}^{4}}{K} & \frac{N_{c}^{2}}{K} & -\frac{F}{K} & -\frac{N_{c}^{2}}{K} & 0 & 0 & 0\\
0 & -\frac{FN_{c}^{2}}{3N_{c}^{2}+1} & \frac{F}{3N_{c}^{2}+1} & \frac{N_{c}^{2}}{3N_{c}^{2}+1} & \frac{N_{c}^{2}}{3N_{c}^{2}+1} & 0 & 0\\
\frac{N_{c}^{2}}{L} & 0 & -\frac{1}{L} & \frac{DN_{c}^{2}}{L} & 0 & 0 & 0\\
0 & \frac{N_{c}^{2}}{K} & -\frac{1}{K} & \frac{DN_{c}^{2}}{K} & 0 & 0 & 0\\
\frac{FN_{c}^{2}}{K} & 0 & -\frac{F}{K} & -\frac{2N_{c}^{2}}{K} & 0 & 0 & 0\\
0 & -\frac{FN_{c}^{2}}{3N_{c}^{2}+1} & \frac{F}{3N_{c}^{2}+1} & \frac{2N_{c}^{2}}{3N_{c}^{2}+1} & 0 & 0 & 0\\
0 & \frac{N_{c}^{2}}{K} & -\frac{1}{K} & 0 & \frac{DN_{c}^{2}}{K} & 0 & 0\\
0 & -\frac{FN_{c}^{2}}{3N_{c}^{2}+1} & \frac{F}{3N_{c}^{2}+1} & 0 & \frac{2N_{c}^{2}}{3N_{c}^{2}+1} & 0 & 0\\
0 & \frac{N_{c}^{2}}{L} & -\frac{F}{L} & 0 & 0 & 0 & \frac{N_{c}^{4}}{L}
\end{array}\right)$\tabularnewline
\bottomrule
\end{tabular}
\par\end{centering}
\caption{Matrices $\mathbf{M}$ of structures appearing in the six-parton processes
(part II) ($D=N_{c}^{2}-2$, $F=N_{c}^{2}+1$, $K=N_{c}^{4}-2N_{c}^{2}-1$,
$L=N_{c}^{4}-N_{c}^{2}-1$). The subscripts in the sub-process indication
correspond to the momenta enumeration. \label{tab:Mmatrices_6part2}}
\end{table}

\begin{table}[H]
\begin{centering}
\begin{tabular}{c}
\toprule
$g_{1}q_{6}\rightarrow g_{2}g_{3}g_{4}q_{5}$\tabularnewline
\midrule
$\left(\begin{array}{ccc}
-\frac{1}{N_{A}^{3}} & \frac{N_{c}^{2}\left(N_{c}^{4}-3N_{c}^{2}+3\right)}{N_{A}^{3}} & 0\\
\frac{F}{N_{A}^{2}} & \frac{N_{c}^{2}\left(N_{c}^{2}-3\right)}{N_{A}^{2}} & 0\\
\frac{K}{N_{A}} & -\frac{N_{c}^{2}\left(N_{c}^{2}-3\right)}{N_{A}} & 0\\
\frac{3N_{c}^{2}+1}{1-N_{c}^{4}} & \frac{N_{c}^{2}\left(N_{c}^{2}+3\right)}{N_{c}^{4}-1} & 0\\
\frac{1}{N_{A}^{2}} & \frac{DN_{c}^{2}}{N_{A}^{2}} & 0\\
-\frac{F}{N_{A}} & \frac{2N_{c}^{2}}{N_{A}} & 0\\
-\frac{1}{N_{A}} & \frac{N_{c}^{2}}{N_{A}} & 0\\
1 & 0 & 0\\
1 & -N_{c}^{2} & N_{c}^{2}\\
-\frac{1}{N_{A}} & \frac{2N_{c}^{2}}{N_{A}} & -\frac{N_{c}^{2}}{N_{A}}\\
F & -2N_{c}^{2} & N_{c}^{2}\\
-\frac{1}{L} & \frac{N_{c}^{2}}{L} & \frac{DN_{c}^{2}}{L}\\
1 & DN_{c}^{2} & -DN_{c}^{2}\\
1 & -\frac{2N_{c}^{2}}{F} & \frac{2N_{c}^{2}}{F}\\
1 & -\frac{N_{c}^{2}}{F} & \frac{N_{c}^{2}}{F}\\
-\frac{F}{K} & \frac{FN_{c}^{2}}{K} & -\frac{2N_{c}^{2}}{K}\\
\frac{1}{F} & 0 & \frac{N_{c}^{2}}{F}\\
\frac{1}{1-N_{c}^{4}} & \frac{N_{c}^{2}}{N_{A}} & \frac{N_{c}^{2}}{1-N_{c}^{4}}\\
\frac{1}{F} & -\frac{N_{c}^{2}}{F} & \frac{2N_{c}^{2}}{F}\\
-\frac{1}{K} & 0 & \frac{DN_{c}^{2}}{K}\\
-\frac{1}{K} & \frac{N_{c}^{2}}{K} & \frac{N_{c}^{2}\left(N_{c}^{2}-3\right)}{K}\\
\frac{F}{3N_{c}^{2}+1} & 0 & \frac{2N_{c}^{2}}{3N_{c}^{2}+1}\\
-\frac{F}{K} & \frac{N_{c}^{4}}{K} & -\frac{N_{c}^{2}}{K}
\end{array}\right)$\tabularnewline
\bottomrule
\end{tabular}
\par\end{centering}
\caption{Matrices $\mathbf{M}$ of structures appearing in the six-parton processes
(part III) ($D=N_{c}^{2}-2$, $F=N_{c}^{2}+1$, $K=N_{c}^{4}-2N_{c}^{2}-1$,
$L=N_{c}^{4}-N_{c}^{2}-1$). The subscripts in the sub-process indication
correspond to the momenta enumeration. \label{tab:Mmatrices_6part3}}
\end{table}

\begin{table}[H]
\begin{centering}
\begin{tabular}{c|c}
\toprule
$g_{1}g_{6}\rightarrow q_{2}\bar{q}_{3}q_{4}\bar{q}_{5}$  & $g_{1}q_{6}\rightarrow g_{2}q_{3}\bar{q}_{4}q_{5}$\tabularnewline
\midrule 
$\left(\begin{array}{ccccccc}
\frac{N_{c}^{2}}{N_{A}} & 0 & -\frac{1}{N_{A}} & 0 & 0 & 0 & 0\\
0 & 0 & 0 & 1 & 0 & 0 & 0\\
0 & -N_{c}^{2} & 1 & 0 & 0 & 0 & 0\\
0 & 0 & -\frac{1}{N_{A}} & 0 & 0 & \frac{N_{c}^{2}}{N_{A}} & 0\\
0 & 0 & 0 & 0 & 1 & 0 & 0\\
0 & 0 & -\frac{1}{N_{A}} & 0 & 0 & 0 & \frac{N_{c}^{2}}{N_{A}}
\end{array}\right)$ & $\left(\begin{array}{ccc}
-\frac{1}{N_{A}} & \frac{N_{c}^{2}}{N_{A}} & 0\\
0 & 0 & 1\\
1 & 0 & 0\\
0 & 1 & 0\\
\frac{1}{4N_{c}^{2}} & -\frac{1}{4N_{c}^{2}} & 0\\
0 & -\frac{1}{4N_{c}^{2}} & \frac{1}{4N_{c}^{2}}\\
0 & \frac{N_{c}^{2}}{N_{A}} & -\frac{1}{N_{A}}\\
\frac{1}{4} & -\frac{1}{4} & 0\\
0 & -\frac{1}{4} & \frac{1}{4}
\end{array}\right)$\tabularnewline
\bottomrule
\end{tabular}
\par\end{centering}
\caption{Matrices $\mathbf{M}$ of structures appearing in the six-parton processes
(part IV). The subscripts in the sub-process indication correspond
to the momenta enumeration. \label{tab:Mmatrices_6part4}}
\end{table}

\section{Block matrices for six parton processes}

\subsection{$g\left(k_{1}\right)g\left(k_{6}\right)\rightarrow g\left(k_{2}\right)g\left(k_{3}\right)g\left(k_{4}\right)g\left(k_{5}\right)$}
\label{ssec:6g_Ts}
\begin{equation}
T_{1}=\left(\begin{array}{cccccc}
\Phi_{1} & \Phi_{2} & \Phi_{2} & \Phi_{3} & \Phi_{3} & \Phi_{4}^{*}\\
\Phi_{2} & \Phi_{1} & \Phi_{3} & \Phi_{4}^{*} & \Phi_{2} & \Phi_{3}\\
\Phi_{2} & \Phi_{3} & \Phi_{1} & \Phi_{2} & \Phi_{4}^{*} & \Phi_{3}\\
\Phi_{3} & \Phi_{4}^{*} & \Phi_{2} & \Phi_{1} & \Phi_{3} & \Phi_{2}\\
\Phi_{3} & \Phi_{2} & \Phi_{4}^{*} & \Phi_{3} & \Phi_{1} & \Phi_{2}\\
\Phi_{4}^{*} & \Phi_{3} & \Phi_{3} & \Phi_{2} & \Phi_{2} & \Phi_{1}
\end{array}\right),\,\,\,\,T_{2}=\left(\begin{array}{cccccc}
\Phi_{2} & \Phi_{5} & \Phi_{3} & \Phi_{6} & \Phi_{7} & \Phi_{8}^{*}\\
\Phi_{5} & \Phi_{2} & \Phi_{7} & \Phi_{8}^{*} & \Phi_{3} & \Phi_{6}\\
\Phi_{3} & \Phi_{7} & \Phi_{4}^{*} & \Phi_{8}^{*} & \Phi_{9} & \Phi_{10}\\
\Phi_{6} & \Phi_{8}^{*} & \Phi_{8}^{*} & \Phi_{10} & \Phi_{10} & \Phi_{11}\\
\Phi_{7} & \Phi_{3} & \Phi_{9} & \Phi_{10} & \Phi_{4}^{*} & \Phi_{8}^{*}\\
\Phi_{8}^{*} & \Phi_{6} & \Phi_{10} & \Phi_{11} & \Phi_{8}^{*} & \Phi_{10}
\end{array}\right),
\label{eq:6g_T1}
\end{equation}

\begin{equation}
T_{3}=\left(\begin{array}{cccccc}
\Phi_{3} & \Phi_{7} & \Phi_{4}^{*} & \Phi_{8}^{*} & \Phi_{9} & \Phi_{10}\\
\Phi_{6} & \Phi_{8}^{*} & \Phi_{8}^{*} & \Phi_{10} & \Phi_{10} & \Phi_{11}\\
\Phi_{2} & \Phi_{5} & \Phi_{3} & \Phi_{6} & \Phi_{7} & \Phi_{8}^{*}\\
\Phi_{5} & \Phi_{2} & \Phi_{7} & \Phi_{8}^{*} & \Phi_{3} & \Phi_{6}\\
\Phi_{8}^{*} & \Phi_{6} & \Phi_{10} & \Phi_{11} & \Phi_{8}^{*} & \Phi_{10}\\
\Phi_{7} & \Phi_{3} & \Phi_{9} & \Phi_{10} & \Phi_{4}^{*} & \Phi_{8}^{*}
\end{array}\right),\,\,\,\,T_{4}=\left(\begin{array}{cccccc}
\Phi_{6} & \Phi_{8}^{*} & \Phi_{8}^{*} & \Phi_{10} & \Phi_{10} & \Phi_{11}\\
\Phi_{3} & \Phi_{7} & \Phi_{4}^{*} & \Phi_{8}^{*} & \Phi_{9} & \Phi_{10}\\
\Phi_{8}^{*} & \Phi_{6} & \Phi_{10} & \Phi_{11} & \Phi_{8}^{*} & \Phi_{10}\\
\Phi_{7} & \Phi_{3} & \Phi_{9} & \Phi_{10} & \Phi_{4}^{*} & \Phi_{8}^{*}\\
\Phi_{2} & \Phi_{5} & \Phi_{3} & \Phi_{6} & \Phi_{7} & \Phi_{8}^{*}\\
\Phi_{5} & \Phi_{2} & \Phi_{7} & \Phi_{8}^{*} & \Phi_{3} & \Phi_{6}
\end{array}\right),
\end{equation}

\begin{equation}
T_{5}=\left(\begin{array}{cccccc}
\Phi_{4}^{*} & \Phi_{8}^{*} & \Phi_{3} & \Phi_{7} & \Phi_{10} & \Phi_{9}\\
\Phi_{8}^{*} & \Phi_{10} & \Phi_{6} & \Phi_{8}^{*} & \Phi_{11} & \Phi_{10}\\
\Phi_{3} & \Phi_{6} & \Phi_{2} & \Phi_{5} & \Phi_{8}^{*} & \Phi_{7}\\
\Phi_{7} & \Phi_{8}^{*} & \Phi_{5} & \Phi_{2} & \Phi_{6} & \Phi_{3}\\
\Phi_{10} & \Phi_{11} & \Phi_{8}^{*} & \Phi_{6} & \Phi_{10} & \Phi_{8}^{*}\\
\Phi_{9} & \Phi_{10} & \Phi_{7} & \Phi_{3} & \Phi_{8}^{*} & \Phi_{4}^{*}
\end{array}\right),\,\,\,\,T_{6}=\left(\begin{array}{cccccc}
\Phi_{8}^{*} & \Phi_{10} & \Phi_{6} & \Phi_{8}^{*} & \Phi_{11} & \Phi_{10}\\
\Phi_{4}^{*} & \Phi_{8}^{*} & \Phi_{3} & \Phi_{7} & \Phi_{10} & \Phi_{9}\\
\Phi_{10} & \Phi_{11} & \Phi_{8}^{*} & \Phi_{6} & \Phi_{10} & \Phi_{8}^{*}\\
\Phi_{9} & \Phi_{10} & \Phi_{7} & \Phi_{3} & \Phi_{8}^{*} & \Phi_{4}^{*}\\
\Phi_{3} & \Phi_{6} & \Phi_{2} & \Phi_{5} & \Phi_{8}^{*} & \Phi_{7}\\
\Phi_{7} & \Phi_{8}^{*} & \Phi_{5} & \Phi_{2} & \Phi_{6} & \Phi_{3}
\end{array}\right),
\end{equation}

\begin{equation}
T_{7}=T_{2}^{M}\,.
 \label{eq:6g_T7}
\end{equation}
$T_{2}^{M}$ denotes a mirror reflection of the matrix $T_{2}$ with
respect to the anti-diagonal, which can be written as a similarity
transformation 

\begin{equation}
T_{2}^{M}=JT_{2}J\,,
\end{equation}
with
\begin{equation}
J=\left(\begin{array}{cccccc}
0 & 0 & 0 & 0 & 0 & 1\\
0 & 0 & 0 & 0 & 1 & 0\\
0 & 0 & 0 & 1 & 0 & 0\\
0 & 0 & 1 & 0 & 0 & 0\\
0 & 1 & 0 & 0 & 0 & 0\\
1 & 0 & 0 & 0 & 0 & 0
\end{array}\right)\,.
\end{equation}
The same relation holds between block matrices for color factors (Eq.~\ref{eq:6g_colorf}).
Nonetheless, for convenience we list explicitly elements of $T_{7}$
matrix:

\begin{equation}
T_{7}=\left(\begin{array}{cccccc}
\Phi_{10} & \Phi_{8}^{*} & \Phi_{11} & \Phi_{10} & \Phi_{6} & \Phi_{8}^{*}\\
\Phi_{8}^{*} & \Phi_{4}^{*} & \Phi_{10} & \Phi_{9} & \Phi_{3} & \Phi_{7}\\
\Phi_{11} & \Phi_{10} & \Phi_{10} & \Phi_{8}^{*} & \Phi_{8}^{*} & \Phi_{6}\\
\Phi_{10} & \Phi_{9} & \Phi_{8}^{*} & \Phi_{4}^{*} & \Phi_{7} & \Phi_{3}\\
\Phi_{6} & \Phi_{3} & \Phi_{8}^{*} & \Phi_{7} & \Phi_{2} & \Phi_{5}\\
\Phi_{8}^{*} & \Phi_{7} & \Phi_{6} & \Phi_{3} & \Phi_{5} & \Phi_{2}
\end{array}\right)\,.
\end{equation}

\subsection{$g\left(k_{1}\right)g\left(k_{6}\right)\rightarrow q\left(k_{2}\right)\bar{q}\left(k_{3}\right)g\left(k_{4}\right)g\left(k_{5}\right)$}

\label{ssec:gg_qqbgg_Ts}

\begin{equation}
T_{1}=\left(\begin{array}{cccccc}
\Phi_{1} & \Phi_{2} & \Phi_{3} & \Phi_{4} & \Phi_{5} & \Phi_{5}\\
\Phi_{2} & \Phi_{2} & \Phi_{4} & \Phi_{6} & \Phi_{5} & \Phi_{5}\\
\Phi_{3} & \Phi_{4} & \Phi_{1} & \Phi_{2} & \Phi_{5} & \Phi_{5}\\
\Phi_{4} & \Phi_{6} & \Phi_{2} & \Phi_{2} & \Phi_{5} & \Phi_{5}\\
\Phi_{5} & \Phi_{5} & \Phi_{5} & \Phi_{5} & \Phi_{5} & \Phi_{5}\\
\Phi_{5} & \Phi_{5} & \Phi_{5} & \Phi_{5} & \Phi_{5} & \Phi_{5}
\end{array}\right),\,\,\,\,T_{2}=\left(\begin{array}{cccccc}
\Phi_{2} & \Phi_{5} & \Phi_{5} & \Phi_{7} & \Phi_{7} & \Phi_{8}\\
\Phi_{5} & \Phi_{5} & \Phi_{9} & \Phi_{10} & \Phi_{7} & \Phi_{7}\\
\Phi_{4} & \Phi_{11} & \Phi_{5} & \Phi_{7} & \Phi_{12} & \Phi_{13}\\
\Phi_{14} & \Phi_{15} & \Phi_{15} & \Phi_{16} & \Phi_{16} & \Phi_{17}\\
\Phi_{9} & \Phi_{9} & \Phi_{18} & \Phi_{19} & \Phi_{10} & \Phi_{10}\\
\Phi_{15} & \Phi_{9} & \Phi_{20} & \Phi_{21} & \Phi_{10} & \Phi_{16}
\end{array}\right),
\label{eq:gg_qqbgg_T1}
\end{equation}

\begin{equation}
T_{3}=\left(\begin{array}{cccccc}
\Phi_{4} & \Phi_{11} & \Phi_{5} & \Phi_{7} & \Phi_{12} & \Phi_{13}\\
\Phi_{14} & \Phi_{15} & \Phi_{15} & \Phi_{16} & \Phi_{16} & \Phi_{17}\\
\Phi_{2} & \Phi_{5} & \Phi_{5} & \Phi_{7} & \Phi_{7} & \Phi_{8}\\
\Phi_{5} & \Phi_{5} & \Phi_{9} & \Phi_{10} & \Phi_{7} & \Phi_{7}\\
\Phi_{15} & \Phi_{9} & \Phi_{20} & \Phi_{21} & \Phi_{10} & \Phi_{16}\\
\Phi_{9} & \Phi_{9} & \Phi_{18} & \Phi_{19} & \Phi_{10} & \Phi_{10}
\end{array}\right),\,\,\,\,T_{4}=\left(\begin{array}{cccccc}
\Phi_{7} & \Phi_{7} & \Phi_{8} & \Phi_{22} & \Phi_{13} & \Phi_{23}\\
\Phi_{7} & \Phi_{7} & \Phi_{8} & \Phi_{8} & \Phi_{24} & \Phi_{13}\\
\Phi_{7} & \Phi_{7} & \Phi_{13} & \Phi_{23} & \Phi_{8} & \Phi_{22}\\
\Phi_{7} & \Phi_{7} & \Phi_{24} & \Phi_{13} & \Phi_{8} & \Phi_{8}\\
\Phi_{7} & \Phi_{7} & \Phi_{7} & \Phi_{7} & \Phi_{7} & \Phi_{7}\\
\Phi_{7} & \Phi_{7} & \Phi_{7} & \Phi_{7} & \Phi_{7} & \Phi_{7}
\end{array}\right),
\end{equation}

\begin{equation}
T_{5}=\left(\begin{array}{cccccc}
\Phi_{2} & \Phi_{5} & \Phi_{5} & \Phi_{7} & \Phi_{7} & \Phi_{8}\\
\Phi_{5} & \Phi_{5} & \Phi_{9} & \Phi_{10} & \Phi_{7} & \Phi_{7}\\
\Phi_{5} & \Phi_{9} & \Phi_{5} & \Phi_{7} & \Phi_{10} & \Phi_{7}\\
\Phi_{7} & \Phi_{10} & \Phi_{7} & \Phi_{5} & \Phi_{9} & \Phi_{5}\\
\Phi_{7} & \Phi_{7} & \Phi_{10} & \Phi_{9} & \Phi_{5} & \Phi_{5}\\
\Phi_{8} & \Phi_{7} & \Phi_{7} & \Phi_{5} & \Phi_{5} & \Phi_{2}
\end{array}\right),\,\,\,\,T_{6}=\left(\begin{array}{cccccc}
\Phi_{6} & \Phi_{15} & \Phi_{5} & \Phi_{7} & \Phi_{16} & \Phi_{24}\\
\Phi_{15} & \Phi_{20} & \Phi_{9} & \Phi_{10} & \Phi_{21} & \Phi_{16}\\
\Phi_{5} & \Phi_{9} & \Phi_{5} & \Phi_{7} & \Phi_{10} & \Phi_{7}\\
\Phi_{7} & \Phi_{10} & \Phi_{7} & \Phi_{5} & \Phi_{9} & \Phi_{5}\\
\Phi_{16} & \Phi_{21} & \Phi_{10} & \Phi_{9} & \Phi_{20} & \Phi_{15}\\
\Phi_{24} & \Phi_{16} & \Phi_{7} & \Phi_{5} & \Phi_{15} & \Phi_{6}
\end{array}\right),
\end{equation}

\begin{equation}
T_{7}=\left(\begin{array}{cccccc}
\Phi_{10} & \Phi_{16} & \Phi_{7} & \Phi_{8} & \Phi_{17} & \Phi_{13}\\
\Phi_{10} & \Phi_{10} & \Phi_{7} & \Phi_{7} & \Phi_{16} & \Phi_{12}\\
\Phi_{19} & \Phi_{21} & \Phi_{10} & \Phi_{7} & \Phi_{16} & \Phi_{7}\\
\Phi_{18} & \Phi_{20} & \Phi_{9} & \Phi_{5} & \Phi_{15} & \Phi_{5}\\
\Phi_{9} & \Phi_{9} & \Phi_{5} & \Phi_{5} & \Phi_{15} & \Phi_{11}\\
\Phi_{9} & \Phi_{15} & \Phi_{5} & \Phi_{2} & \Phi_{14} & \Phi_{4}
\end{array}\right),\,\,\,\,T_{8}=\left(\begin{array}{cccccc}
\Phi_{16} & \Phi_{10} & \Phi_{17} & \Phi_{13} & \Phi_{7} & \Phi_{8}\\
\Phi_{10} & \Phi_{10} & \Phi_{16} & \Phi_{12} & \Phi_{7} & \Phi_{7}\\
\Phi_{21} & \Phi_{19} & \Phi_{16} & \Phi_{7} & \Phi_{10} & \Phi_{7}\\
\Phi_{20} & \Phi_{18} & \Phi_{15} & \Phi_{5} & \Phi_{9} & \Phi_{5}\\
\Phi_{9} & \Phi_{9} & \Phi_{15} & \Phi_{11} & \Phi_{5} & \Phi_{5}\\
\Phi_{15} & \Phi_{9} & \Phi_{14} & \Phi_{4} & \Phi_{5} & \Phi_{2}
\end{array}\right),
\end{equation}

\begin{equation}
T_{8}^{\intercal}=T_{2}^{M}\,,\qquad T_{9}=\left(\begin{array}{cccccc}
\Phi_{5} & \Phi_{5} & \Phi_{5} & \Phi_{5} & \Phi_{5} & \Phi_{5}\\
\Phi_{5} & \Phi_{5} & \Phi_{5} & \Phi_{5} & \Phi_{5} & \Phi_{5}\\
\Phi_{5} & \Phi_{5} & \Phi_{2} & \Phi_{2} & \Phi_{6} & \Phi_{4}\\
\Phi_{5} & \Phi_{5} & \Phi_{2} & \Phi_{1} & \Phi_{4} & \Phi_{3}\\
\Phi_{5} & \Phi_{5} & \Phi_{6} & \Phi_{4} & \Phi_{2} & \Phi_{2}\\
\Phi_{5} & \Phi_{5} & \Phi_{4} & \Phi_{3} & \Phi_{2} & \Phi_{1}
\end{array}\right)=T_{1}^{M}\,.
\label{eq:gg_qqbgg_T9}
\end{equation}

\subsection{$g\left(k_{1}\right)q\left(k_{6}\right)\rightarrow g\left(k_{2}\right)g\left(k_{3}\right)g\left(k_{4}\right)q\left(k_{5}\right)$}
\label{ssec:gq_gggq_Ts}
\begin{equation}
T_{1}=\left(\begin{array}{cccccc}
\Phi_{1} & \Phi_{2} & \Phi_{2} & \Phi_{3} & \Phi_{3} & \Phi_{4}\\
\Phi_{2} & \Phi_{1} & \Phi_{3} & \Phi_{4} & \Phi_{2} & \Phi_{3}\\
\Phi_{2} & \Phi_{3} & \Phi_{1} & \Phi_{2} & \Phi_{4} & \Phi_{3}\\
\Phi_{3} & \Phi_{4} & \Phi_{2} & \Phi_{1} & \Phi_{3} & \Phi_{2}\\
\Phi_{3} & \Phi_{2} & \Phi_{4} & \Phi_{3} & \Phi_{1} & \Phi_{2}\\
\Phi_{4} & \Phi_{3} & \Phi_{3} & \Phi_{2} & \Phi_{2} & \Phi_{1}
\end{array}\right),\,\,\,\,T_{2}=\left(\begin{array}{cccccc}
\Phi_{5} & \Phi_{6} & \Phi_{7} & \Phi_{8} & \Phi_{9} & \Phi_{8}\\
\Phi_{6} & \Phi_{5} & \Phi_{9} & \Phi_{8} & \Phi_{7} & \Phi_{8}\\
\Phi_{10} & \Phi_{11} & \Phi_{7} & \Phi_{8} & \Phi_{12} & \Phi_{8}\\
\Phi_{13} & \Phi_{14} & \Phi_{15} & \Phi_{8} & \Phi_{16} & \Phi_{8}\\
\Phi_{11} & \Phi_{10} & \Phi_{12} & \Phi_{8} & \Phi_{7} & \Phi_{8}\\
\Phi_{14} & \Phi_{13} & \Phi_{16} & \Phi_{8} & \Phi_{15} & \Phi_{8}
\end{array}\right),
\label{eq:gq_gggq_T1}
\end{equation}

\begin{equation}
T_{3}=\left(\begin{array}{cccccc}
\Phi_{10} & \Phi_{11} & \Phi_{7} & \Phi_{8} & \Phi_{12} & \Phi_{8}\\
\Phi_{13} & \Phi_{14} & \Phi_{15} & \Phi_{8} & \Phi_{16} & \Phi_{8}\\
\Phi_{5} & \Phi_{6} & \Phi_{7} & \Phi_{8} & \Phi_{9} & \Phi_{8}\\
\Phi_{6} & \Phi_{5} & \Phi_{9} & \Phi_{8} & \Phi_{7} & \Phi_{8}\\
\Phi_{14} & \Phi_{13} & \Phi_{16} & \Phi_{8} & \Phi_{15} & \Phi_{8}\\
\Phi_{11} & \Phi_{10} & \Phi_{12} & \Phi_{8} & \Phi_{7} & \Phi_{8}
\end{array}\right),\,\,\,\,T_{4}=\left(\begin{array}{cccccc}
\Phi_{13} & \Phi_{14} & \Phi_{15} & \Phi_{8} & \Phi_{16} & \Phi_{8}\\
\Phi_{10} & \Phi_{11} & \Phi_{7} & \Phi_{8} & \Phi_{12} & \Phi_{8}\\
\Phi_{14} & \Phi_{13} & \Phi_{16} & \Phi_{8} & \Phi_{15} & \Phi_{8}\\
\Phi_{11} & \Phi_{10} & \Phi_{12} & \Phi_{8} & \Phi_{7} & \Phi_{8}\\
\Phi_{5} & \Phi_{6} & \Phi_{7} & \Phi_{8} & \Phi_{9} & \Phi_{8}\\
\Phi_{6} & \Phi_{5} & \Phi_{9} & \Phi_{8} & \Phi_{7} & \Phi_{8}
\end{array}\right),
\end{equation}

\begin{equation}
T_{5}=\left(\begin{array}{cccccc}
\Phi_{5} & \Phi_{6} & \Phi_{7} & \Phi_{8} & \Phi_{9} & \Phi_{8}\\
\Phi_{6} & \Phi_{5} & \Phi_{9} & \Phi_{8} & \Phi_{7} & \Phi_{8}\\
\Phi_{7} & \Phi_{9} & \Phi_{7} & \Phi_{8} & \Phi_{17} & \Phi_{8}\\
\Phi_{8} & \Phi_{8} & \Phi_{8} & \Phi_{8} & \Phi_{8} & \Phi_{8}\\
\Phi_{9} & \Phi_{7} & \Phi_{17} & \Phi_{8} & \Phi_{7} & \Phi_{8}\\
\Phi_{8} & \Phi_{8} & \Phi_{8} & \Phi_{8} & \Phi_{8} & \Phi_{8}
\end{array}\right),\,\,\,\,T_{6}=\left(\begin{array}{cccccc}
\Phi_{18} & \Phi_{19} & \Phi_{7} & \Phi_{8} & \Phi_{20} & \Phi_{8}\\
\Phi_{19} & \Phi_{21} & \Phi_{9} & \Phi_{8} & \Phi_{22} & \Phi_{8}\\
\Phi_{7} & \Phi_{9} & \Phi_{7} & \Phi_{8} & \Phi_{17} & \Phi_{8}\\
\Phi_{8} & \Phi_{8} & \Phi_{8} & \Phi_{8} & \Phi_{8} & \Phi_{8}\\
\Phi_{20} & \Phi_{22} & \Phi_{17} & \Phi_{8} & \Phi_{23} & \Phi_{8}\\
\Phi_{8} & \Phi_{8} & \Phi_{8} & \Phi_{8} & \Phi_{8} & \Phi_{8}
\end{array}\right),
\end{equation}

\begin{equation}
T_{7}=\left(\begin{array}{cccccc}
\Phi_{19} & \Phi_{21} & \Phi_{9} & \Phi_{8} & \Phi_{22} & \Phi_{8}\\
\Phi_{18} & \Phi_{19} & \Phi_{7} & \Phi_{8} & \Phi_{20} & \Phi_{8}\\
\Phi_{20} & \Phi_{22} & \Phi_{17} & \Phi_{8} & \Phi_{23} & \Phi_{8}\\
\Phi_{8} & \Phi_{8} & \Phi_{8} & \Phi_{8} & \Phi_{8} & \Phi_{8}\\
\Phi_{7} & \Phi_{9} & \Phi_{7} & \Phi_{8} & \Phi_{17} & \Phi_{8}\\
\Phi_{8} & \Phi_{8} & \Phi_{8} & \Phi_{8} & \Phi_{8} & \Phi_{8}
\end{array}\right),\,\,\,\,T_{8}=\left(\begin{array}{cccccc}
\Phi_{21} & \Phi_{19} & \Phi_{22} & \Phi_{8} & \Phi_{9} & \Phi_{8}\\
\Phi_{19} & \Phi_{18} & \Phi_{20} & \Phi_{8} & \Phi_{7} & \Phi_{8}\\
\Phi_{22} & \Phi_{20} & \Phi_{23} & \Phi_{8} & \Phi_{17} & \Phi_{8}\\
\Phi_{8} & \Phi_{8} & \Phi_{8} & \Phi_{8} & \Phi_{8} & \Phi_{8}\\
\Phi_{9} & \Phi_{7} & \Phi_{17} & \Phi_{8} & \Phi_{7} & \Phi_{8}\\
\Phi_{8} & \Phi_{8} & \Phi_{8} & \Phi_{8} & \Phi_{8} & \Phi_{8}
\end{array}\right).
\label{eq:gq_gggq_T8}
\end{equation}

\subsection{$g\left(k_{1}\right)\bar{q}\left(k_{6}\right)\rightarrow g\left(k_{2}\right)g\left(k_{3}\right)g\left(k_{4}\right)\bar{q}\left(k_{5}\right)$}
\label{ssec:gqb_gggqb_Ts}
\begin{equation}
T_{1}=\left(\begin{array}{cccccc}
\Phi_{8} & \Phi_{8} & \Phi_{8} & \Phi_{8} & \Phi_{8} & \Phi_{8}\\
\Phi_{8} & \Phi_{8} & \Phi_{8} & \Phi_{8} & \Phi_{8} & \Phi_{8}\\
\Phi_{8} & \Phi_{8} & \Phi_{8} & \Phi_{8} & \Phi_{8} & \Phi_{8}\\
\Phi_{8} & \Phi_{8} & \Phi_{8} & \Phi_{8} & \Phi_{8} & \Phi_{8}\\
\Phi_{8} & \Phi_{8} & \Phi_{8} & \Phi_{8} & \Phi_{8} & \Phi_{8}\\
\Phi_{8} & \Phi_{8} & \Phi_{8} & \Phi_{8} & \Phi_{8} & \Phi_{8}
\end{array}\right),\,\,\,\,T_{2}=\left(\begin{array}{cccccc}
\Phi_{7} & \Phi_{7} & \Phi_{7} & \Phi_{7} & \Phi_{7} & \Phi_{7}\\
\Phi_{7} & \Phi_{7} & \Phi_{7} & \Phi_{7} & \Phi_{7} & \Phi_{7}\\
\Phi_{7} & \Phi_{7} & \Phi_{5} & \Phi_{5} & \Phi_{18} & \Phi_{10}\\
\Phi_{7} & \Phi_{7} & \Phi_{5} & \Phi_{1} & \Phi_{10} & \Phi_{2}\\
\Phi_{7} & \Phi_{7} & \Phi_{18} & \Phi_{10} & \Phi_{5} & \Phi_{5}\\
\Phi_{7} & \Phi_{7} & \Phi_{10} & \Phi_{2} & \Phi_{5} & \Phi_{1}
\end{array}\right),
\label{eq:gqb_gggqb_T1}
\end{equation}

\begin{equation}
T_{3}=\left(\begin{array}{cccccc}
\Phi_{17} & \Phi_{17} & \Phi_{9} & \Phi_{9} & \Phi_{20} & \Phi_{12}\\
\Phi_{17} & \Phi_{23} & \Phi_{9} & \Phi_{15} & \Phi_{22} & \Phi_{16}\\
\Phi_{9} & \Phi_{9} & \Phi_{6} & \Phi_{6} & \Phi_{19} & \Phi_{11}\\
\Phi_{9} & \Phi_{15} & \Phi_{6} & \Phi_{2} & \Phi_{13} & \Phi_{3}\\
\Phi_{20} & \Phi_{22} & \Phi_{19} & \Phi_{13} & \Phi_{21} & \Phi_{14}\\
\Phi_{12} & \Phi_{16} & \Phi_{11} & \Phi_{3} & \Phi_{14} & \Phi_{4}
\end{array}\right),\,\,\,\,T_{4}=\left(\begin{array}{cccccc}
\Phi_{17} & \Phi_{23} & \Phi_{9} & \Phi_{15} & \Phi_{22} & \Phi_{16}\\
\Phi_{17} & \Phi_{17} & \Phi_{9} & \Phi_{9} & \Phi_{20} & \Phi_{12}\\
\Phi_{20} & \Phi_{22} & \Phi_{19} & \Phi_{13} & \Phi_{21} & \Phi_{14}\\
\Phi_{12} & \Phi_{16} & \Phi_{11} & \Phi_{3} & \Phi_{14} & \Phi_{4}\\
\Phi_{9} & \Phi_{9} & \Phi_{6} & \Phi_{6} & \Phi_{19} & \Phi_{11}\\
\Phi_{9} & \Phi_{15} & \Phi_{6} & \Phi_{2} & \Phi_{13} & \Phi_{3}
\end{array}\right),
\end{equation}

\begin{equation}
T_{5}=\left(\begin{array}{cccccc}
\Phi_{23} & \Phi_{17} & \Phi_{22} & \Phi_{16} & \Phi_{9} & \Phi_{15}\\
\Phi_{17} & \Phi_{17} & \Phi_{20} & \Phi_{12} & \Phi_{9} & \Phi_{9}\\
\Phi_{22} & \Phi_{20} & \Phi_{21} & \Phi_{14} & \Phi_{19} & \Phi_{13}\\
\Phi_{16} & \Phi_{12} & \Phi_{14} & \Phi_{4} & \Phi_{11} & \Phi_{3}\\
\Phi_{9} & \Phi_{9} & \Phi_{19} & \Phi_{11} & \Phi_{6} & \Phi_{6}\\
\Phi_{15} & \Phi_{9} & \Phi_{13} & \Phi_{3} & \Phi_{6} & \Phi_{2}
\end{array}\right).
\label{eq:gqb_gggqb_T5}
\end{equation}

\section{Large $N_{c}$ limit for the TMD gluon distributions}

\label{sec:App_LargeNc}

For reader's convenience we list the large $N_{c}$ expansions of
the results presented in Section~\ref{sec:Results}.

\begin{table}[H]
\begin{centering}
\begin{tabular}{c|c}
\toprule
$g_{1}g_{5}\rightarrow g_{2}g_{3}g_{4}$ & $\left(\begin{array}{ccccccc}
\frac{1}{4} & 0 & 0 & 0 & 0 & \frac{3}{4} & 0\\
0 & 0 & 0 & 0 & 0 & 1 & 0\\
0 & 0 & 0 & 0 & 0 & 1 & 0\\
0 & -\frac{1}{4}N_{c}^{3} & 0 & 0 & 0 & 0 & 0
\end{array}\right)$\tabularnewline
\midrule
$g_{1}g_{5}\rightarrow q_{2}\bar{q}_{3}g_{4}$ & $\left(\begin{array}{ccccccc}
0 & 0 & 0 & 0 & 0 & 1 & 0\\
1 & 0 & 0 & 0 & 0 & 0 & 0\\
0 & -N_{c}^{2} & 0 & 0 & 0 & 0 & 0\\
0 & -1 & 0 & 0 & 1 & 0 & 0\\
-N_{c}^{2} & 0 & 0 & N_{c}^{2} & 0 & 0 & 0\\
0 & -1 & 0 & 1 & 0 & 0 & 0
\end{array}\right)$\tabularnewline
\midrule
$g_{1}q_{5}\rightarrow g_{2}g_{3}q_{4}$ & $\left(\begin{array}{ccc}
0 & 1 & 0\\
-1 & 2 & 0\\
0 & 1 & 0\\
1 & 0 & 0\\
0 & -N_{c}^{2} & N_{c}^{2}\\
0 & 0 & 1
\end{array}\right)$\tabularnewline
\midrule
$g_{1}q_{5}\rightarrow q_{2}\bar{q}_{3}q_{4}$  & $\left(\begin{array}{ccc}
1 & 0 & 0\\
0 & 1 & 0\\
0 & 0 & 1
\end{array}\right)$\tabularnewline
\bottomrule
\end{tabular}
\par\end{centering}
\caption{The large $N_{c}$ limit of the matrices $\mathbf{M}$ from Table~\ref{tab:Mmatrices_5part}.
\label{tab:Mlimits_5part}}
\end{table}

\begin{table}[H]
\begin{centering}
\begin{tabular}{c}
\toprule
$g_{1}g_{6}\rightarrow g_{2}g_{3}g_{4}g_{5}$\tabularnewline 
\midrule
$\left(\begin{array}{ccccccc}
\frac{1}{8} & 0 & 0 & 0 & 0 & \frac{7}{8} & 0\\
0 & 0 & 0 & 0 & 0 & 1 & 0\\
0 & 0 & 0 & 0 & 0 & 1 & 0\\
\frac{N_{c}^{2}}{2} & \frac{3N_{c}^{2}}{4} & 0 & \frac{1}{8}N_{c}^{2} & \frac{1}{8}N_{c}^{2} & \frac{N_{c}^{2}}{4} & -N_{c}^{2}\\
0 & 0 & 0 & 0 & 0 & 1 & 0\\
0 & 0 & 0 & 0 & 0 & 1 & 0\\
0 & 0 & 0 & 0 & 0 & 1 & 0\\
\frac{N_{c}^{2}}{4} & \frac{3N_{c}^{2}}{4} & 0 & 0 & \frac{1}{4}N_{c}^{2} & \frac{N_{c}^{2}}{4} & -\frac{3N_{c}^{2}}{4}\\
0 & 0 & 0 & 0 & 0 & 1 & 0\\
\frac{1}{3} & 0 & 0 & 0 & \frac{1}{3} & \frac{1}{3} & 0\\
0 & \frac{1}{12}N_{c}^{2} & 0 & 0 & 0 & 0 & 0
\end{array}\right)$\tabularnewline
\bottomrule
\end{tabular}
\par\end{centering}
\caption{The large $N_{c}$ limit of the matrices $\mathbf{M}$ from Table~\ref{tab:Mmatrices_6part1}.
\label{tab:Mlimits_6part1}}
\end{table}

\begin{table}[H]
\begin{centering}
\begin{tabular}{c}
\toprule
$g_{1}g_{6}\rightarrow q_{2}\bar{q}_{3}g_{4}g_{5}$\tabularnewline
\midrule
$\left(\begin{array}{ccccccc}
0 & 0 & 0 & 0 & 0 & 1 & 0\\
0 & 0 & 0 & 0 & 0 & 1 & 0\\
-1 & 0 & 0 & 0 & 0 & 2 & 0\\
0 & 0 & 0 & 0 & 0 & -N_{c}^{2} & N_{c}^{2}\\
1 & 0 & 0 & 0 & 0 & 0 & 0\\
0 & 0 & 0 & 0 & 0 & 0 & 1\\
0 & -N_{c}^{2} & 0 & 0 & 0 & 0 & 0\\
0 & -1 & 0 & 0 & 1 & 0 & 0\\
-N_{c}^{2} & 0 & 0 & N_{c}^{2} & 0 & 0 & 0\\
0 & -1 & 0 & 1 & 0 & 0 & 0\\
0 & -N_{c}^{4} & 0 & 0 & 0 & 0 & 0\\
1 & 0 & 0 & 0 & 0 & 0 & 0\\
0 & 0 & 0 & 0 & 0 & 0 & 1\\
-N_{c}^{4} & 0 & 0 & 0 & 0 & N_{c}^{4} & 0\\
0 & -N_{c}^{2} & 0 & 0 & 0 & 0 & 0\\
1 & 0 & 0 & 0 & 0 & 0 & 0\\
0 & -\frac{1}{3}N_{c}^{2} & 0 & 0 & 0 & 0 & 0\\
0 & 0 & 0 & 1 & 0 & 0 & 0\\
0 & 0 & 0 & 1 & 0 & 0 & 0\\
1 & 0 & 0 & 0 & 0 & 0 & 0\\
0 & -\frac{1}{3}N_{c}^{2} & 0 & 0 & 0 & 0 & 0\\
0 & 0 & 0 & 0 & 1 & 0 & 0\\
0 & -\frac{1}{3}N_{c}^{2} & 0 & 0 & 0 & 0 & 0\\
0 & 0 & 0 & 0 & 0 & 0 & 1
\end{array}\right)$\tabularnewline
\bottomrule
\end{tabular}
\par\end{centering}
\caption{The large $N_{c}$ limit of the matrices $\mathbf{M}$ from Table~\ref{tab:Mmatrices_6part2}.
\label{tab:Mlimits_6part2}}
\end{table}

\begin{table}[H]
\begin{centering}
\begin{tabular}{c}
\toprule
$g_{1}q_{6}\rightarrow g_{2}g_{3}g_{4}q_{5}$\tabularnewline
\midrule
$\left(\begin{array}{ccc}
0 & 1 & 0\\
0 & 1 & 0\\
N_{c}^{2} & -N_{c}^{2} & 0\\
0 & 1 & 0\\
0 & 1 & 0\\
-1 & 2 & 0\\
0 & 1 & 0\\
1 & 0 & 0\\
0 & -N_{c}^{2} & N_{c}^{2}\\
0 & 2 & -1\\
N_{c}^{2} & -2N_{c}^{2} & N_{c}^{2}\\
0 & 0 & 1\\
0 & N_{c}^{4} & -N_{c}^{4}\\
1 & -2 & 2\\
1 & -1 & 1\\
0 & 1 & 0\\
0 & 0 & 1\\
0 & 1 & 0\\
0 & -1 & 2\\
0 & 0 & 1\\
0 & 0 & 1\\
\frac{1}{3} & 0 & \frac{2}{3}\\
0 & 1 & 0
\end{array}\right)$\tabularnewline
\bottomrule
\end{tabular}
\par\end{centering}
\caption{The large $N_{c}$ limit of the matrices $\mathbf{M}$ from Table~\ref{tab:Mmatrices_6part3}.
\label{tab:Mlimits_6part3}}
\end{table}

\begin{table}[H]
\begin{centering}
\begin{tabular}{c|c}
\toprule
$g_{1}g_{6}\rightarrow q_{2}\bar{q}_{3}q_{4}\bar{q}_{5}$ & $g_{1}q_{6}\rightarrow g_{2}q_{3}\bar{q}_{4}q_{5}$
\tabularnewline
\midrule
$\left(\begin{array}{ccccccc}
1 & 0 & 0 & 0 & 0 & 0 & 0\\
0 & 0 & 0 & 1 & 0 & 0 & 0\\
0 & -N_{c}^{2} & 0 & 0 & 0 & 0 & 0\\
0 & 0 & 0 & 0 & 0 & 1 & 0\\
0 & 0 & 0 & 0 & 1 & 0 & 0\\
0 & 0 & 0 & 0 & 0 & 0 & 1
\end{array}\right)$ & 
$\left(\begin{array}{ccc}
0 & 1 & 0\\
0 & 0 & 1\\
1 & 0 & 0\\
0 & 1 & 0\\
0 & 0 & 0\\
0 & 0 & 0\\
0 & 1 & 0\\
\frac{1}{4} & -\frac{1}{4} & 0\\
0 & -\frac{1}{4} & \frac{1}{4}
\end{array}\right)$\tabularnewline
\bottomrule
\end{tabular}
\par\end{centering}
\caption{The large $N_{c}$ limit of the matrices $\mathbf{M}$ from Table~\ref{tab:Mmatrices_6part4}.
\label{tab:Mlimits_6part4}}
\end{table}

\section{Color matrices}

\label{sec:App_Color}

Below, we list the color factors for five and six parton processes.
The convention for the enumerating of the rows and columns, \textit{i.e.} the
order of the partial amplitudes are the same as in Section~\ref{sec:Results}.
These color factors agree with \citep{Kuijf:1991kn},\citep{DelDuca1999},
after a suitable permutation of partial amplitudes is done.

Let us remind, that the actual color factors to be used in factorization formula together with the TMD matrices, are defined in Eq.~(\ref{eq:Cprime}). That is, the zero matrix elements have to be replaced by one.

\subsection*{$\emptyset\rightarrow ggggg$ }

\begin{equation}
\mathbf{C}=\frac{1}{4}N_{c}^{3}N_{A}\left(\begin{array}{cccccc}
1 & \frac{1}{2} & \frac{1}{2} & \frac{1}{4} & \frac{1}{4} & 0\\
\frac{1}{2} & 1 & \frac{1}{4} & 0 & \frac{1}{2} & \frac{1}{4}\\
\frac{1}{2} & \frac{1}{4} & 1 & \frac{1}{2} & 0 & \frac{1}{4}\\
\frac{1}{4} & 0 & \frac{1}{2} & 1 & \frac{1}{4} & \frac{1}{2}\\
\frac{1}{4} & \frac{1}{2} & 0 & \frac{1}{4} & 1 & \frac{1}{2}\\
0 & \frac{1}{4} & \frac{1}{4} & \frac{1}{2} & \frac{1}{2} & 1
\end{array}\right)\,
\end{equation}

\subsection*{$\emptyset\rightarrow q\bar{q}ggg$}

\begin{equation}
\mathbf{C}=\frac{1}{8}\frac{N_{A}}{N_{c}^{2}}\left(\begin{array}{cccccc}
N_{A}^{2} & -N_{A} & -N_{A} & 1 & 1 & F\\
-N_{A} & N_{A}^{2} & 1 & F & -N_{A} & 1\\
-N_{A} & 1 & N_{A}^{2} & -N_{A} & F & 1\\
1 & F & -N_{A} & N_{A}^{2} & 1 & -N_{A}\\
1 & -N_{A} & F & 1 & N_{A}^{2} & -N_{A}\\
F & 1 & 1 & -N_{A} & -N_{A} & N_{A}^{2}
\end{array}\right)\,,
\end{equation}
with $F=N_{c}^{2}+1$.

\subsection*{$\emptyset\rightarrow q\bar{q}r\bar{r}g$ }

\begin{equation}
\mathbf{C}=\frac{1}{2}N_{A}\left(\begin{array}{cccc}
\frac{1}{N_{c}} & 0 & -\frac{1}{N_{c}} & -\frac{1}{N_{c}}\\
0 & \frac{1}{N_{c}} & -\frac{1}{N_{c}} & -\frac{1}{N_{c}}\\
-\frac{1}{N_{c}} & -\frac{1}{N_{c}} & N_{c} & 0\\
-\frac{1}{N_{c}} & -\frac{1}{N_{c}} & 0 & N_{c}
\end{array}\right)\,.
\end{equation}

\subsection*{$\emptyset\rightarrow gggggg$ }

\begin{equation}
\mathbf{C}=\frac{1}{4}N_{c}^{4}N_{A}\left(\begin{array}{cccc}
C_{1} & C_{2} & C_{3} & C_{4}\\
C_{2} & C_{1} & C_{5} & C_{6}\\
C_{3}^{\intercal} & C_{5} & C_{1} & C_{7}\\
C_{4}^{\intercal} & C_{6}^{\intercal} & C_{7} & C_{1}
\end{array}\right)\,,
\label{eq:6g_colorf}
\end{equation}
where

\begin{equation}
C_{1}=\left(\begin{array}{cccccc}
1 & \frac{1}{2} & \frac{1}{2} & \frac{1}{4} & \frac{1}{4} & 0\\
\frac{1}{2} & 1 & \frac{1}{4} & 0 & \frac{1}{2} & \frac{1}{4}\\
\frac{1}{2} & \frac{1}{4} & 1 & \frac{1}{2} & 0 & \frac{1}{4}\\
\frac{1}{4} & 0 & \frac{1}{2} & 1 & \frac{1}{4} & \frac{1}{2}\\
\frac{1}{4} & \frac{1}{2} & 0 & \frac{1}{4} & 1 & \frac{1}{2}\\
0 & \frac{1}{4} & \frac{1}{4} & \frac{1}{2} & \frac{1}{2} & 1
\end{array}\right)\,,\qquad C_{2}=\left(\begin{array}{cccccc}
\frac{1}{2} & \frac{1}{4} & \frac{1}{4} & \frac{1}{8} & \frac{1}{8} & 0\\
\frac{1}{4} & \frac{1}{2} & \frac{1}{8} & 0 & \frac{1}{4} & \frac{1}{8}\\
\frac{1}{4} & \frac{1}{8} & 0 & 0 & a+\frac{1}{8} & a\\
\frac{1}{8} & 0 & 0 & a & a & a\\
\frac{1}{8} & \frac{1}{4} & a+\frac{1}{8} & a & 0 & 0\\
0 & \frac{1}{8} & a & a & 0 & a
\end{array}\right)\,,
\end{equation}

\begin{equation}
C_{3}=\left(\begin{array}{cccccc}
\frac{1}{4} & \frac{1}{8} & 0 & 0 & a+\frac{1}{8} & a\\
\frac{1}{8} & 0 & 0 & a & a & a\\
\frac{1}{2} & \frac{1}{4} & \frac{1}{4} & \frac{1}{8} & \frac{1}{8} & 0\\
\frac{1}{4} & \frac{1}{2} & \frac{1}{8} & 0 & \frac{1}{4} & \frac{1}{8}\\
0 & \frac{1}{8} & a & a & 0 & a\\
\frac{1}{8} & \frac{1}{4} & a+\frac{1}{8} & a & 0 & 0
\end{array}\right)\,,\qquad C_{4}=\left(\begin{array}{cccccc}
\frac{1}{8} & 0 & 0 & a & a & a\\
\frac{1}{4} & \frac{1}{8} & 0 & 0 & a+\frac{1}{8} & a\\
0 & \frac{1}{8} & a & a & 0 & a\\
\frac{1}{8} & \frac{1}{4} & a+\frac{1}{8} & a & 0 & 0\\
\frac{1}{2} & \frac{1}{4} & \frac{1}{4} & \frac{1}{8} & \frac{1}{8} & 0\\
\frac{1}{4} & \frac{1}{2} & \frac{1}{8} & 0 & \frac{1}{4} & \frac{1}{8}
\end{array}\right)\,,
\end{equation}

\begin{equation}
C_{5}=\left(\begin{array}{cccccc}
0 & 0 & \frac{1}{4} & \frac{1}{8} & a & a+\frac{1}{8}\\
0 & a & \frac{1}{8} & 0 & a & a\\
\frac{1}{4} & \frac{1}{8} & \frac{1}{2} & \frac{1}{4} & 0 & \frac{1}{8}\\
\frac{1}{8} & 0 & \frac{1}{4} & \frac{1}{2} & \frac{1}{8} & \frac{1}{4}\\
a & a & 0 & \frac{1}{8} & a & 0\\
a+\frac{1}{8} & a & \frac{1}{8} & \frac{1}{4} & 0 & 0
\end{array}\right)\,,\qquad C_{6}=\left(\begin{array}{cccccc}
0 & a & \frac{1}{8} & 0 & a & a\\
0 & 0 & \frac{1}{4} & \frac{1}{8} & a & a+\frac{1}{8}\\
a & a & 0 & \frac{1}{8} & a & 0\\
a+\frac{1}{8} & a & \frac{1}{8} & \frac{1}{4} & 0 & 0\\
\frac{1}{4} & \frac{1}{8} & \frac{1}{2} & \frac{1}{4} & 0 & \frac{1}{8}\\
\frac{1}{8} & 0 & \frac{1}{4} & \frac{1}{2} & \frac{1}{8} & \frac{1}{4}
\end{array}\right)\,,
\end{equation}

\begin{equation}
C_{7}=\left(\begin{array}{cccccc}
a & 0 & a & a & \frac{1}{8} & 0\\
0 & 0 & a & a+\frac{1}{8} & \frac{1}{4} & \frac{1}{8}\\
a & a & a & 0 & 0 & \frac{1}{8}\\
a & a+\frac{1}{8} & 0 & 0 & \frac{1}{8} & \frac{1}{4}\\
\frac{1}{8} & \frac{1}{4} & 0 & \frac{1}{8} & \frac{1}{2} & \frac{1}{4}\\
0 & \frac{1}{8} & \frac{1}{8} & \frac{1}{4} & \frac{1}{4} & \frac{1}{2}
\end{array}\right)\,,
\end{equation}
with $a=\frac{3}{2N_{c}^{2}}$.

\subsection*{$\emptyset\rightarrow q\bar{q}gggg$ }

\begin{equation}
\mathbf{C}=\frac{1}{16}\frac{N_{A}}{N_{c}^{3}}\left(\begin{array}{cccc}
C_{1} & C_{2} & C_{3} & C_{4}\\
C_{2} & C_{1} & C_{5} & C_{6}\\
C_{3}^{\intercal} & C_{5} & C_{1} & C_{7}\\
C_{4}^{\intercal} & C_{6}^{\intercal} & C_{7} & C_{1}
\end{array}\right)\,,
\end{equation}
where

\begin{equation}
C_{1}=\left(\begin{array}{cccccc}
N_{A}^{3} & -N_{A}^{2} & -N_{A}^{2} & N_{A} & N_{A} & N_{c}^{4}-1\\
-N_{A}^{2} & N_{A}^{3} & N_{A} & N_{c}^{4}-1 & -N_{A}^{2} & N_{A}\\
-N_{A}^{2} & N_{A} & N_{A}^{3} & -N_{A}^{2} & N_{c}^{4}-1 & N_{A}\\
N_{A} & N_{c}^{4}-1 & -N_{A}^{2} & N_{A}^{3} & N_{A} & -N_{A}^{2}\\
N_{A} & -N_{A}^{2} & N_{c}^{4}-1 & N_{A} & N_{A}^{3} & -N_{A}^{2}\\
N_{c}^{4}-1 & N_{A} & N_{A} & -N_{A}^{2} & -N_{A}^{2} & N_{A}^{3}
\end{array}\right)\,,
\end{equation}

\begin{equation}
C_{2}=\left(\begin{array}{cccccc}
-N_{A}^{2} & N_{A} & N_{A} & -1 & -1 & -N_{c}^{2}-1\\
N_{A} & -N_{A}^{2} & -1 & -N_{c}^{2}-1 & N_{A} & -1\\
N_{A} & -1 & N_{c}^{4}-1 & -N_{c}^{2}-1 & L & K\\
-1 & -N_{c}^{2}-1 & -N_{c}^{2}-1 & K & K & -3N_{c}^{2}-1\\
-1 & N_{A} & L & K & N_{c}^{4}-1 & -N_{c}^{2}-1\\
-N_{c}^{2}-1 & -1 & K & -3N_{c}^{2}-1 & -N_{c}^{2}-1 & K
\end{array}\right)\,,
\end{equation}

\begin{equation}
C_{3}=\left(\begin{array}{cccccc}
N_{A} & -1 & N_{c}^{4}-1 & -N_{c}^{2}-1 & L & K\\
-1 & -N_{c}^{2}-1 & -N_{c}^{2}-1 & K & K & -3N_{c}^{2}-1\\
-N_{A}^{2} & N_{A} & N_{A} & -1 & -1 & -N_{c}^{2}-1\\
N_{A} & -N_{A}^{2} & -1 & -N_{c}^{2}-1 & N_{A} & -1\\
-N_{c}^{2}-1 & -1 & K & -3N_{c}^{2}-1 & -N_{c}^{2}-1 & K\\
-1 & N_{A} & L & K & N_{c}^{4}-1 & -N_{c}^{2}-1
\end{array}\right)\,,
\end{equation}

\begin{equation}
C_{4}=\left(\begin{array}{cccccc}
-1 & -N_{c}^{2}-1 & -N_{c}^{2}-1 & K & K & -3N_{c}^{2}-1\\
N_{A} & -1 & N_{c}^{4}-1 & -N_{c}^{2}-1 & L & K\\
-N_{c}^{2}-1 & -1 & K & -3N_{c}^{2}-1 & -N_{c}^{2}-1 & K\\
-1 & N_{A} & L & K & N_{c}^{4}-1 & -N_{c}^{2}-1\\
-N_{A}^{2} & N_{A} & N_{A} & -1 & -1 & -N_{c}^{2}-1\\
N_{A} & -N_{A}^{2} & -1 & -N_{c}^{2}-1 & N_{A} & -1
\end{array}\right)\,,
\end{equation}

\begin{equation}
C_{5}=\left(\begin{array}{cccccc}
N_{c}^{4}-1 & -N_{c}^{2}-1 & N_{A} & -1 & K & L\\
-N_{c}^{2}-1 & K & -1 & -N_{c}^{2}-1 & -3N_{c}^{2}-1 & K\\
N_{A} & -1 & -N_{A}^{2} & N_{A} & -N_{c}^{2}-1 & -1\\
-1 & -N_{c}^{2}-1 & N_{A} & -N_{A}^{2} & -1 & N_{A}\\
K & -3N_{c}^{2}-1 & -N_{c}^{2}-1 & -1 & K & -N_{c}^{2}-1\\
L & K & -1 & N_{A} & -N_{c}^{2}-1 & N_{c}^{4}-1
\end{array}\right)\,,
\end{equation}

\begin{equation}
C_{6}=\left(\begin{array}{cccccc}
-N_{c}^{2}-1 & K & -1 & -N_{c}^{2}-1 & -3N_{c}^{2}-1 & K\\
N_{c}^{4}-1 & -N_{c}^{2}-1 & N_{A} & -1 & K & L\\
K & -3N_{c}^{2}-1 & -N_{c}^{2}-1 & -1 & K & -N_{c}^{2}-1\\
L & K & -1 & N_{A} & -N_{c}^{2}-1 & N_{c}^{4}-1\\
N_{A} & -1 & -N_{A}^{2} & N_{A} & -N_{c}^{2}-1 & -1\\
-1 & -N_{c}^{2}-1 & N_{A} & -N_{A}^{2} & -1 & N_{A}
\end{array}\right)\,,
\end{equation}

\begin{equation}
C_{7}=\left(\begin{array}{cccccc}
K & -N_{c}^{2}-1 & -3N_{c}^{2}-1 & K & -1 & -N_{c}^{2}-1\\
-N_{c}^{2}-1 & N_{c}^{4}-1 & K & L & N_{A} & -1\\
-3N_{c}^{2}-1 & K & K & -N_{c}^{2}-1 & -N_{c}^{2}-1 & -1\\
K & L & -N_{c}^{2}-1 & N_{c}^{4}-1 & -1 & N_{A}\\
-1 & N_{A} & -N_{c}^{2}-1 & -1 & -N_{A}^{2} & N_{A}\\
-N_{c}^{2}-1 & -1 & -1 & N_{A} & N_{A} & -N_{A}^{2}
\end{array}\right)\,,
\end{equation}

We find the following symmetry transformations 

\begin{equation}
C_{3}=\left(C_{6}^{M}\right)^{\intercal} \,,\qquad C_{7}=C_{2}^{M}\,,
\end{equation}
where 

\begin{equation}
A^{M}=JAJ\,,
\end{equation}
with

\begin{equation}
J=\left(\begin{array}{cccccc}
0 & 0 & 0 & 0 & 0 & 1\\
0 & 0 & 0 & 0 & 1 & 0\\
0 & 0 & 0 & 1 & 0 & 0\\
0 & 0 & 1 & 0 & 0 & 0\\
0 & 1 & 0 & 0 & 0 & 0\\
1 & 0 & 0 & 0 & 0 & 0
\end{array}\right)\,.
\end{equation}

\subsection*{$\emptyset\rightarrow q\bar{q}r\bar{r}gg$ }

\begin{equation}
\mathbf{C}=\frac{1}{4}N_{A}\left(\begin{array}{cc}
C_{1} & C_{2}\\
C_{2} & C_{3}
\end{array}\right)\,,
\end{equation}
where

\begin{equation}
C_{1}=\left(\begin{array}{cccccc}
\frac{N_{A}}{N_{c}^{2}} & 0 & \frac{1}{N_{c}^{2}} & -\frac{1}{N_{c}^{2}} & 0 & \frac{1}{N_{c}^{2}}\\
0 & \frac{N_{A}}{N_{c}^{2}} & 0 & 0 & \frac{1}{N_{c}^{2}} & 0\\
\frac{1}{N_{c}^{2}} & 0 & \frac{N_{A}}{N_{c}^{2}} & \frac{1}{N_{c}^{2}} & 0 & -\frac{1}{N_{c}^{2}}\\
-\frac{1}{N_{c}^{2}} & 0 & \frac{1}{N_{c}^{2}} & \frac{N_{A}}{N_{c}^{2}} & 0 & \frac{1}{N_{c}^{2}}\\
0 & \frac{1}{N_{c}^{2}} & 0 & 0 & \frac{N_{A}}{N_{c}^{2}} & 0\\
\frac{1}{N_{c}^{2}} & 0 & -\frac{1}{N_{c}^{2}} & \frac{1}{N_{c}^{2}} & 0 & \frac{N_{A}}{N_{c}^{2}}
\end{array}\right)\,,
\end{equation}
\[
C_{2}=\left(\begin{array}{cccccc}
-\frac{N_{A}}{N_{c}^{2}} & \frac{1}{N_{c}^{2}} & -\frac{N_{A}}{N_{c}^{2}} & \frac{1}{N_{c}^{2}} & -\frac{N_{A}}{N_{c}^{2}} & \frac{1}{N_{c}^{2}}\\
\frac{1}{N_{c}^{2}} & -\frac{N_{A}}{N_{c}^{2}} & -\frac{N_{A}}{N_{c}^{2}} & -\frac{N_{A}}{N_{c}^{2}} & -\frac{N_{A}}{N_{c}^{2}} & \frac{1}{N_{c}^{2}}\\
-\frac{N_{A}}{N_{c}^{2}} & -\frac{N_{A}}{N_{c}^{2}} & -\frac{N_{A}}{N_{c}^{2}} & \frac{1}{N_{c}^{2}} & \frac{1}{N_{c}^{2}} & \frac{1}{N_{c}^{2}}\\
\frac{1}{N_{c}^{2}} & -\frac{N_{A}}{N_{c}^{2}} & \frac{1}{N_{c}^{2}} & -\frac{N_{A}}{N_{c}^{2}} & \frac{1}{N_{c}^{2}} & -\frac{N_{A}}{N_{c}^{2}}\\
-\frac{N_{A}}{N_{c}^{2}} & -\frac{N_{A}}{N_{c}^{2}} & \frac{1}{N_{c}^{2}} & \frac{1}{N_{c}^{2}} & -\frac{N_{A}}{N_{c}^{2}} & -\frac{N_{A}}{N_{c}^{2}}\\
\frac{1}{N_{c}^{2}} & \frac{1}{N_{c}^{2}} & \frac{1}{N_{c}^{2}} & -\frac{N_{A}}{N_{c}^{2}} & -\frac{N_{A}}{N_{c}^{2}} & -\frac{N_{A}}{N_{c}^{2}}
\end{array}\right)\,,
\]

\begin{equation}
C_{3}=N_{c}^{2}C_{1}\,.
\end{equation}




\bibliographystyle{spphys}       
\bibliography{references}   


\end{document}